\documentclass[useAMS,fleqn,usenatbib]{mnras}

\usepackage{txfonts}

\usepackage[T1]{fontenc}
\usepackage{ae,aecompl}

\usepackage{graphicx}	
\usepackage{amssymb}	

\title[Apodized Kepler Periodogram]{An Apodized Kepler Periodogram for Separating Planetary and Stellar Activity Signals}
\author[P. C. Gregory]{Philip C. Gregory$^{1}$\thanks{E-mail:
gregory@phas.ubc.ca}
\\
$^{1}$Physics and Astronomy Department, University of British Columbia, 6224 Agricultural Rd., Vancouver, BC V6T 1Z1, Canada}

\date{Accepted XXX. Received YYY; in original form ZZZ}

\pubyear{2015}

\begin{document}
\date{MNRAS acceptance 13 Jan. 2016}
\label{firstpage}
\pagerange{\pageref{firstpage}--\pageref{lastpage}}
\maketitle

\begin{abstract}
A new apodized Keplerian (AK) model is proposed for the analysis of precision radial velocity (RV) data to model both planetary and stellar activity (SA) induced RV signals. A symmetrical Gaussian apodization function with unknown width and center can distinguish planetary signals from SA signals on the basis of the span of the apodization window. The general model for $m$ apodized Keplerian signals includes a linear regression term between RV and the stellar activity diagnostic $\log(R'hk)$, as well as an extra Gaussian noise term with unknown standard deviation. The model parameters are explored using a Bayesian fusion MCMC code. A differential version of the Generalized Lomb-Scargle periodogram that employs a control diagnostic provides an additional way of distinguishing SA signals and helps guide the choice of new periods. Results are reported for a recent international RV blind challenge which included multiple state of the art simulated data sets supported by a variety of stellar activity diagnostics. In the current implementation, the AK method achieved a reduction in SA noise by a factor of approximately 6. Final parameter estimates for the planetary candidates are derived from fits that include AK signals to model the SA components and simple Keplerians to model the planetary candidates. Preliminary results are also reported for AK models augmented by a moving average component that allows for correlations in the residuals. 
\end{abstract}


\section{Introduction}

At the ``Towards Other Earths II'' meeting held in Porto Portugal in September 2014, one theme emerged that bares directly on the subject of this paper. Intrinsic stellar activity (SA) has become the main limiting factor for planet searches in both transit and radial velocity (RV) data. SA includes stellar oscillations, granulation, spots and faculae/plages and long-term magnetic activity cycles (see, e.g., \cite{Saar1997}, \cite{Santos2010}, \cite{Boisse2011}; \cite{Dumusque2011a}). Observational strategy \citep{Dumusque2011b} can mitigate or average out some of these activity-induced signal. The presence of active regions on the stellar surface can show periodicities and amplitudes similar to those induced by real planetary signals. Indeed, these signals can mimic planetary signals (e.g., \cite{Figueira2010}, \cite{Santos2014}, \cite{Haywood2014}, \cite{Robertson2015}). 

	Currently the best RV precision is at the m/s level but new spectrographs are under development like ESPRESSO and EXPRES that aim to improve RV precision by a factor of approximately 10 over the current best spectrographs, HARPS and HARPS-N. Clearly, the success of these developments hinges on our ability to distinguish true planetary signals from SA induced signals. At the same time good progress has been made in simulating stellar activity signals. At the Porto meeting Xavier Dumusque challenged the community to a large scale blind test using simulated RV data at the 0.7 m/s level of precision, to understand the limitations of present solutions to deal with stellar signals and to select the best approach. The RV challenge data~\footnote{Details of the RV challenge data sets and results (Dumusque in preparation 2016) are available from: http://rv-challenge.wikispaces.com/} were generated using a modification of the SOAP code, which is described in \cite{Boisse2012} and \cite{Dumusque2014}. 

This paper describes a new approach to RV analysis using apodized Keplerian models to distinguish between planetary and SA induced RV signals. The methodology is developed in the next section using the initial Challenge test data set for which the signals present were known in advance. A summary of the method is given in Section~\ref{sec:Summary}. This is followed by our analysis and results for the first five RV Challenge data sets.

\section{Methodology}
\label{sec:analysis}

\subsection{Apodized Keplerian models}
\label{sec:AKM}

The first step in the new approach employs apodized Keplerian (AK) models in which the semi-amplitude of the Kepler RV model is multiplied by a symmetrical Gaussian of unknown width $\tau$ and with an unknown center of the apodizing window $t_a$. The same model is used to fit both the planetary and SA induced RV signals. Since a true planetary signal spans the duration of the data the apodization time,  $\tau$, will be large while SA induced signals generally vary on shorter time scales. One exception to this is the rapidly rotating (P = 0.59 d) active M4 dwarf GJ 1243 which exhibits a star spot signal that remains stable over the entire baseline of the Kepler mission \citep{Davenport2015}. As an additional check on the validity of planetary candidate signals, we employed a control diagnostic periodogram which is explained in Section~\ref{sec:ControlPeriodogram}.

In addition to the RV measurements, each of the 15 challenge data sets includes simultaneous observation of three stellar activity diagnostics. Two of these come from additional information on the spectral line shape that are extracted from the cross correlation function (CCF), the average shape of all spectral lines of the star. These two shape parameters are the CCF width (FWHM) and  bisector span (BIS). The third diagnostic $\log(R'hk)$ is based the Ca II H \& K line flux that is sensitive to active regions on the stellar surface. A preliminary analysis of the first 5 data sets plus test data indicated a strong correlation between RV and the $\log(R'hk)$ diagnostic and slightly reduced correlation with the FWHM diagnostic. For the data sets analyzed in this study, the correlation with the BIS diagnostic was greatly reduced and not considered further.

The general model for $m$ apodized Keplerian signals also included a linear regression term~\footnote{Our treatment of this regression is only approximate because we only have measured values of the $\log(R'hk)$ diagnostic not the true values. In general, the effect of the measurement errors is to reduce any correlation. A better approach would be to treat the true values of the diagnostic as hidden parameters that can be marginalized in a Bayesian hierarchical analysis. The virtues of the current approximate treatment are simplicity and speed.} between RV and the SA diagnostic $\log(R'hk)$  as well as an extra Gaussian noise term with unknown standard deviation $s$. The correlation term is expected to be particularly useful in removing SA signals associated with the star's long-term magnetic cycle (Dumusque2011c). It is also expected to partially remove SA associated with the rotation of active regions (spots and plages).  

For an $m$ signal apodized Kepler model the predicted radial velocity is given by
\begin{eqnarray}
y(t_i) & = & V +\sum_{j=1}^m \Big[ K_j \exp[-\frac{(t_i - t_{ja})^2}{2 \tau_j^2}] \times \nonumber\\*
&  &  (\cos\{\theta(t_i+\chi P_j)+\omega_j\} + e_j \cos \omega_j) \Big] \nonumber\\*
& + & \beta \ rhk(t_i),
\label{eq:orbit1}
\end{eqnarray}
where $rhk \equiv \log(R'hk)$. The model also includes an extra Gaussian white noise term with unknown standard deviation $s$ which can account for stellar jitter.
The equation involves $7 \times m + 3$  unknown parameters
\begin{itemize}
\item[] $V =$ a constant velocity.
\item[] $\beta =$ the $\log(R'hk)$ correlation parameter.
\item[] $K_j =$ velocity semi-amplitude for $j^{\rm th}$ signal. 
\item[] $\tau_j =$ apodization time constant for $j^{\rm th}$ signal. 
\item[] $ t_{ja}=$ center of apodizing window for $j^{\rm th}$ signal. 
\item[] $P_j =$ the period for the $j^{\rm th}$ signal.
\item[] $e_j =$ the eccentricity for the $j^{\rm th}$ signal.
\item[] $\omega_j =$ the longitude of periastron for $j^{\rm th}$ signal.
\item[] $\chi_j =$ the fraction of the period, prior to the start of data taking, that periastron occurred at. For a planetary signal, $\chi_j P_j =$ the number of days prior to $t_i = 0$ that the star was at periastron, for an orbital period of $P_j$ days. The time $t_i = 0$ corresponds to the unweighted average of the observation times. 
\item[] $s =$ standard deviation of extra Gaussian noise term.
\end{itemize}
The item $\theta(t_i+\chi_j P_j) =$ the true anomaly, the angle of the star in its orbit relative to periastron at time $t_i$.

As described in more detail in \citealt{Gregory2011}, we employed a re-parameterization of $\chi$ and $\omega$ to improve the Markov chain Monte Carlo (MCMC) convergence speed motivated by the work of Ford (2006). The two new parameters are $\psi=2\pi\chi+\omega$ and $\phi=2 \pi\chi-\omega$. Parameter $\psi$ is well determined for all eccentricities. Although $\phi$ is not well determined for low eccentricities, it is at least orthogonal to the $\psi$ parameter. We use a uniform prior for $\psi$ in the interval 0 to $4 \pi$ and uniform prior for $\phi$ in the interval $-2 \pi$ to $+2 \pi$. This insures that a prior that is wraparound continuous in $(\chi,\omega)$, maps into a wraparound continuous distribution in $(\psi,\phi)$. To account for the Jacobian of this re-parameterization it is necessary to multiply the Bayesian integrals by a factor of $(4 \pi)^{-np}$, where $np =$ the number of periods in the model.

The AK models were explored using an automated fusion MCMC algorithm (FMCMC), a general purpose tool for nonlinear model fitting and regression analysis (\citealt{Gregory2011, Gregory2013}). FMCMC is a special version of the Metropolis MCMC algorithm that incorporates parallel tempering, genetic crossover operations, and an automatic simulated annealing. Each of these features facilitate the detection of a global minimum in chi-squared in a highly multi-modal environment. By combining all three, the algorithm greatly increases the probability of realizing this goal. The FMCMC is controlled by a unique adaptive control system that automates the tuning of the MCMC proposal distributions for efficient exploration of the model parameter space even when the parameters are highly correlated. The AK models combined with the FMCMC algorithm constitute a multi-signal apodized Keplerian periodogram. 

\subsection{Priors}
\label{sec:priors}

A more detailed description of FMCMC is available in Chapter 1 of  the `Supplement to Bayesian Logical Data Analysis for the Physical Sciences,' a free supplement available in the resources section of the Cambridge University Press website for my Textbook `Bayesian Logical Data Analysis for the Physical Sciences: A Comparative Approach with {\it Mathematica} Support.' A {\it Mathematica} implementation of fusion MCMC is also available from the resource section.  Chapter 1 also includes a comparison of three marginal likelihood estimators used for Bayesian model comparison and concludes in favor of the Nested Restricted Monte Carlo (NRMC) estimator which is is used in this work. The supplement includes a detailed discussion of the priors adopted by this author for exoplanet RV analysis. Two different frequency priors were discussed: (a) $p(f|I) = 1/\sqrt{f} \times 0.5/(\sqrt{f_H} -\sqrt{f_L})$ and (b) $p(f|I) = 1/f \times 1/\ln(f_H/f_L)$, where $f_H$ and $f_L$ are the prior upper and lower frequency bounds, respectively. In this work (a) was used primarily for parameter estimation purposes which helps in the detection of shorter period signals while the scale invariant form (b) was adopted for model comparison purposes which utilized NRMC method for marginal likelihood estimation.

A new eccentricity prior was utilized in this work intermediate between Kipping's low period ($P < 382.3$ d) case and his average for all 396 planets \citep{Kipping2013}. The equation for the new eccentricity prior is 
\begin{equation}
p(e|I) = 7.354*(1 - e^{0.3})^{1.5}.
\label{eq:eccPrior}
\end{equation} 

The two additional parameters for a one signal AK model are the apodization time,  $\tau$, and the apodization function center time $t_a$. We used a scale invariant prior for $\tau$ and a uniform prior for $t_a$. The upper limit on $\tau$ was set at 4 times the duration of the data set which corresponds to a a maximum variation in the window height of approximately 1\% across the data set for a window with a center at the middle of the data or by 3\% for one centered at either end of the data. The lower limit was set to the data duration divided by 40 which corresponds to 37 d. Some of the calculations were redone using an upper limit of 8 times the duration and a lower limit of 10 d with no noticeable improvement in our ability to distinguish SA and planetary signals. Much more analysis would be needed to select an optimum lower limit for $\tau$. The upper and lower limits on the uniform $t_a$ prior were $-T$ to $+T$ relative to the mean data sample time, where T is the data duration. 

As we will shall see the primary role of the AK models is to distinguish planetary signal candidates from SA signals. Suppose the results indicate that $k$ of the signals are planetary and $m-k$ are SA signals. Final model parameter estimates and model comparisons are based on subsequent runs using a model of $k$ Keplerians and $m-k$ apodized Keplerians.

\subsection{Challenge test data set}
\label{sec:test}

\begin{figure*}
\includegraphics[width=140mm]{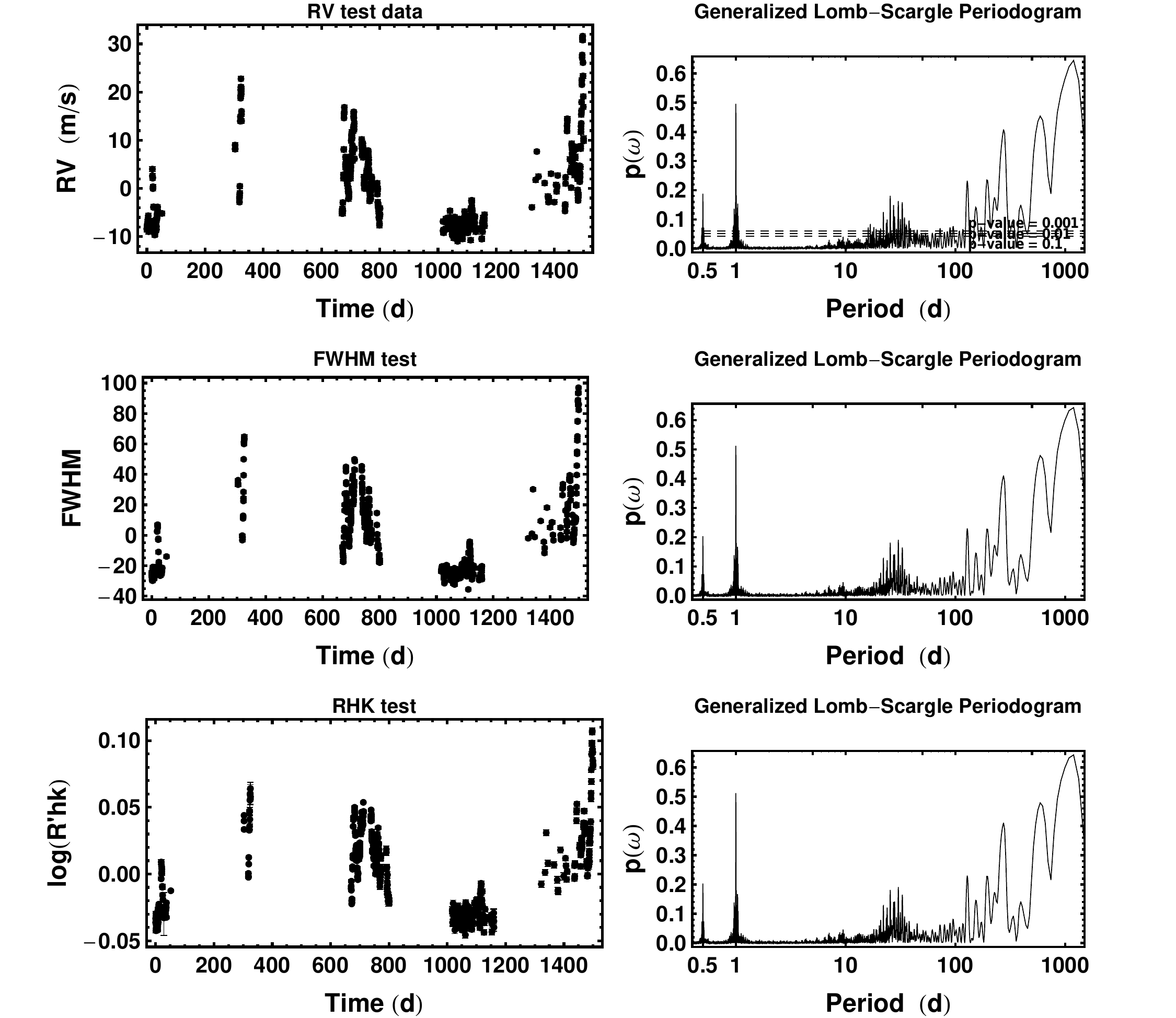}
\caption{The left column of panels are plots of the raw RV test data and two diagnostics, the FWHM and $\log(R'hk)$. The test data set is known to have a single planetary signal with a period of 16 d. For each plot the mean value has been subtracted. The right hand column of panels show the corresponding Generalized Lomb-Scargle (GLS) periodograms. For all the GLS periodogram plots the abscissa is expressed as a period where $P = \frac{2 \pi}{\omega}$.\label{fig:rawCor}}
\end{figure*}
The test data set contained a single planetary signal in a circular orbit with $P = 16.0$ d, $K = 1.5$ m/s and phase $=0.0$, together with a variety of stellar activity signals. Fig.~\ref{fig:rawCor} shows a comparison between the raw RV test data and two diagnostics, the FWHM and $\log(R'hk)$, and their Generalized Lomb-Scargle (GLS) periodograms \citep{Zechmeister2009}. To first oder the three signals appear highly correlated but close inspection of the periodograms indicate differences between the RV data and the diagnostics in the vicinity of the known 16 d planetary signal. The GLS periodogram $p(\omega)$ measures the relative $\chi^2$ -reduction, $p(\omega)$ , as a function of frequency $\omega$ and is normalised to unity by $\chi_0^2$ (the  $\chi^2$ for the weighted mean of the data). The quantity $p(\omega)$ is normalized to unity with $p(\omega) = 0$ indicating no improvement of the fit and $p(\omega) = 1$ a ``perfect'' fit.  For all the GLS periodogram plots the abscissa is expressed as a period where $P = \frac{2 \pi}{\omega}$. 
Frequentist p-values~\footnote{It is common practice to refer to the p-value (e.g., \cite{Zechmeister2009} as the false alarm probabilities (FAP) of the highest peak in the periodogram. This is incorrect, the p-value says nothing about the probability of this particular signal, rather it's the fraction of possible hypothetical signals, generated under the null hypothesis that there is no real signal present, that would have a $p(\omega) \ge$ the actual value observed at the peak of the periodogram. Every one of these hypothetical signals has a FAP = 1 for producing the p-value. For any signal to have a FAP $\ne 1$, alternatives to the null must sometimes act. We cannot calculate the FAP of the actual peak without specifying the alternatives to the null hypothesis and how often they act. We should stick to calling the quantity a p-value and, following Tom Loredo's suggestion, think of it as a measure of how surprised we are to observe such a peak under the null hypothesis. The difficulty in interpreting p-values has been highlighted in many papers (e.g., Berger and Sellke 1987; Sellke {\it et al.} 2001). The focus of these works is that p-values are commonly considered to imply considerably greater evidence against the null hypothesis than is actually warranted.} of 0.1, 0.01, 0.001 are shown in the upper RH periodogram~\footnote{The interesting region of $p(\omega)$ is where the frequentist p-value is small ($\ll 1$). In this region p-value$  \approx M * (1-z))^{(N-3)/2}$ \citep{Zechmeister2009}, where $z$ is the corresponding maximum $p(\omega)$, $M =$ the number of independent frequencies and $N$ is the number of data points. \cite{Cumming2004} recommends setting $M = \Delta f/\delta f$, where $\Delta f$ is the frequency range examined $\approx f_{\rm max}$, and $\delta f =$ the resolution of the periodogram $\approx 1/T$ where $T$ is the duration of the data.}.

In Fig.~\ref{fig:RVcorFWHMcor}, the upper left panel shows the RV data after removing the best linear regression fit with $\log(R'hk)$ as the independent variable. This is referred to as RV (rhk corrected) in the panel caption. Removing the correlation with $\log(R'hk)$ reduced the standard deviation of the RV data from 8.55 to 2.72 m/s. We also tried removing a linear regression model with $\log(R'hk)$ and the FWHM as two independent regression variables. This resulted in only a very slight reduction in the standard deviation of the residuals to 2.71 m/s. For this reason, in the apodized Kepler model fitting described below we only include a single $\log(R'hk)$ regression term as indicated in Equation~(\ref{eq:orbit1}). 

\subsection{Control periodogram}
\label{sec:ControlPeriodogram}

\begin{figure*}
\begin{center}
\includegraphics[width=140mm]{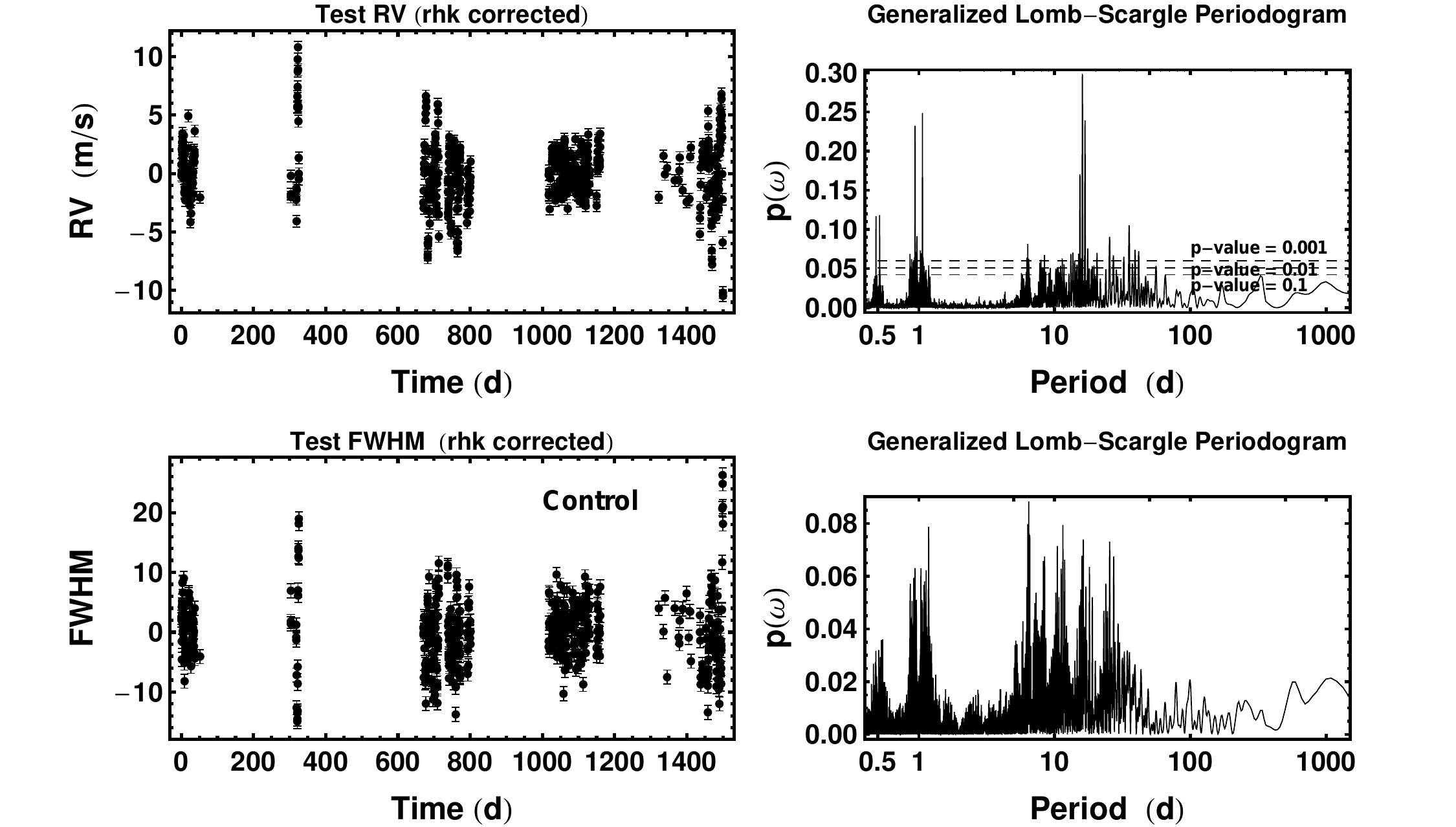}
\caption{The left column of panels are plots of the test data RV and FWHM after removing the best linear regression fits to $\log(R'hk)$ (rhk corrected). The right hand column of panels show the corresponding GLS periodograms.\label{fig:RVcorFWHMcor}}
\end{center}
\end{figure*}

It proved useful to employ the FWHM diagnostic data as a control after removing the best linear regression fit to $\log(R'hk)$. This reduced the standard deviation of the raw FWHM diagnostic from 26.2 to 5.6 m/s. This control data appears in the lower left panel of Fig.~\ref{fig:RVcorFWHMcor} and is referred to as FWHM (rhk corrected) in the caption. It provides another check on validity of planetary candidate signals identified with the apodized Keplerian model. A valid planetary signal is not normally expected to have a significant counterpart in the GLS periodogram of the control data~\footnote{\cite{Cuntz2000} suggested that there may be an observable interaction between a parent star and a close-in giant planet, specifically, an external heating of the star's outer atmosphere. \cite{Shkolnik2003}  detected the synchronous enhancement of CaII H and K emission with the short-period (3.093 d , $M_p \sin i = 0.84 M_J$) planetary orbit in HD 179949.}. The GLS periodograms of the RV (rhk corrected) data and control are shown in the two RHS panels. The periodogram of RV (rhk corrected) exhibits a dominant peak with a period of 16 d and a forest of smaller peaks. The periodogram of the control exhibits multiple lower level peaks which were interpreted as residual SA signals or instrumental systematics.

\subsection{Differential periodograms}
\label{sec:AKM}

\begin{figure*}
\begin{center}
\includegraphics[width=120mm]{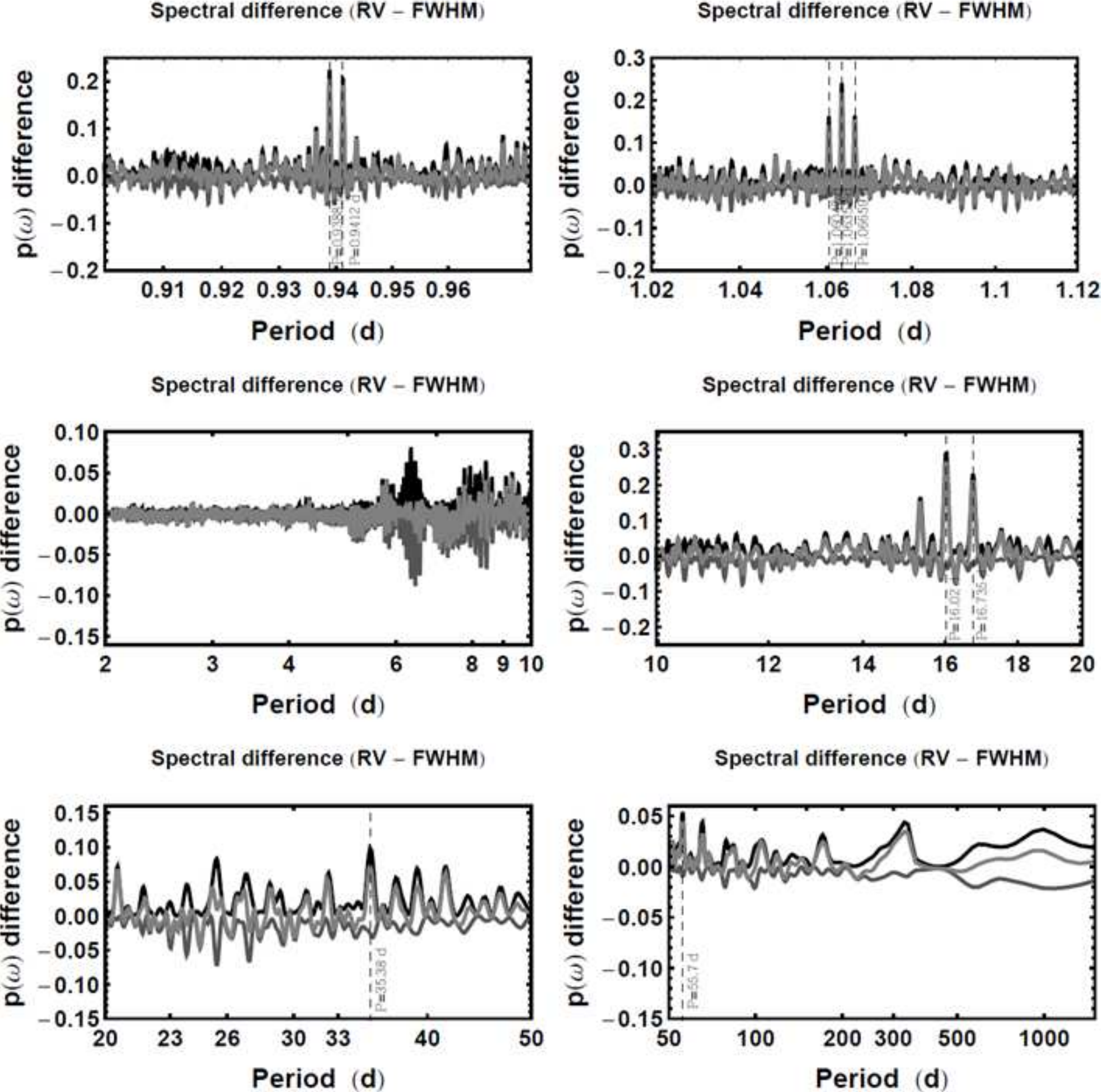}
\caption{A differential GLS periodogram of selected period regions for the test data based on the RV and FWHM control (both rhk corrected).\label{fig:RVdifSpec}}
\end{center}
\end{figure*}

It also proved useful to construct a differential form of the GLS periodogram of the RV residuals for selected period regions as is shown in Fig.~\ref{fig:RVdifSpec}. The black trace is from the upper right periodogram in Fig.~\ref{fig:RVcorFWHMcor}. The dark gray trace is the negative of the periodogram of the control and the light gray trace shows the difference, the black trace plus the gray trace. The differential GLS is shown for 6 period intervals. The strongest feature occurs at 16 d and on either side are the two one year aliases. One day aliases can be seen in the vicinity of $P = 0.94$ d and $P = 1.06$ d. For all three, the light gray trace and the black trace coincide closely near the positive peaks, consistent with a planetary signal or its alias. Signals in common to both black and dark gray traces (e.g., as occurs near $P = 6.3$ d) indicates stellar activity.
At each stage in the analysis the GLS periodogram (Fig.~\ref{fig:RVcorFWHMcor}) and the differential GLS periodogram (Fig.~\ref{fig:RVdifSpec}) of the current RV residuals provides guidance on the choice of the next period to include in the AK model. 

\begin{figure*}
\begin{center}
\includegraphics[width=100mm]{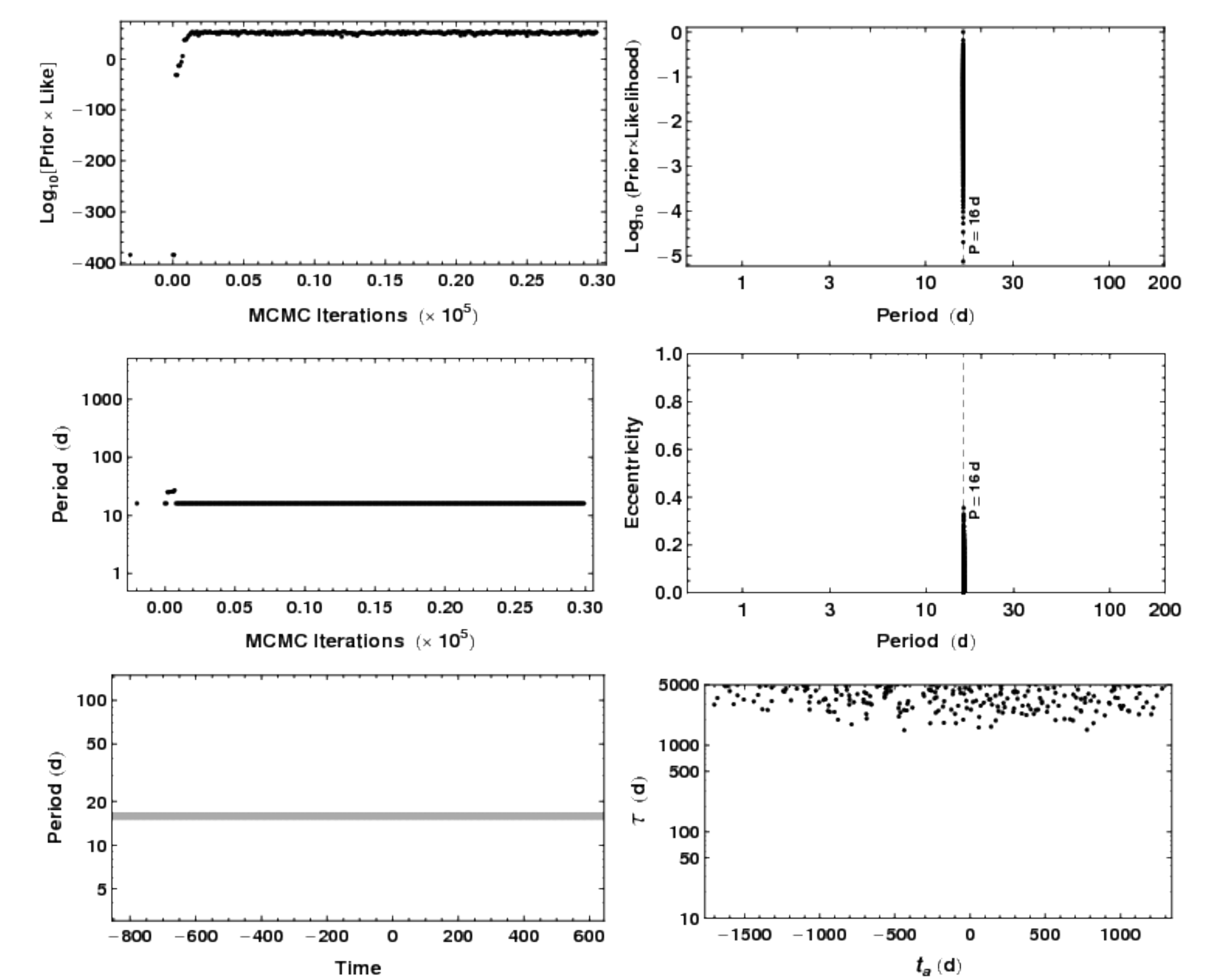}
\caption{The upper left panel is a plot of the Log$_{10}$[Prior $\times$ Likelihood] versus iteration for the 1 signal apodized Kepler (AK) periodogram of the test data. The upper right shows Log$_{10}$[Prior $\times$ Likelihood] versus period showing the 1 period detected. The middle left shows the values of the 1 unknown period parameter versus iteration number. The middle right shows the eccentricity parameter versus period parameter. The lower left shows the apodization window for each signal (gray trace for MAP values of $\tau$ and $t_a$, black for a representative set of samples which is hidden below the gray). The lower right is a plot of the apodization time constant, $\tau$, versus apodization window center time, $t_a$.\label{fig:testApodKep1grams}}
\end{center}
\end{figure*}

Fig.~\ref{fig:testApodKep1grams} shows sample FMCMC results for a one AK signal model. The lower left panel shows the span of the apodization window within the overall data window for each signal (gray trace for MAP values of $\tau$ and $t_a$, black for a representative set of samples which in this case is completely hidden below the gray). The lower and upper boundaries of the apodization window are defined by $t_{a} -\tau$ and $t_{a} + \tau$, respectively.

\begin{figure*}
\begin{center}
\includegraphics[width=150mm]{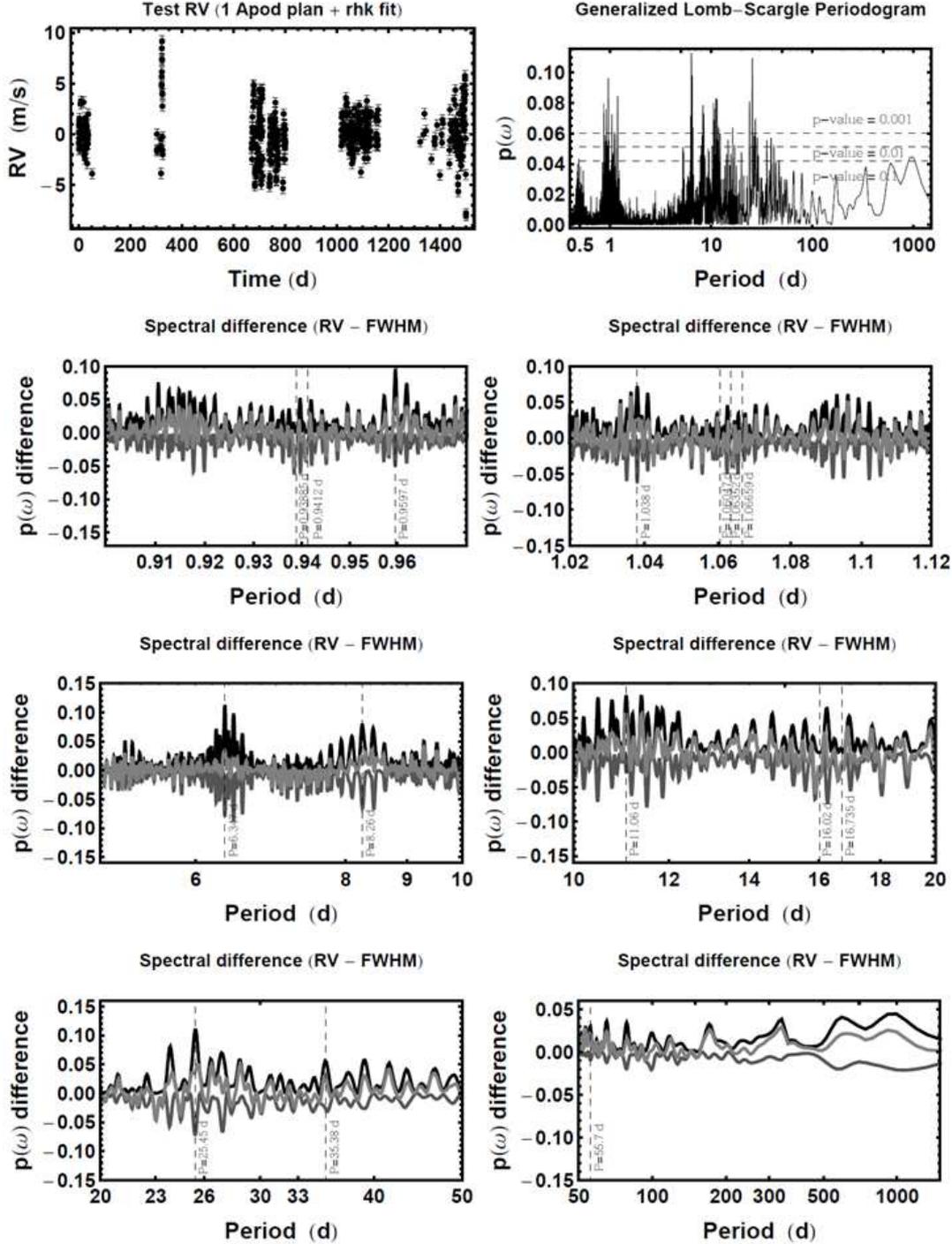}
\caption{The panels show the one signal AK model residuals for the test data set (top left), its GLS periodogram (top right), and the differential periodogram for selected period ranges.\label{fig:test_RVdifSpec1}}
\end{center}
\end{figure*}

The top left panel of Fig.~\ref{fig:test_RVdifSpec1} shows the 1 signal AK model residuals (i.e., $m = 1$ in Equation~(\ref{eq:orbit1})). The top right panel shows the corresponding GLS periodogram. The remaining panels show the differential periodogram of the residuals for selected period intervals. The $P = 16$ day signal and its one year and daily aliases have been successfully removed. The feature at $P = 6.3$ d is now the highest peak in the residuals.

The analysis proceeds by repeatedly adding an additional AK signal to the model with an initial period estimate based on the previous residuals and examining the the FMCMC results plus the GLS and differential periodograms of the residuals. Note: the the linear $\log(R'hk)$ correlation is not removed prior to each fit but is accomplished by the $\beta \times \ rhk(t_i)$ term in Equ.~\ref{eq:orbit1} as part of each fit. This process continues until the p-value $>0.01$ for the highest peak in the GLS periodogram of the residuals and the FMCMC results for the next more complicated model indicates there is no well defined solution for an additional signal. In the interest of space we jump to our final 5 AK signal model results.

\begin{figure*}
\begin{center}
\includegraphics[width=110mm]{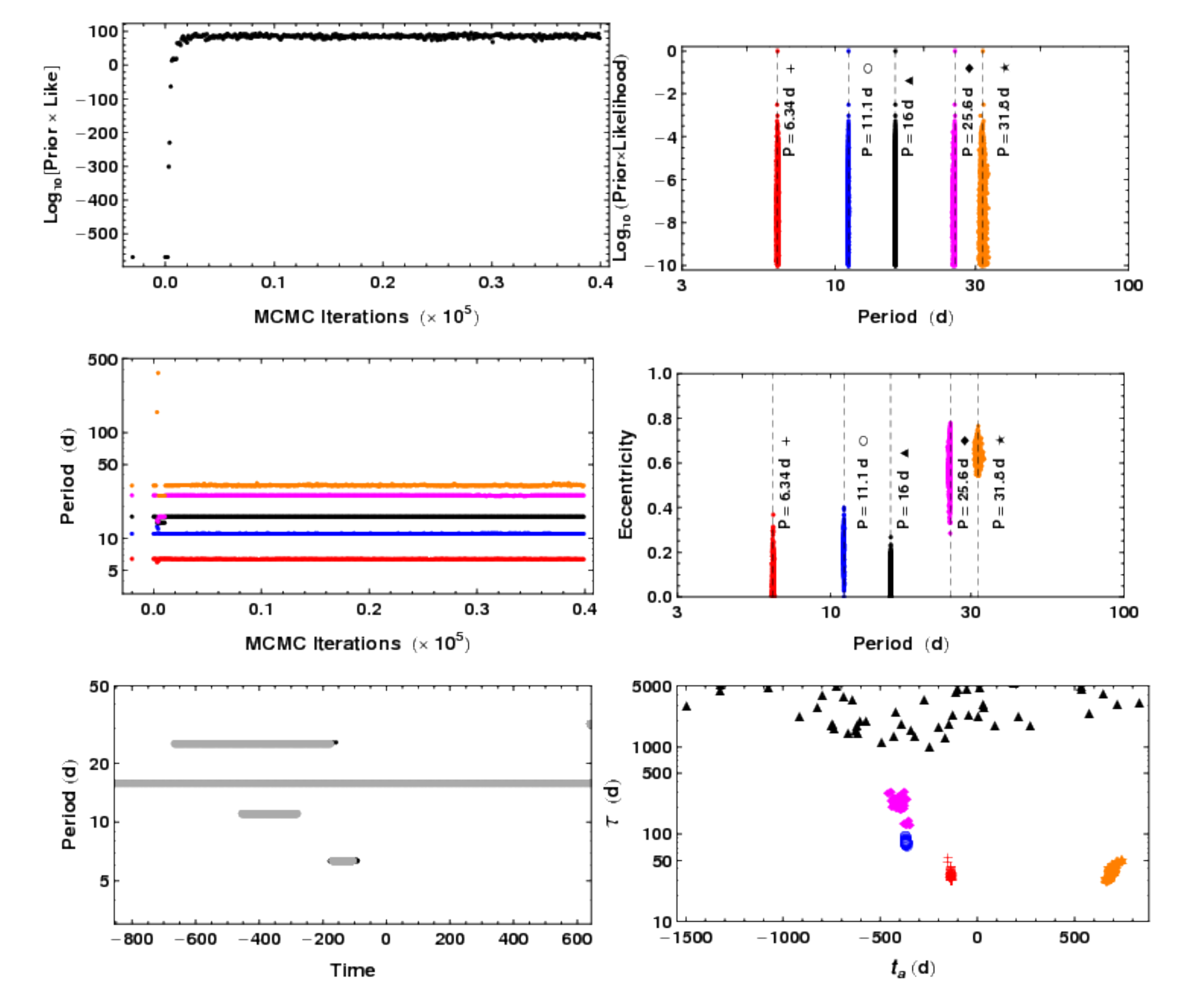}
\caption{The upper left panel is a plot of the Log$_{10}$[Prior $\times$ Likelihood] versus iteration for the 5 signal AK periodogram of the test data. The upper right shows Log$_{10}$[Prior $\times$ Likelihood] versus period showing the 5 periods detected. The middle left shows the values of the 5 unknown period parameter versus iteration number. The middle right shows the eccentricity parameters versus period parameters. The lower left shows the apodization window for each signal (gray trace for MAP values of $\tau$ and $t_a$, black for a representative set of samples which is hidden below the gray). The lower right is a plot of the apodization time constant, $\tau$, versus apodization window center time, $t_a$, where the symbols refer to the different signal periods specified in the panels above.\label{fig:test_ApodKep5_grams}}
\end{center}
\end{figure*}

Fig.~\ref{fig:test_ApodKep5_grams} shows sample FMCMC results for a five AK signal model. The upper left panel is a plot of the Log$_{10}$[Prior $\times$ Likelihood] versus iteration for the 5 signal AK periodogram of the test data. The upper right shows Log$_{10}$[Prior $\times$ Likelihood] versus period showing the 5 periods detected. The middle left shows the values of the 5 unknown period parameter versus iteration number. The middle right shows the eccentricity parameters versus period parameters. The lower left panel shows the span of the apodization window within the overall data window for each signal (gray trace for MAP values of $\tau$ and $t_a$, black for a representative set of samples which is mainly hidden below the gray). The lower and upper boundaries of the $j^{\rm th}$ apodization windows are defined by $t_{ja} -\tau_j$ and $t_{ja} + \tau_j$, respectively. The lower right panel is a plot of the apodization time constant, $\tau$, versus apodization window center time, $t_a$.
Based on the span of the apodization window, all of the signals, with the exception of $P = 16$ d, are consistent with SA. As a check on the planetary interpretation of the 16 d signal, Fig.~\ref{fig:RVdifSpec} indicates that $p(\omega)$ for the control trace is within 0.03 of zero for a period of 16 d.  
At this point in the analysis, the peak in the GLS periodogram of the 5 signal AK fit residuals at $P = 0.95$ d was down to $p(\omega) = 0.045$. The p-value was $> 0.01$ and no further signals were extracted. 
\begin{figure*}
\begin{center}
\includegraphics[width=80mm]{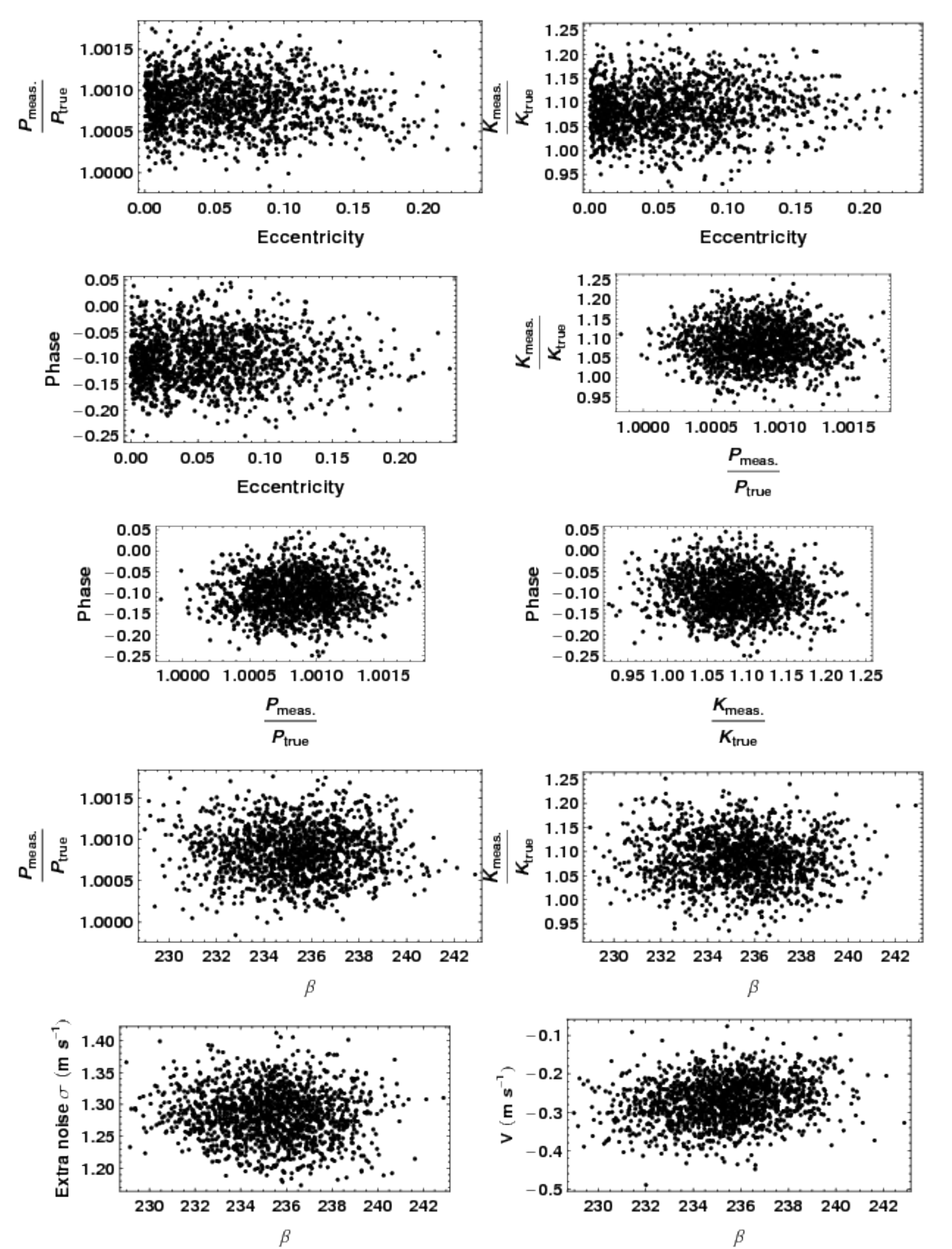}
\caption{Parameter correlation plots for the  $P = 16$ d planetary signal in the test data set based on a model consisting of four apodized Keplerians plus one ordinary Keplerian.\label{fig:ParCorr}}
\end{center}
\end{figure*}
\begin{figure}
\includegraphics[width=87mm]{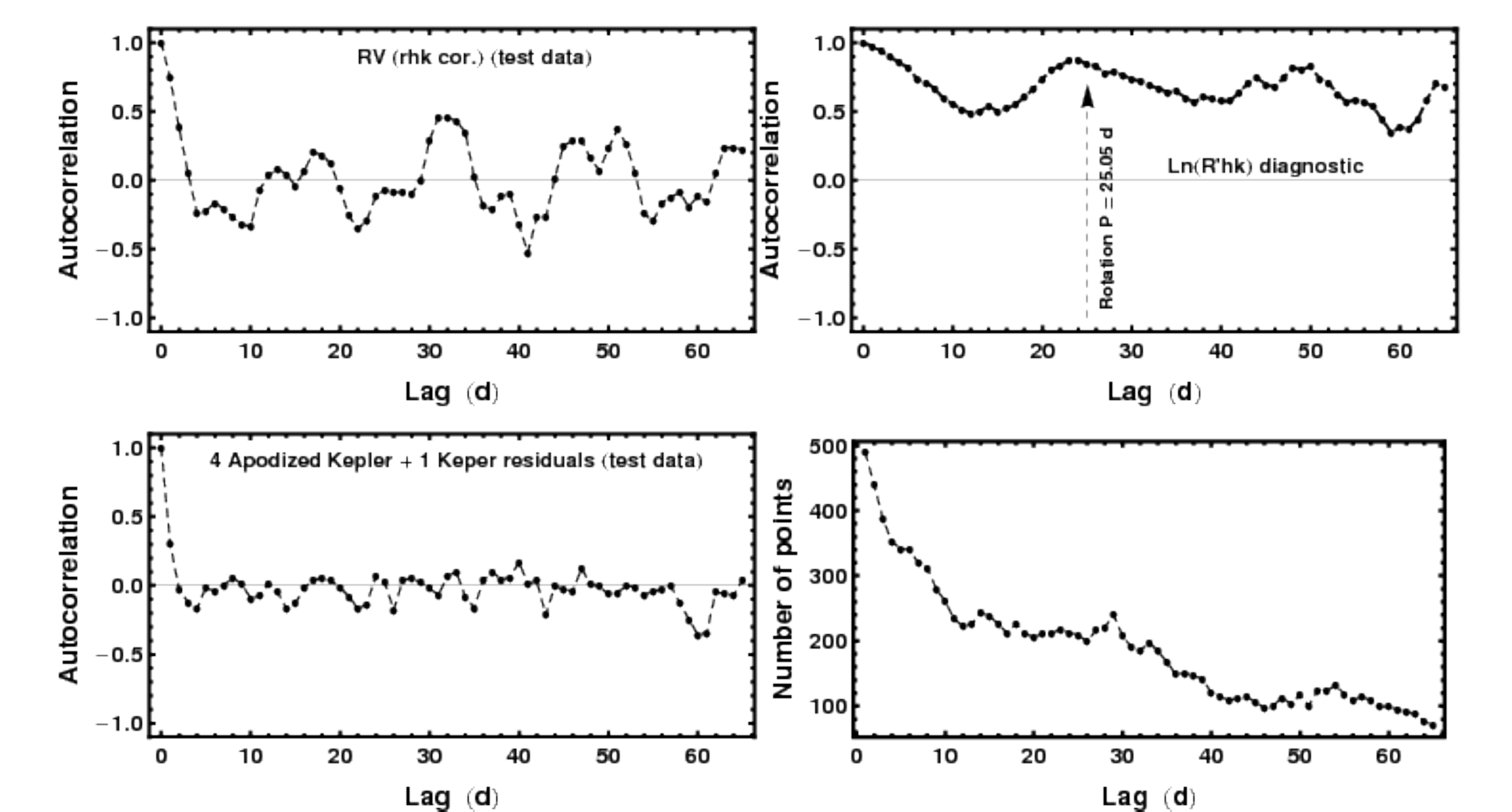}
\caption{The top row shows the autocorrelation function of the (rhk corrected) RV test data  and the raw $\log(R'hk)$ diagnostic. The left panel in the second row shows the autocorrelation function of the RV residuals after removing a model consisting of 4 apodized Keplerian signals plus 1 Kepler signal and the $\log(R'hk)$ regression. The bottom panel  shows a plot of the number of such sample pairs as a function of the lag.\label{fig:test_AutoCor4AK1Kplot}}
\end{figure}

\subsection{Apodized Keplerian plus ordinary Keplerian models}
\label{sec:AKOKM}

The obtain the best $P = 16$ d parameter estimates the $m$ AK model terms in Equation~(\ref{eq:orbit1}) were replaced by four apodized Keplerians plus one ordinary Keplerian to model the 16 d planetary signal, and the data refit. The starting FMCMC parameter values were taken from the best fitting 5 AK signal model. Fig.~\ref{fig:ParCorr} shows a sample of resulting 16 d parameter correlation plots, which in a Bayesian analysis reflect our current state of knowledge for the 16 d planetary signal in the test data set.

The $P$ and $K$ values have been normalized by dividing by their true values to convert to ratios. Since the true eccentricity $=0$ the relevant phase for comparison is our re-parameterized phase term $\psi=2 \pi\chi+\omega$. In our analysis the phase is referenced to the mean observation time. This was converted to a phase referenced to the time of the first sample and added to an additional $\pi/2$ to compare with the Xavier Dumusque's definition of phase which refers to a sine function while ours refers to a cosine function. 

It is not too meaningful to draw any conclusions about parameter biases on the basis of this one data set. Later in Section~\ref{sec:Discussion}, we compare the $K$ and $P$ parameter estimates derived from the $m$ AK model fits to the $k$ AK plus $(m-k)$ Kepler model fits for the six data sets analyzed (including the test data). Nine planets were correctly identified in these six data sets. From this it is clear that AK model estimates of the $K$ parameter are significantly biased to larger values compared to their known true values. As a result, our final parameter estimates for candidate planets were based on the $k$ AK plus $(m-k)$ Kepler model fits, where $k$ is the number of signals attributed to SA. 

Ideally, we would like to see the Bayesian marginals span the true values of $P, K$, and phase slightly better than indicated by Fig.~\ref{fig:ParCorr}. This might be an indication that the true scatter is underestimated possibly as a result of correlated noise in the residuals. One way of exploring the residuals is to compute the autocorrelation function, $\rho(j)$  equation~(\ref{eq:ACF1}). 
\begin{equation}
\rho(j) = \frac{\sum_{\rm overlap} [(x_{i}-\overline{x})\ (x_{i+j}-\overline{x})]}{\sqrt{\sum_{\rm overlap} (x_{i}-\overline{x})^2} \times \sqrt{\sum_{\rm overlap} (x_{i+j}-\overline{x})^2}},
\label{eq:ACF1}
\end{equation}
where $x_i$ is the $i^{\rm th}$ residual, $j$ is the lag and $\overline{x}$ is the mean of the samples in the overlap region.

The top row of Fig.~\ref{fig:test_AutoCor4AK1Kplot} shows the autocorrelation function of the (rhk corrected) RV test data and the raw $\log(R'hk)$ diagnostic. Note: the highest peak in the autocorrelation function for the $\log(R'hk)$ diagnostic corresponds roughly with the star's true rotation period of 25.05 d (vertical dashed line). The rotation period was not provided during the competition. SA signals frequently occur at the rotation period and harmonics. The left panel in the second row shows the autocorrelation function for the RV residuals for the 4 apodized Kepler plus 1 Kepler model including the $\log(R'hk)$ regression. Clearly, after after removing a model consisting of  5 signals and the $\log(R'hk)$ regression, the autocorrelation of the residuals is looking much close to white noise but there is still evidence for a small positive autocorrelation for a lag of one day. The remaining panel shows a plot of the number of such sample pairs as a function of the lag. Because the data are not uniformly sampled, for each lag all sample pairs that differed in time by this lag $\pm 0.1$ d were utilized.

\subsection{Summary and Discussion of Methodology}
\label{sec:Summary}

In this section we summarize the steps in the methodology and illustrate its success in modeling the both long-term and short-term SA induced RV variations. The method consists of the following steps:
\begin{enumerate}
\item[1.] Remove the best linear regression fit from the raw RV data to obtain the RV (rhk corrected) data, using the SA diagnostic $\log(R'hk)$ as the independent regression variable. 
\item[2.] Compute the GLS periodogram of RV (rhk corrected) and p-value levels corresponding to 0.1, 0.01, 0.001. The highest peak in this periodogram is used as the initial estimate for the period of the AK model signal.
\item[3.] Remove the best linear regression fit from the raw FWHM data to obtain the control FWHM  (rhk corrected) data, using the $\log(R'hk)$ diagnostic as the independent regression variable. 
\item[4.] Compute the GLS periodogram of control FWHM (rhk corrected).
\item[5.] Compute the differential GLS periodogram for selected period regions, showing on the same plot, the periodogram of RV (rhk corrected) in black, the periodogram of the control FWHM (rhk corrected) plotted negatively and shown in dark gray, and the their sum shown in gray. Planetary candidate signals for which the gray trace and black trace are significantly different (i. e., the dark gray trace has a $p(\omega)$ more negative than $-0.05$) indicates that the signal is possibly SA in origin. 
\item[6.] Fit the model described by Equ.~\ref{eq:orbit1} to the raw RV data, in our case using the FMCMC algorithm. Use the highest peak in the GLS periodogram of the rhk corrected RV data as the initial estimate for the period of the AK model signal, regardless of whether the differential GLS periodogram indicates it is SA in origin. We are using the apodized Keplerians to model both types of signals. 
\item[7.] Compute the GLS and differential periodograms of the fit residuals.
\item[8.] Repeat the last two items by including an additional AK signal with an initial period estimated from the largest peak in the GLS periodogram of the previous fit residuals.
\item[9.] Repeat the previous step until the p-value of the highest peak in the residuals is greater than 0.01 and the FMCMC fit indicates there is no well defined solution for an additional signal. 
\item[10.] Examine the span of the apodization window for the MAP parameter values and representative samples for each signal extracted. The lower and upper boundaries of the $j^{\rm th}$ apodization windows were chosen to be $t_{ja} -\tau_j$ and $t_{ja} + \tau_j$, respectively. 
\item[11.] If the apodization window spans the data duration and, $p(\omega)$ for the dark gray trace in the differential periodogram of the residuals is less negative than $-0.05$, then treat the signal as a candidate planetary signal. If the apodization window spans the data duration and $p(\omega)$ for the dark gray trace is more negative than $-0.05$, then treat the signal as a possible candidate planetary signal. If the apodization window does not span the data duration classify it as a SA signal.
\item[12.] Maintain a list of the order in which the signals were extracted and the signal designation as P for planetary candidate, P? for possible planetary candidate, and SA for stellar activity signal. 
\item[13.] Suppose there are $m$ signals in total, $k$ of which are SA signals and $(m-k)$ are planetary candidates or possible planetary. Fit a model with $k$ AK signals and $(m-k)$ ordinary Keplerian signals together with the $\beta \times \ rhk(t_i)$ term of Equ.~\ref{eq:orbit1} to derive final parameter estimates for the planetary and possible planetary candidates.
\item[14.] Compute Bayes factors as a further test of the validity of any planetary candidate. This is discussed further starting in Section~\ref{sec:ModComp}.  

\end{enumerate}

\begin{figure*}
\begin{center}
\includegraphics[width=130mm]{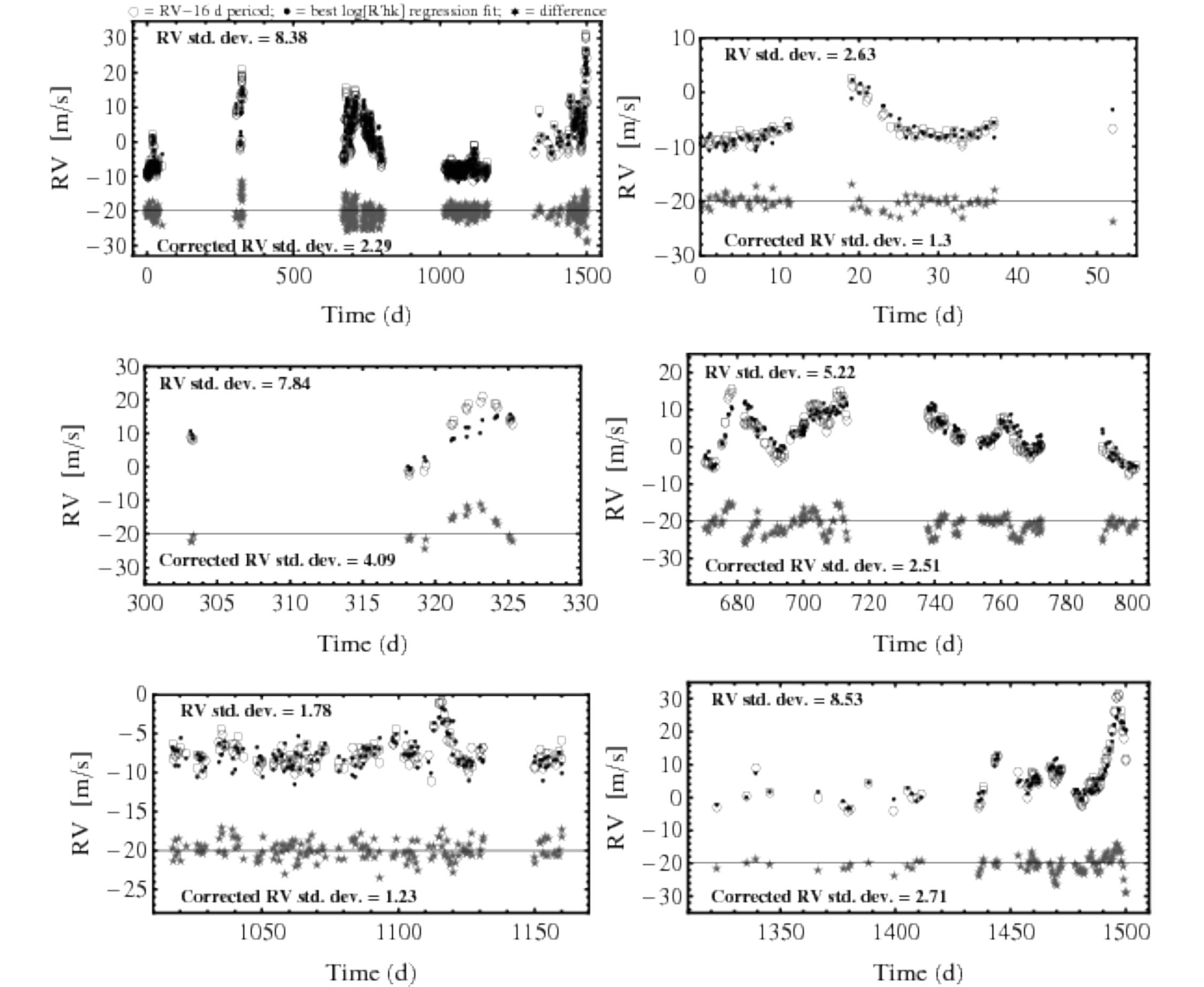}
\caption{The open circles in the upper left panel show the raw RV test data minus the MAP fit for the 16 d planet. The dots show how much of this remainder can be accounted for by a correlation with the stellar activity diagnostic $\log(R'hk)$. The starred points show the difference between the open circles and the dots. The other panels show expanded versions, one for each observing season.\label{fig:RVminus16DcorSeason}}
\end{center}
\end{figure*}

\begin{figure*}
\begin{center}
\includegraphics[width=130mm]{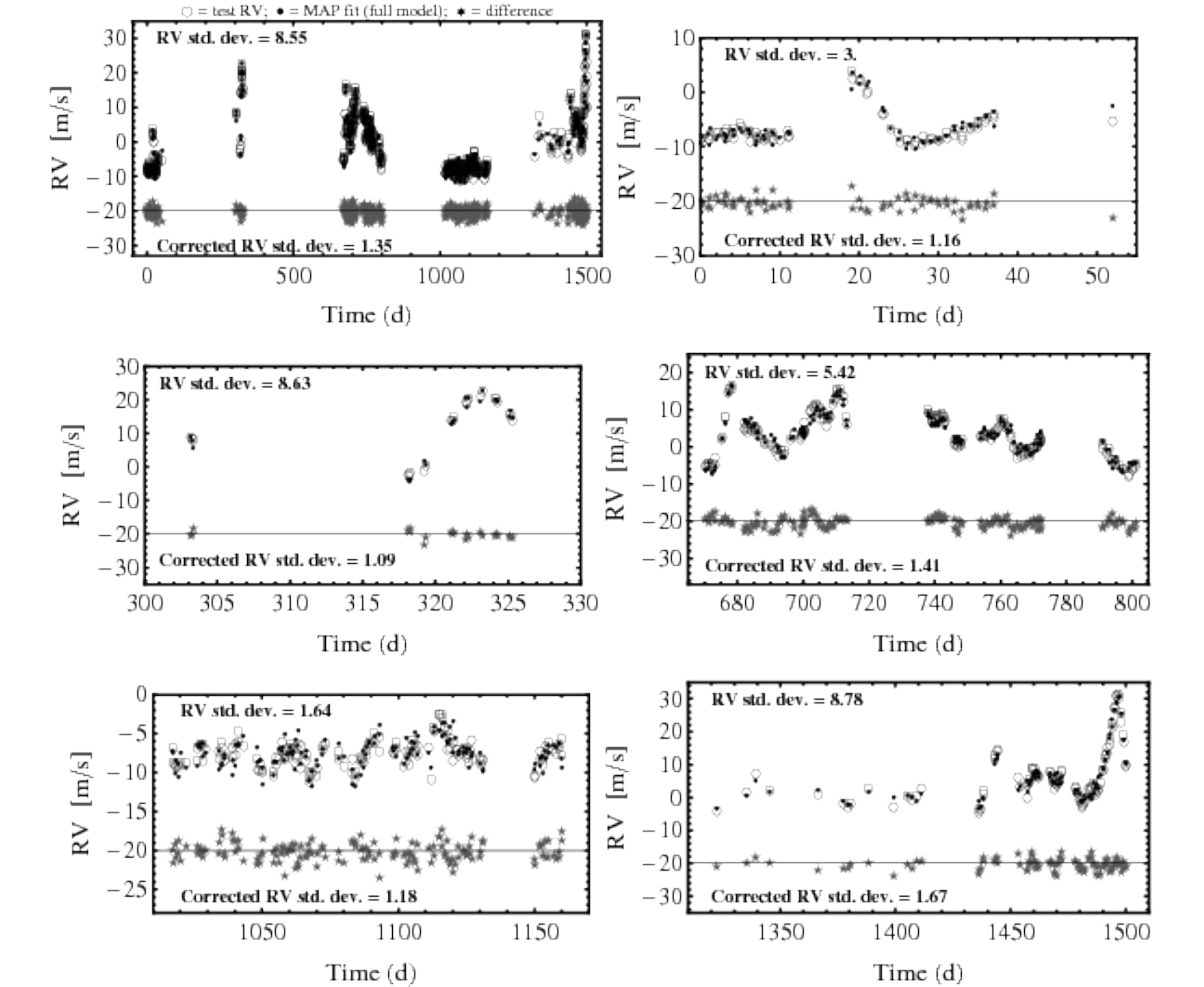}
\caption{The open circles in the upper left panel show the raw RV test data. The dotted points show how much of each RV measurement can be accounted for by the full model (Equ. 1) fit. The starred points show the difference between the open circles and the dots. The other panels show expanded versions, one for each observing season.\label{fig:RVfmcmcfitSeason}}
\end{center}
\end{figure*}
\begin{figure*}
\begin{center}
\includegraphics[width=140mm]{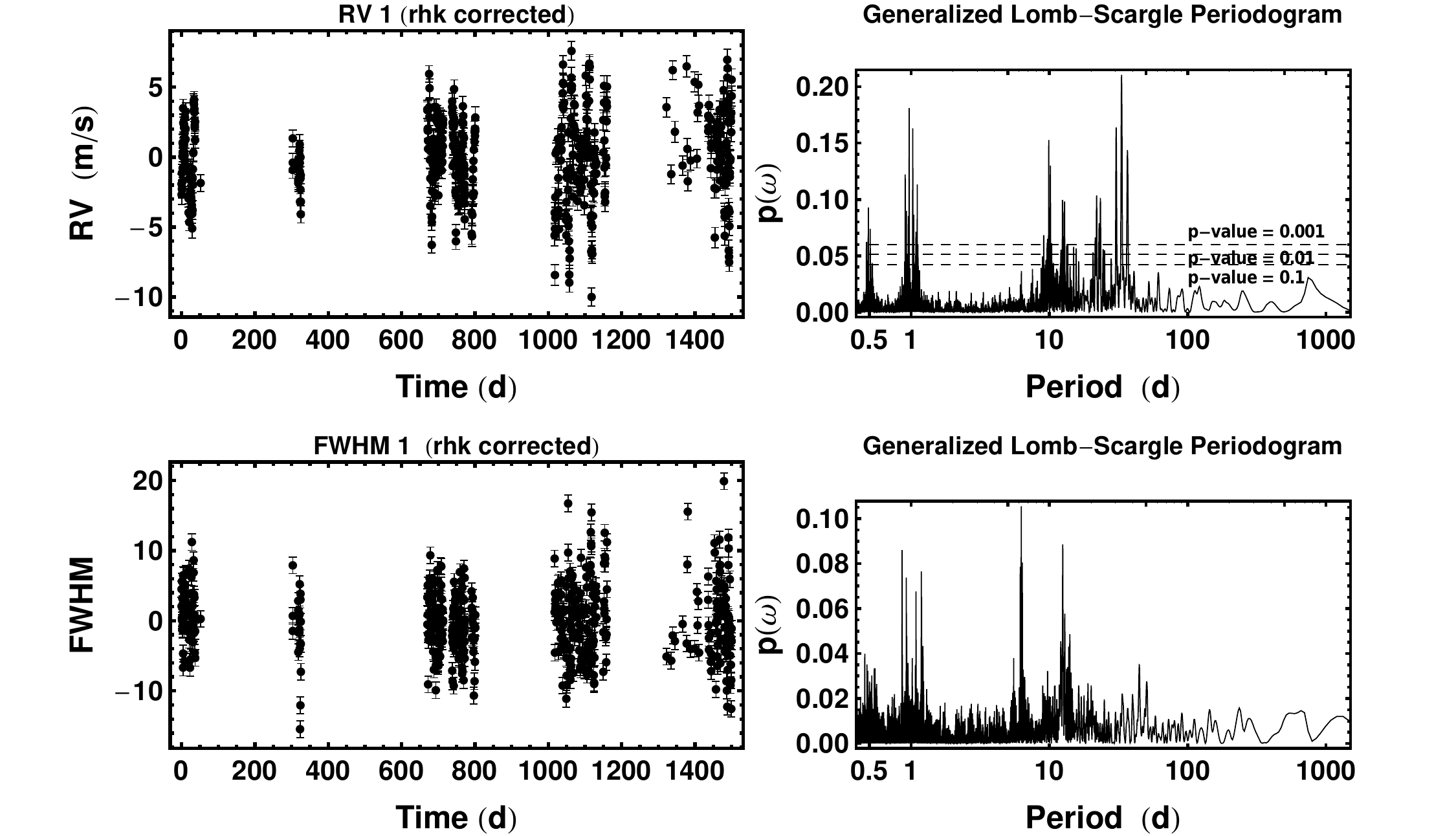}
\caption{The RV 1 data and FWHM (control)  after removing the best linear regression fits to $\log(R'hk)$ (rhk corrected) together with their GLS periodograms on the right.\label{fig:RV1rhkCor}}
\end{center}
\end{figure*}

In the upper left panel of Fig.~\ref{fig:RVminus16DcorSeason}, the (open circles) represent the raw RV test data minus the MAP fit for the 16 d planet which we designate as RV$_{\rm 1planCorr}$.  The dots show how much of the RV$_{\rm 1planCorr}$ can be accounted for by a correlation with the SA diagnostic $\log(R'hk)$, based on a weighted least squares linear regression fit using $\log(R'hk)$ as the independent regression variable. The starred points show the difference between the open circles and the dots. Removing the correlation with $\log(R'hk)$ reduced the standard deviation of RV$_{\rm 1planCorr}$ from 8.38 to 2.29 m/s. The other panels show expanded versions, one for each observing season. Clearly, the $\log(R'hk)$ regression fit tracks the long-term SA very well. Close inspection of the seasonal panels indicates the regression fit only provides a first order tracking of the short term variations in RV arising from individual active regions (spots and plages) .

In the upper left panel of Fig.~\ref{fig:RVfmcmcfitSeason}, the (open circles) represent the raw RV data.  The dots show the MAP fit of the full model consisting of four apodized Keplerians plus one ordinary Keplerian (to model the 16 d planetary signal), together with the linear regression term using $\log(R'hk)$ as the independent regression variable. The starred points show the difference between the open circles and the dots. Employing the full model reduced the standard deviation of from 8.55 to 1.35 m/s. The other panels show expanded versions, one for each observing season. Clearly, the full model provides a much better fit to the RV data including the short term variations arising from individual active regions.

\section{Results}
\label{sec:results}

The procedures outlined in Section~\ref{sec:analysis} were applied to the first five RV challenge data sets and the results are given below.
Table~\ref{tab:signalsRV1} shows the order in which the signals emerged  in successive AK models together with the signal designation as planetary (P) or stellar activity (SA) in the final column. 
\begin{table}
 \begin{center}
  \caption{Order of signals extracted from RV 1 together with approximate parameter values and designation of signal type as planetary (P) or stellar activity (SA) .}
  \label{tab:signalsRV1}
  \begin{tabular}{@{}llllll@{}}
  \hline
   Period   & K  & ecc & $\tau$ & $t_a$ & Signal\\
 (d) & (m/s) & & (d) & (d)& type\\
\hline
33.35 & 2.5 & 0.32 & 1356 & -133 & P \\
& & & & & \\
9.89 & 1.8 & 0.0 & 1274 & -341 & P \\
& & & & & \\
23.4 & 1.6 & 0.24 & 1747 & -1204 & P \\
& & & & & \\
12.5 & 2.5 & 0.37 & 557 & 410 & SA \\
& & & & & \\
10.6 & 1.0 & 0.0 & 494 & -199 & SA \\
& & & & & \\
8.38 & 2.1 & 0.39 & 38 & -93 & SA \\
& & & & & \\
\hline
\end{tabular}
 \end{center}
\end{table}

\subsection{First data set, RV 1}
\label{sec:RV1}

\begin{figure*}
\begin{center}
\includegraphics[width=110mm]{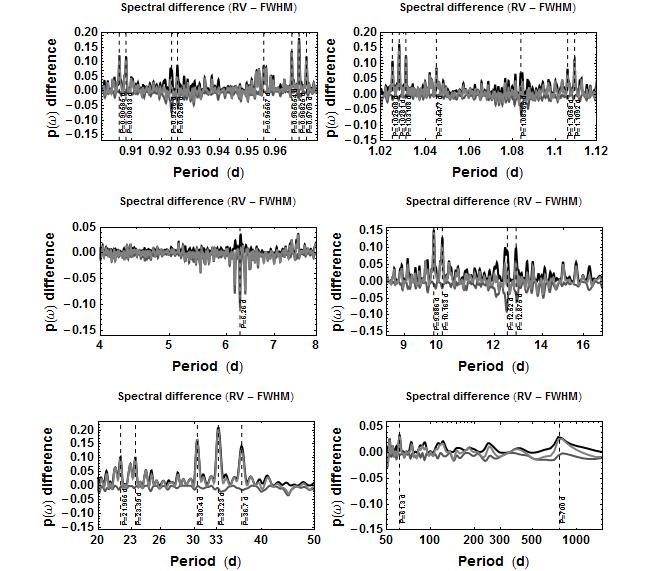}
\caption{The differential GLS periodogram for the RV 1 data (rhk corrected) for selected period ranges.\label{fig:RV1rhkCorDetail}}
\end{center}
\end{figure*}

Comparison of the raw RV and FWHM and $\log(R'hk)$ diagnostics indicated that there was a clear correlation between the three. The raw RV 1 data had a standard deviation of 5.6 m/s. After removing the best linear regression fit with $\log(R'hk)$ as the independent variable, the standard deviation was reduced to 3.0 m/s. The panels on the left of Fig.~\ref{fig:RV1rhkCor} show the RV data and FWHM (control) with the best linear regression fits to $\log(R'hk)$ removed (rhk corrected). Their GLS periodograms appear on the right. Note: for each apodized Keplerian model fit, the correlation parameter $\beta$ between RV and the stellar activity diagnostic $\log(R'hk)$ was included as a fit parameter according to Equation~(\ref{eq:orbit1}). Fig.~\ref{fig:RV1rhkCorDetail} show the differential GLS periodogram for the RV 1 data (rhk corrected) for selected period ranges. The strongest signal at $P = 33$ d has no significant counterpart in the control. Strong yearly and daily aliases of the 33 d signal are also evident. 
\begin{figure*}
\begin{center}
\includegraphics[width=110mm]{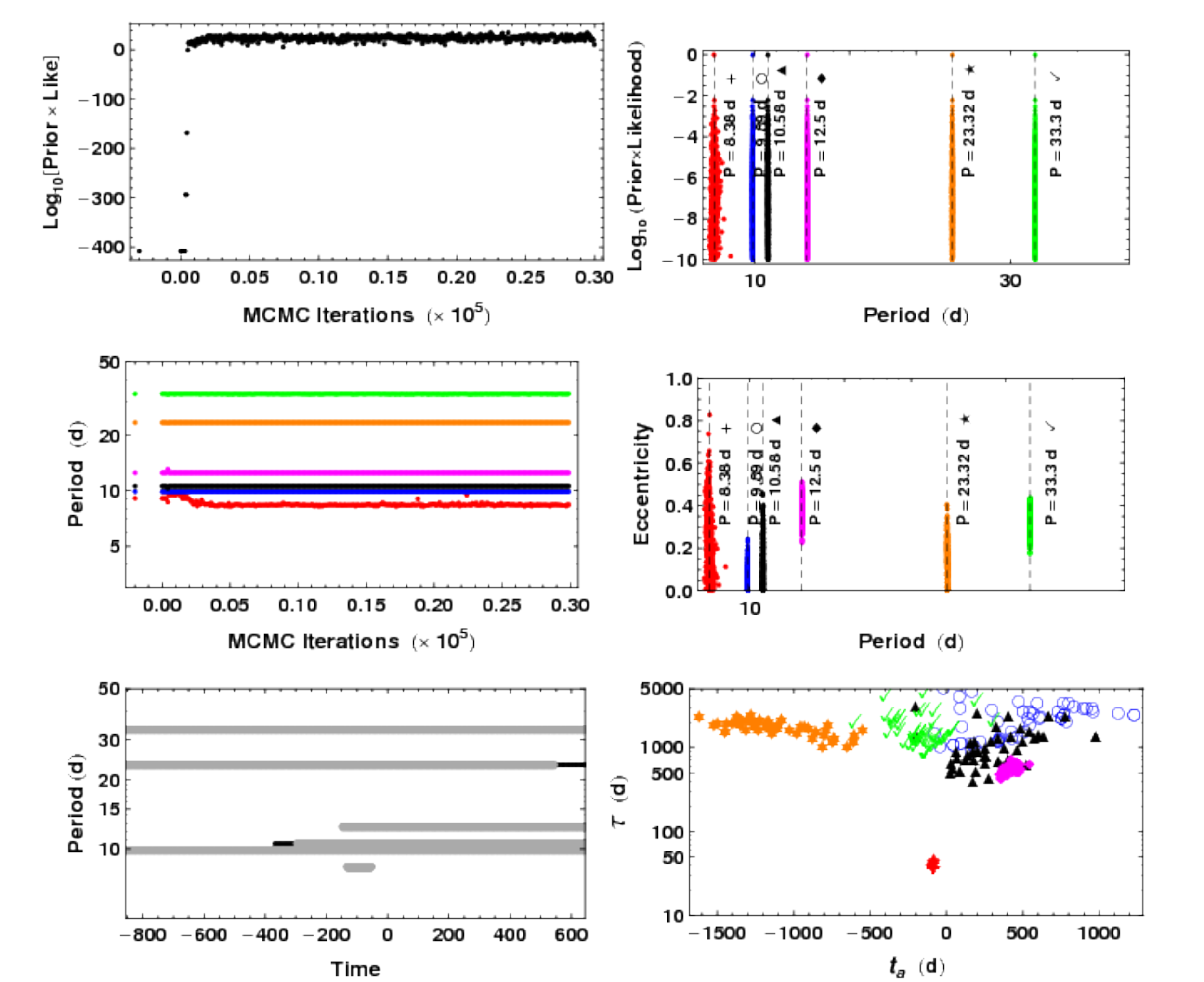}
\caption{The upper left panel is a plot of the Log$_{10}$[Prior $\times$ Likelihood] versus iteration for the 6 signal AK periodogram of the RV 1 data. The upper right shows Log$_{10}$[Prior $\times$ Likelihood] versus period showing the 6 periods detected. The middle left shows the values of the 6 unknown period parameters versus iteration number. The middle right shows the eccentricity parameters versus period parameters. The lower left shows the apodization window for each signal (gray trace for MAP values of $\tau$ and $t_a$, black for a representative set of samples which is mainly hidden below the gray). The lower right is a plot of the apodization time constant, $\tau$, versus apodization window center time, $t_a$, where the symbols refer to the different signal periods specified in the panels above.\label{fig:RV1_ApodKep6_grams}}
\end{center}
\end{figure*}

Fig.~\ref{fig:RV1_ApodKep6_grams} shows sample FMCMC results for a six AK signal model. The lower left panel shows the span of the apodization window within the overall data window for each signal (gray trace for MAP values of $\tau$ and $t_a$, black for a representative set of samples which is mainly hidden below the gray). The lower and upper boundaries of the $j^{\rm th}$ apodization window are defined by $t_{ja} -\tau_j$ and $t_{ja} + \tau_j$, respectively. It is clear from this that there are three planetary candidates with periods of 9.89, 23.4, 33.3 d whose apodization windows span the duration of the data.

\begin{figure*}
\begin{center}
\includegraphics[width=110mm]{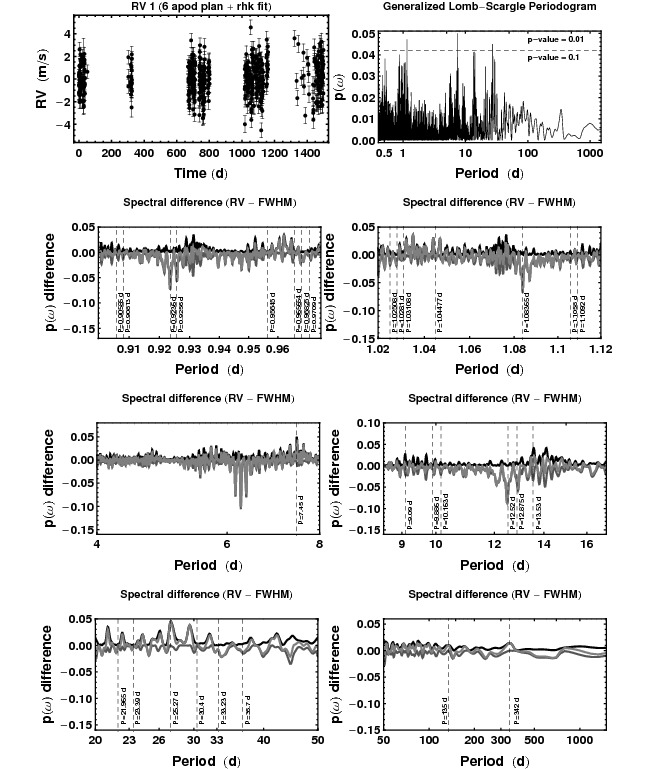}
\caption{The panels show the RV 1 six signal AK model residuals (top left), its GLS periodogram (top right), and the differential periodogram for selected period ranges.\label{fig:RV1_6ApodRVdifSpecGray}}
\end{center}
\end{figure*}

The weighted RMS residual $= 1.44$ m/s for the 6 AK signal model. The standard deviation of the extra Gaussian white noise term $s = 1.26$ m/s. The mean measurement uncertainty was 0.674 m/s.

\begin{table}
 \centering
  \begin{center}
  \caption{ RV 1 parameter estimates and 68\% credible boundaries for 3 planetary candidates discussed in the text. The quoted eccentricity values are for the mode while for the other parameters we quote the median. The value immediately below in parenthesis is the MAP estimate. The errors on semi-major axis, $a$, and $ M \ sin i$ do not include the uncertainty in the mass of the host star. See the note on periastron passage in the text.}
  \label{tab:parerrorsRV1}
  \begin{tabular}{@{}lllllll@{}}
  \hline
   Parameter  & planet 1 & planet 2 & planet 3  \\
\hline
$P$  (d) & $9.899_{-.002}^{+.002}$ & $23.34_{-.01}^{+.01}$ & $33.32_{-.02}^{+.02}$  \\
& (9.896)& (23.34) & (33.33)\\
& & & \\
$K$ (m/s) & $1.8_{-0.1}^{+0.1}$ & $1.6_{-0.1}^{+0.1}$ & $2.5_{-0.1}^{+0.1}$  \\
& (1.8) & (1.6) & (2.5) \\
& & & \\
$e$ & $0.00_{-0.00}^{+0.09}$ & $0.24_{-0.08}^{+0.08}$ & $0.32_{-0.04}^{+0.04}$   \\
& (0.12) & (0.29) & (0.34)  \\
& & &&\\
$\omega$  (rad) & $3.4_{-0.9}^{+0.9}$ & $1.7_{-0.3}^{+0.3}$  & $1.46_{-0.14}^{+0.13}$ \\
& (3.4) & (1.8) & (1.47)  \\
& & &\\
$a$  (au) & $0.08306_{-.00001}^{+.00001}$ & $0.14713_{-.00005}^{+.00005}$ & $0.18654_{-.00008}^{+.00008}$\\
& (0.08304) & (0.14711)  & (0.18656) \\
& & &\\
$M \sin i$  & $5.2_{-0.3}^{+0.3}$ & $5.6_{-0.4}^{+0.4}$  & $10.0_{-0.4}^{+0.4}$\\
($M_E$) & (5.0) & (5.8) & (10.1) \\
& & &\\
Periastron & $55850_{-1}^{+1}$ & $55837.8_{-0.9}^{+0.9}$  & $55845.8_{-0.7}^{+0.7}$ \\
\ passage &  (55850) & (55837.8) & (55845.8) \\
\hline
\end{tabular}
\end{center}
\end{table}

Fig.~\ref{fig:RV1_6ApodRVdifSpecGray} shows the RV 1 six signal AK model residuals (i.e., $m = 6$ in Equation~(\ref{eq:orbit1})), its GLS periodogram (top right), and the differential periodogram for selected period ranges. The highest spectral peak in the residuals has a p-value $> 0.01$. The dashed line representing the  p-value $= 0.01$ is not seen as it is just above the top of the graph. These results are consistent with the conclusion that the $P = 9.89, 33.35, 23.4$ d signals are bona-fide planetary candidates.

Table~\ref{tab:SIMsignalsRV1} shows the true parameters of the 5 planetary signals that were employed in the RV 1 simulation. Clearly the three planetary signals with $K > 1$ were recovered in our analysis but there is no clear evidence for the two weaker planetary signals at $P = 112\ \& \ 273$ d in the GLS residuals of Fig.~\ref{fig:RV1_6ApodRVdifSpecGray}.
\begin{table}
 \begin{center}
  \caption{The true $P, K, e$ parameters of the 5 planetary signals that were employed in the RV 1 simulation.}
  \label{tab:SIMsignalsRV1}
  \begin{tabular}{@{}llll@{}}
  \hline
   Period (d) & M (Me) & K (m/s) & ecc \\
\hline
9.8916 & 4.128 & 1.45 & 0.0960 \\
& & & \\
23.3678 & 6.283 & 1.67 & 0.1236 \\
& & & \\
33.2757 & 8.736 & 2.05 & 0.0832 \\
& & & \\
112.4589 & 2.375 & 0.38 & 0.2090 \\
& & & \\
273.2000 & 1.900 & 0.22 & 0.1600 \\
& & & \\
\hline
\end{tabular}
\end{center}
\end{table}

\begin{figure*}
\begin{center}
\includegraphics[width=160mm]{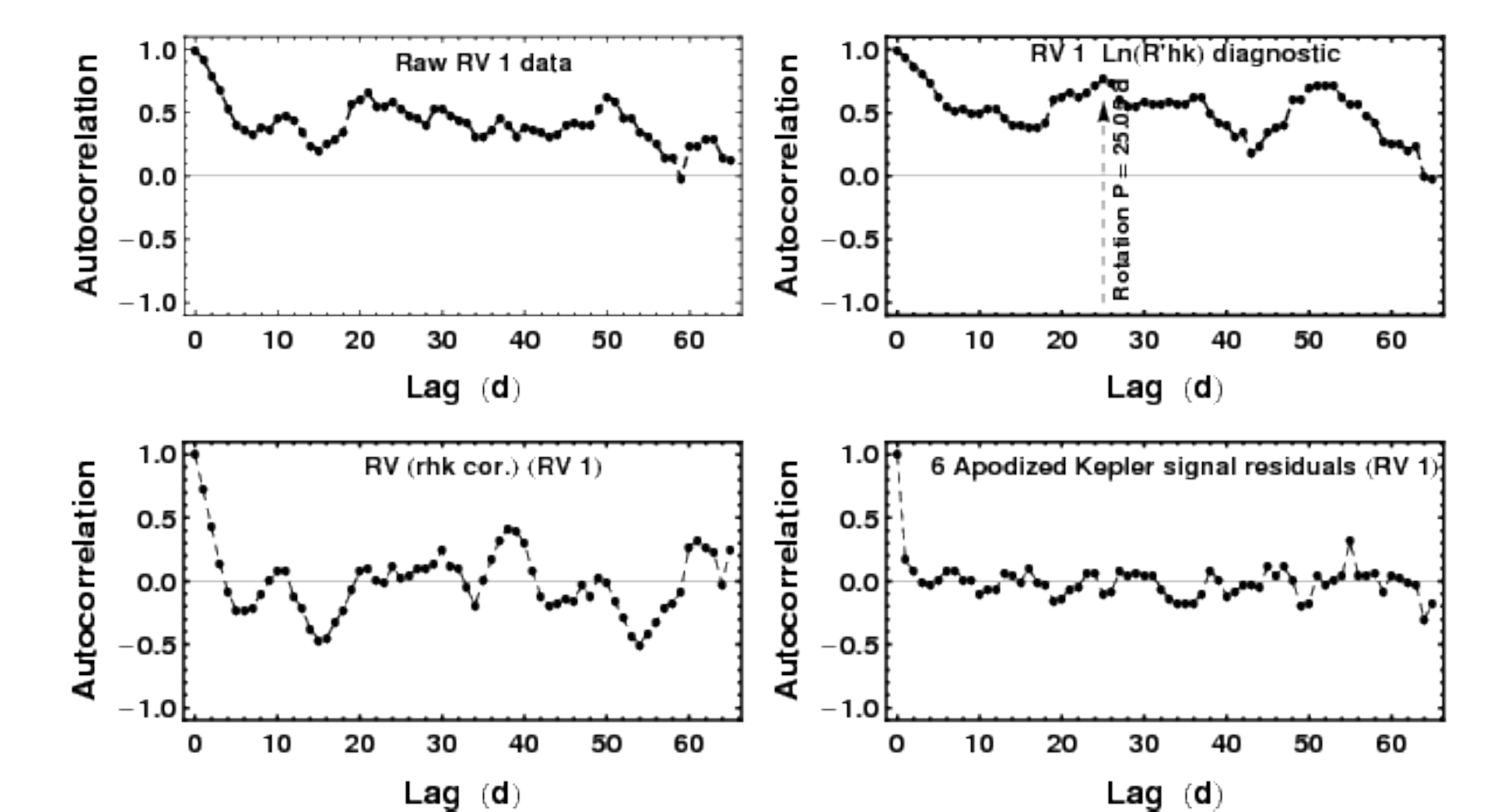}
\caption{The top row shows the autocorrelation function of the raw RV 1 data and the $\log(R'hk)$ diagnostic. The second row shows the autocorrelation function for the RV data with $\log(R'hk)$ regression removed (rhk corrected) and the 6 apodized Kepler model residuals.\label{fig:RV1_AutoCor5plot}}
\end{center}
\end{figure*}

The top row of Fig.~\ref{fig:RV1_AutoCor5plot} shows the autocorrelation function of the raw RV data and the $\log(R'hk)$ diagnostic. In this particular case, the highest peak in the autocorrelation function for the $\log(R'hk)$ diagnostic corresponds closely with the star's rotation period of 25.05 d. SA signals frequently occur at the rotation period and harmonics. The second row shows the autocorrelation function for the RV data with just the $ln(R’hk)$ regression removed (rhk corrected) and the 6 apodized Kepler model residuals. In this case, after the 6 apodized Kepler signals and $\log(R'hk)$ regression are removed, the autocorrelation of the residuals is looking close to white noise.

Table~\ref{tab:parerrorsRV1} gives the final parameter estimates for the 3 planetary candidates derived from a model consisting of 3 apodized Keplerians for the SA signals and 3 Keplerian for the planetary candidates. This analysis yielded the same set of periods found in the 6 AK model fit. The quoted eccentricity values are for the mode while for the other parameters we quote the median. The value immediately below in parenthesis is the MAP estimate. The errors on semi-major axis, $a$, and $ M \ sin i$ do not include the uncertainty in the mass of the host star. The last row gives the date of the periastron passage the occurs just prior to our reference date jdb $= 55855.666$ d which is the unweighted mean observation time for RV 1. 

\subsection{Model comparison}
\label{sec:ModComp}

To test the significance of the 3 planetary candidates we computed Bayes factors comparing the following four models: (a) zero Keplerian signals with only the $\log(R'hk)$ regression removed, (b) one 33.35 d Keplerian signal plus the $\log(R'hk)$ regression, (c) two (33.35 and 9.89 d) Keplerian signals plus $\log(R'hk)$ regression, and (d) three (33.35, 9.89, 23.4 d) Keplerian signals~\footnote{Note: these were the first three signals to emerge in the analysis.} plus $\log(R'hk)$ regression. Relative to model (d) the Bayes factors are 

$$3.8 \times 10^{-66}:7.2 \times 10^{-45}:1.6 \times 10^{-26}:1.0$$

The Bayes factors were computed from model marginal likelihoods based on the NRMC method (\citealt{Gregory2010, Gregory2011, Gregory2013}, and in more detail in Section 1.6 of the `Supplement to Bayesian Logical Data Analysis for the Physical Sciences,' available in the resources section of the Cambridge University Press website for my Textbook `Bayesian Logical Data Analysis for the Physical Sciences: A Comparative Approach with {\it Mathematica} Support').

\begin{figure*}
\begin{center}
\includegraphics[width=160mm]{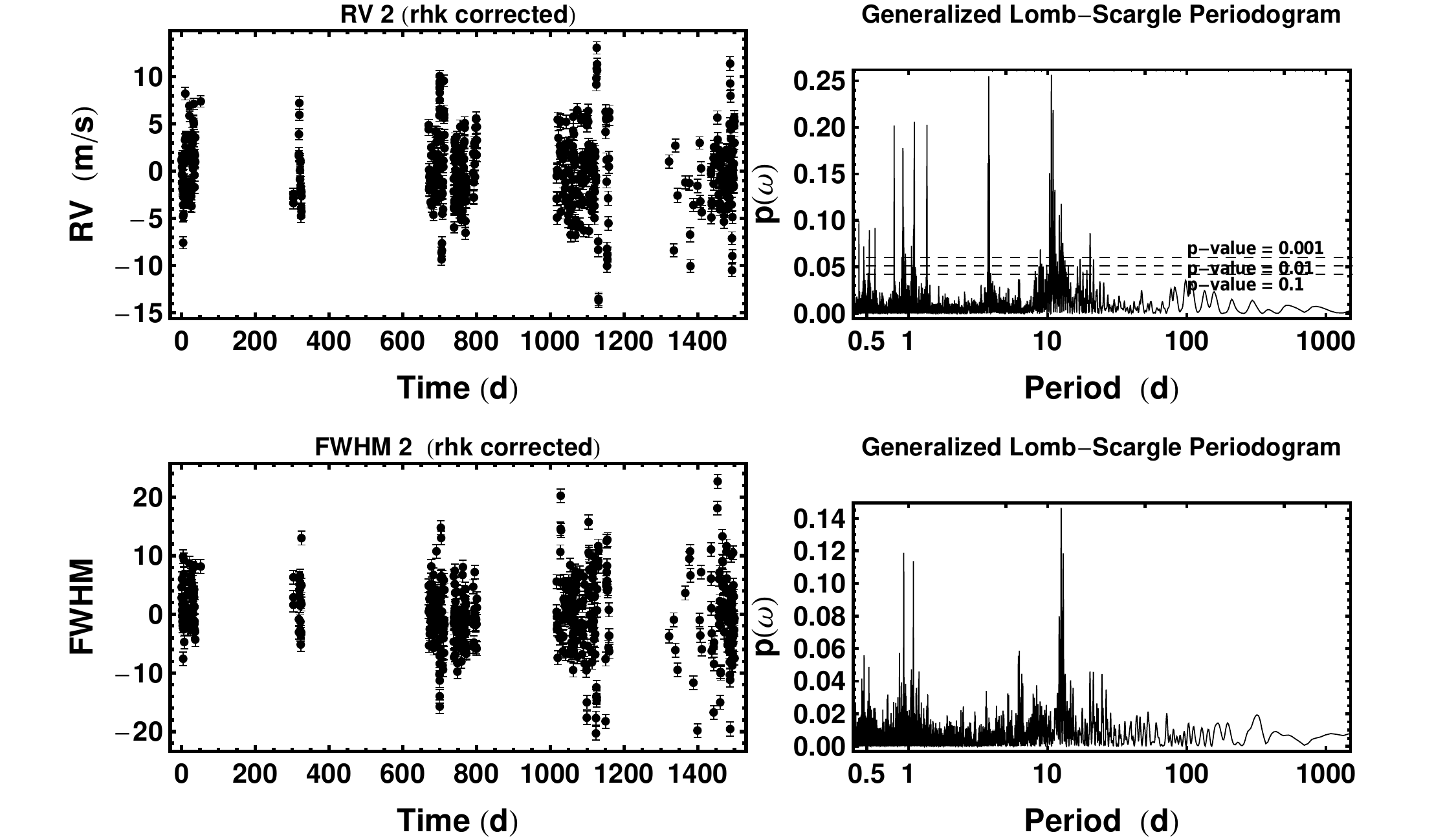}
\caption{The RV 2 data and FWHM (control)  after removing the best linear regression fits to $\log(R'hk)$ together with their GLS periodograms on the right.\label{fig:RV2rhkCor}}
\end{center}
\end{figure*}

\subsection{Second data set, RV 2}
\label{sec:RV2}

\begin{figure*}
\begin{center}
\includegraphics[width=110mm]{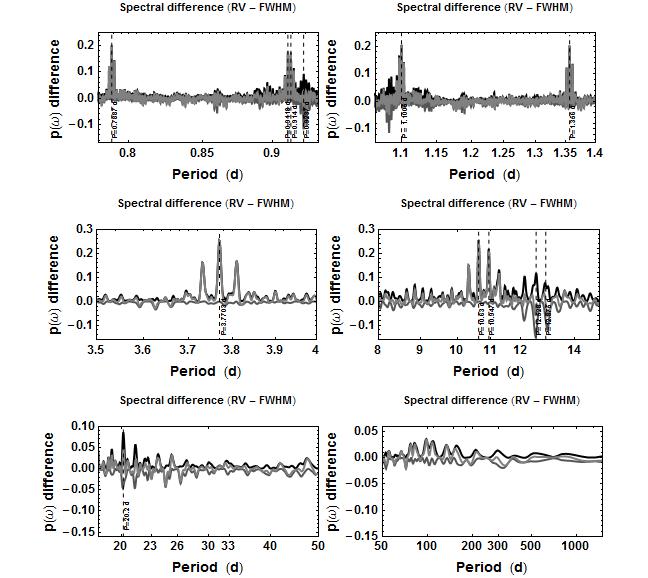}
\caption{The differential GLS periodogram for the rhk corrected RV 2 data for selected period ranges.\label{fig:RV2rhkCorDetail}}
\end{center}
\end{figure*}

Comparison of the raw RV and FWHM and $\log(R'hk)$ diagnostics indicated that there was a clear correlation between the three. The raw RV 2 data had a standard deviation of 8.58 m/s. After removing the best linear regression fit with $\log(R'hk)$ as the independent variable, the standard deviation was reduced to 3.95 m/s. The top two rows of  Fig.~\ref{fig:RV2rhkCor} show the rhk corrected RV and FWHM control together with their GLS periodogram on the right. Fig.~\ref{fig:RV2rhkCorDetail} show the differential periodogram for the rhk corrected RV data for selected period ranges. The strongest signal at $P = 10.64$ d has no significant counterpart in the control. Strong yearly and daily aliases of the signal are also evident. 

\begin{figure*}
\begin{center}
\includegraphics[width=110mm]{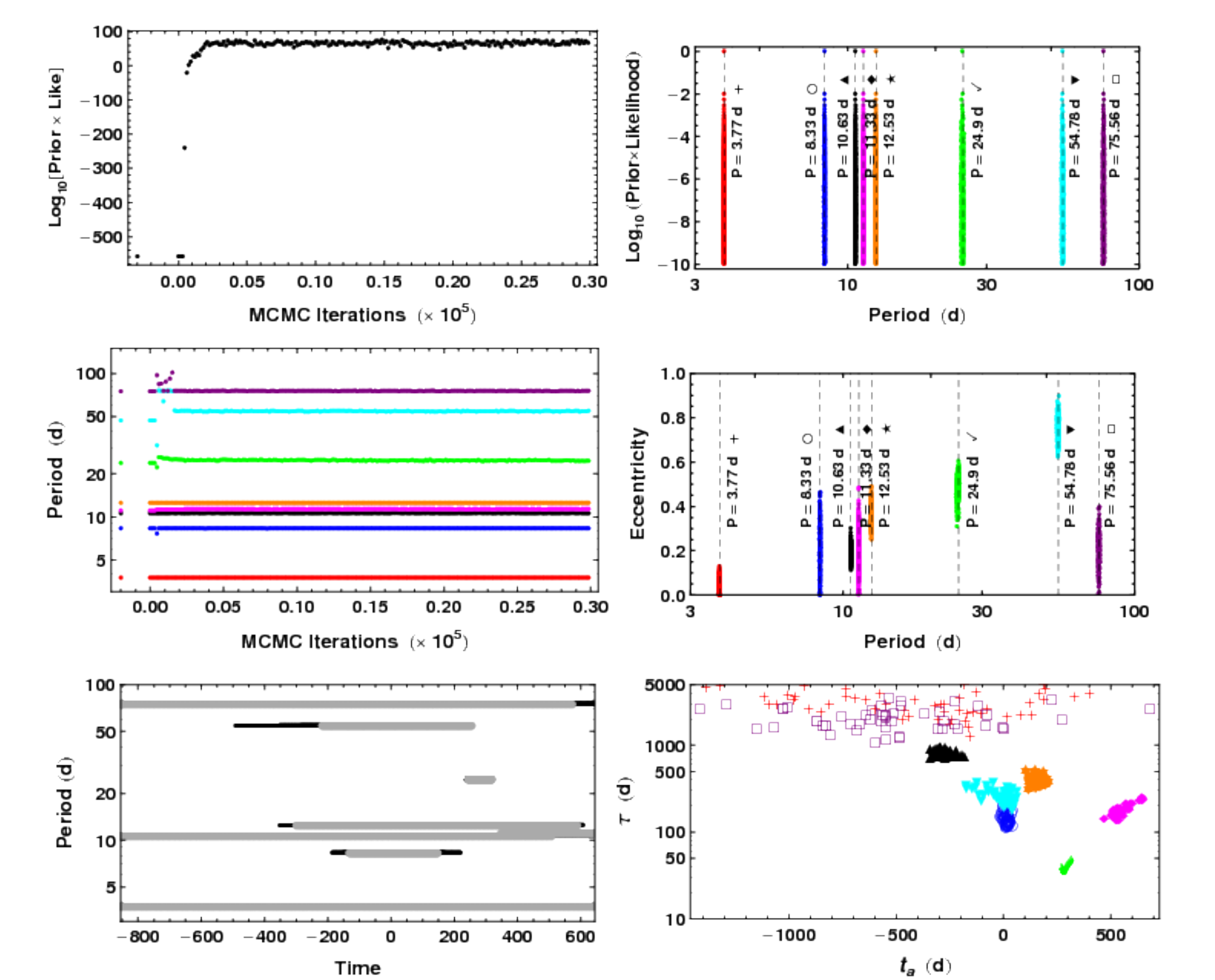}
\caption{The upper left panel is a plot of the Log$_{10}$[Prior $\times$ Likelihood] versus iteration for the 8 signal AK periodogram of the RV 2 data. The upper right shows Log$_{10}$[Prior $\times$ Likelihood] versus period showing the 8 periods detected. The middle left shows the values of the 8 unknown period parameters versus iteration number. The middle right shows the eccentricity parameters versus period parameters. The lower left shows the apodization window for each signal (gray trace for MAP values of $\tau$ and $t_a$, black for a representative set of samples which is mainly hidden below the gray). The lower right is a plot of the apodization time constant, $\tau$, versus apodization window center time, $t_a$.\label{fig:RV2_ApodKep8_grams}}
\end{center}
\end{figure*}

Fig.~\ref{fig:RV2_ApodKep8_grams} shows sample FMCMC results for a 8 AK signal model. The lower left panel shows  the span of the apodization window within the overall data window for each signal (gray trace for MAP values of $\tau$ and $t_a$, black for a representative set of samples which is mainly hidden below the gray). It is clear from this that there are three planetary candidates with periods of 3.77, 10.64, and 75.56 d whose apodization windows span the duration of the data. All three were initially classified as planetary on the basis of Fig.~\ref{fig:RV2_ApodKep8_grams} although the apodization time constant of $\sim 800$ d for the 10.64 d signal make it a borderline P candidate. Also, the presence of SA activity at a period of 11.32 d, in the close vicinity of the 10.64 d signal, perhaps called into question a planetary interpretation of the 10.64 d signal. For the competition, the 10.64 d signal was reported as a probable planetary signal. 

The weighted RMS residual $= 1.42$ m/s for the 8 AK signal model.  The standard deviation of the extra Gaussian white noise term $s = 1.25$ m/s. The mean measurement uncertainty was 0.674 m/s.

\begin{figure*}
\begin{center}
\includegraphics[width=90mm]{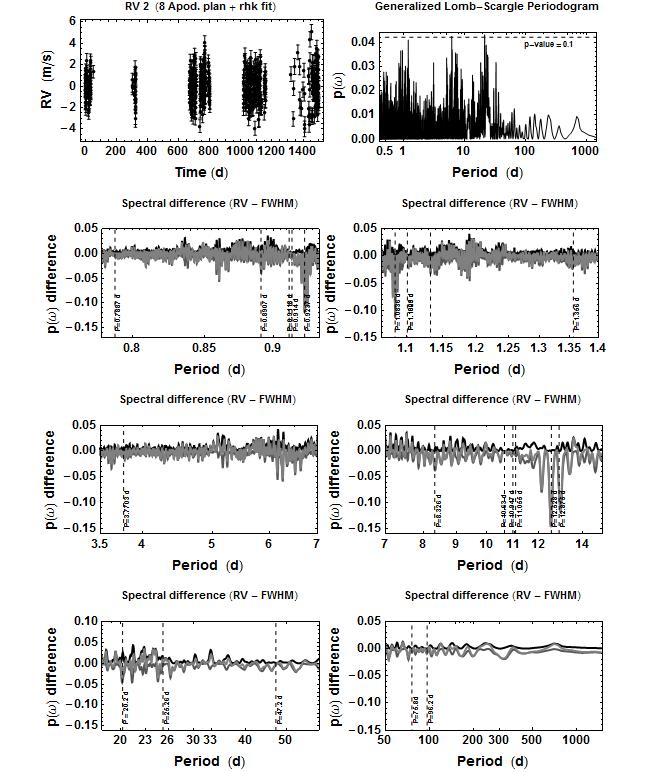}
\caption{The panels show the RV 2 eight signal AK model residuals (top left), its GLS periodogram (top right), and the differential periodogram for selected period ranges.\label{fig:RV2_8ApodRVdifSpecGray}}
\end{center}
\end{figure*}

Fig.~\ref{fig:RV2_8ApodRVdifSpecGray} shows the RV 2 eight signal AK model residuals  (i.e., $m = 8$ in Equation~(\ref{eq:orbit1})), its GLS periodogram (top right), and the differential periodogram for selected period ranges. The highest spectral peak in the residuals has a p-value just $< 0.1$. The dashed line representing the  p-value $= 0.01$ is not seen as it is above the top of the graph. These results are consistent with the conclusion that the $P =  3.77, 10.64, 75.56$ d signals are bona-fide planetary candidates. 

Table~\ref{tab:signalsRV2} lists (in order of extraction) the signals, their nominal parameter values and designation of signal type as planetary (P) or stellar activity (SA) based on the MCMC parameter estimates.  For the competition, the 10.64 d signal was reported as a probable planetary signal for the reasons mentioned above.  A Bayes factor analysis concluded that all three signals were highly significant compared to the next simpler model. 
  
\begin{table}
 \begin{center}
  \caption{Order of signals extracted from RV 2 together with approximate parameter values and designation of signal type as planetary (P) or stellar activity (SA) .}
  \label{tab:signalsRV2}
  \begin{tabular}{@{}llllll@{}}
  \hline
   Period   & K  & ecc & $\tau$ & $t_a$ & Signal\\
 (d) & (m/s) & & (d) & (d)& type\\
\hline
3.770 & 2.8 & 0.036 & 3572 & -296 & P \\
& & & & & \\
10.64 & 3.4 & 0.19 & 848 & -284 & P? \\
& & & & & \\
12.51 & 2.7 & 0.37 & 445 & 25 & SA \\
& & & & & \\
8.33 & 2.2 & 0.21 & 143 & 14 & SA \\
& & & & & \\
24.8 & 5.3 & 0.47 & 38 & 283 & SA \\
& & & & & \\
75.55 & 1.2 & 0.13 & 1181 & -1121 & P \\
& & & & & \\
11.324 & 3.0 & 0.18 & 181 & 542 & SA \\
& & & & & \\
54.6 & 1.4 & 0.70 & 226 & -100 & SA \\
& & & & & \\
\hline
\end{tabular}
\end{center}
\end{table}

Table~\ref{tab:SIMsignalsRV2} shows the true $P, K, e$ parameters of the 5 planetary signals that were employed in the RV 2 simulation. 
Clearly the three planetary signals with $K > 1$ were recovered in our analysis but there is no clear evidence for the two weaker planetary signals at $P = 5.79\ \& \ 20.16$ d in the GLS residuals of Fig.~\ref{fig:RV2_8ApodRVdifSpecGray}.
\begin{table}
 \begin{center}
  \caption{The true $P, K, e$ parameters of the 5 planetary signals that were employed in the RV 2 simulation.}
  \label{tab:SIMsignalsRV2}
  \begin{tabular}{@{}llll@{}}
  \hline
   Period (d)  & M (Me) & K (m/s) & ecc \\
\hline
3.7700 & 5.6675 & 2.75 & 0.0510 \\
& & & \\
5.7936 & 0.6296 & 0.27 & 0.1140 \\
& & & \\
10.6370 & 8.2432 & 2.85 & 0.1372 \\
& & & \\
20.1644 & 1.2318 & 0.34 & 0.0824 \\
& & & \\
75.2835 & 7.4140 & 1.35 & 0.1940 \\
& & & \\
\hline
\end{tabular}
\end{center}
\end{table}
\begin{table}
 \begin{center}
  \caption{ RV 2 parameter estimates and 68\% credible boundaries for 3 planetary candidates discussed in the text. The value immediately below in parenthesis is the MAP estimate. The errors on semi-major axis, $a$, and $ M \ sin i$ do not include the uncertainty in the mass of the host star.}
  \label{tab:parerrorsRV2}
  \begin{tabular}{@{}lllllll@{}}
  \hline
   Parameter  & planet 1 & planet 2 & planet 3  \\
\hline
$P$  (d) & $3.7702_{-.00016}^{+.00018}$ & $10.6395_{-.0015}^{+.0015}$ & $75.76_{-.16}^{+.18}$  \\
& (3.7702)& (10.6397) & (75.78)\\
& & &\\
$K$ (m/s) & $2.7_{-0.1}^{+0.1}$ & $2.9_{-0.11}^{+0.11}$ & $1.37_{-0.12}^{+0.13}$  \\
& (2.72) & (2.95) & (1.37) \\
& & &\\
$e$ & $0.000_{-0.000}^{+0.017}$ & $0.18_{-0.03}^{+0.03}$ & $0.24_{-0.07}^{+0.07}$   \\
& (0.04) & (0.20) & (0.25)  \\
& & &\\
$\omega$  (rad) & $5.9_{-1.1}^{+1.1}$ & $0.70_{-0.18}^{+0.16}$  & $1.3_{-0.3}^{+0.4}$ \\
& (5.96) & (0.68) & (1.32)  \\
& & &\\
$a$  (au) & $0.0436398_{-.0000013}^{+.0000013}$ & $0.087147_{-.000008}^{+.000008}$ & $0.32254_{-.00044}^{+.00051}$\\
& (0.0436396) & (0.087148)  & (0.32260) \\
& & &\\
$M \sin i$  & $5.6_{-0.2}^{+0.2}$ & $8.3_{-0.3}^{+0.3}$  & $7.5_{-0.6}^{+0.6}$\\
($M_E$) & (5.6) & (8.4) & (7.4) \\
& & &\\
Periastron & $55850.5_{-0.7}^{+0.6}$ & $55852.0_{-0.3}^{+0.3}$  & $55812_{-4}^{+4}$ \\
\ passage &  (55854.3) & (55851.9) & (55812.2) \\
\hline
\end{tabular}
\end{center}
\end{table}

The same set of periods was obtained using a model consisting of 5 AK signals and 3 Keplerian signals. Final parameter estimates for the 3 planetary candidates were derived from this latter model and are given in Table~\ref{tab:parerrorsRV2}. The quoted eccentricity values are for the mode while for the other parameters we quote the median. The value immediately below in parenthesis is the MAP estimate. The errors on semi-major axis, $a$, and $ M \ sin i$ do not include the uncertainty in the mass of the host star. The last row gives the date of the periastron passage the occurs just prior to our reference date jdb $= 55855.6693$ d which is the unweighted mean observation time for RV 2.

To test the significance of the first two planetary candidates we computed Bayes factors comparing the following models: (a) zero Keplerian signals with only the $\log(R'hk)$ regression removed, (b) one 3.77 d Keplerian signal plus the $\log(R'hk)$ regression, (c) two (3.77 and 10.64 d) Keplerian signals plus $\log(R'hk)$ regression. The Bayes factors relative to model (c) are 

$$5.6 \times 10^{-60}:9.3 \times 10^{-37}:1.0$$

The three planet model was compared to the 2 planet model by comparing the two models: (a) three apodized Keplerians ($P = 8.33, 12.5, 24.8$ d) plus 2 Keplerians (3.77, 10.64 d) plus the $\log(R'hk)$ regression to (b) three apodized Keplerians ($P = 8.33, 12.5, 24.8$ d) plus 3 Keplerians (3.77, 10.64, 75.7 d) plus the $\log(R'hk)$ regression. The Bayes factor relative to model (b) is $3.4 \times 10^{-9}$:1.0.

\begin{figure*}
\begin{center}
\includegraphics[width=150mm]{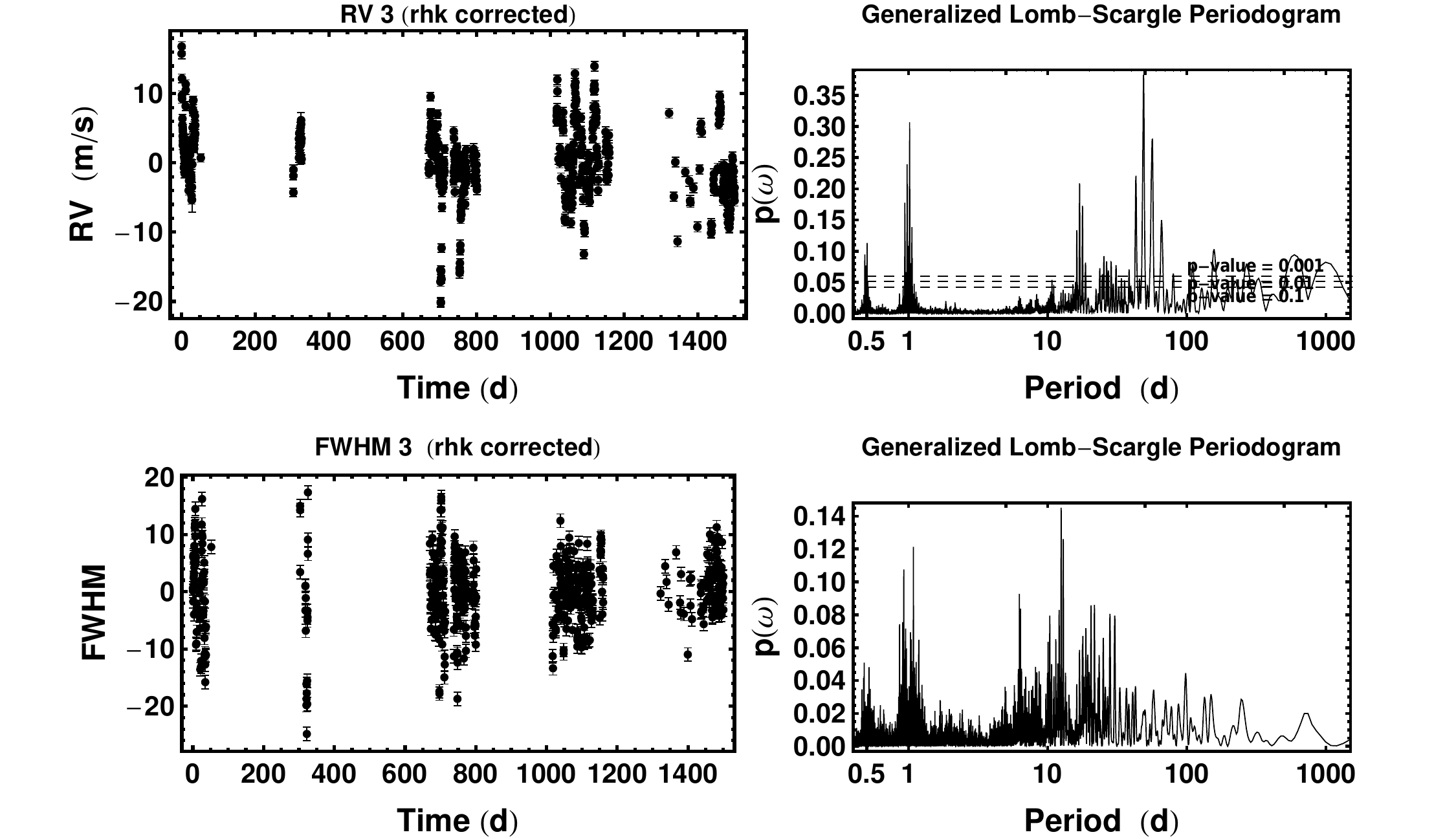}
\caption{The RV 3 data and FWHM (control) after removing the $\log(R'hk)$ diagnostics (rhk corrected) together with their GLS periodograms on the right.\label{fig:RV3rhkCor}}
\end{center}
\end{figure*}

\subsection{Third data set, RV 3}
\label{sec:RV3}

\begin{figure*}
\begin{center}
\includegraphics[width=150mm]{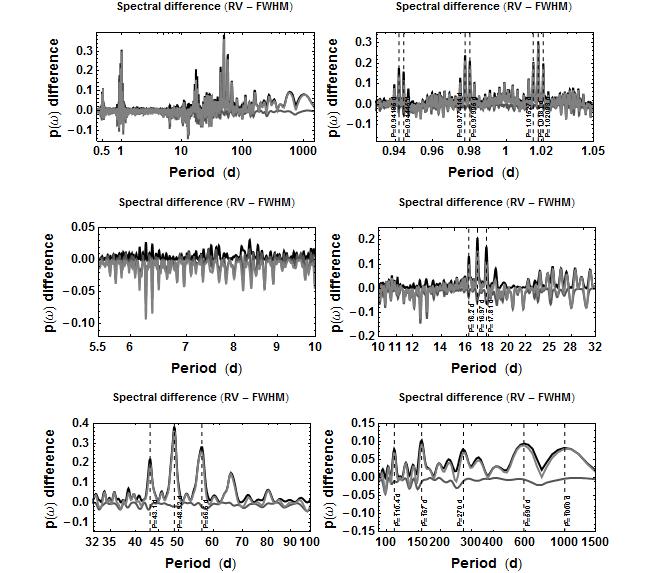}
\caption{The differential GLS periodogram for the rhk corrected RV 3 data for selected period ranges.\label{fig:RV3rhkCorDetail}}
\end{center}
\end{figure*}

Comparison of the raw RV and FWHM and $\log(R'hk)$ diagnostics indicated that there was a clear correlation between the three. The raw RV 3 data had a standard deviation of 10.85 m/s. After removing the best linear regression fit with $\log(R'hk)$ as the independent variable, the standard deviation was reduced to 5.49 m/s. The top two rows of  Fig.~\ref{fig:RV3rhkCor} show the rhk corrected RV data and FWHM control together with their GLS periodogram on the right. Fig.~\ref{fig:RV3rhkCorDetail} show the differential periodogram for the rhk corrected RV data for selected period ranges. The strongest signal at $P = 48.8$ d has no significant counterpart in the control. Strong yearly and daily aliases of the 48.8 d signal are also evident. 

Table~\ref{tab:signalsRV3} lists (in order of extraction) the signals, their nominal parameter values and designation of signal type as planetary (P) or stellar activity (SA) based on the MCMC parameter estimates. The parameter values were taken from the 6 apodized Kepler signal fit. The weighted RMS residual $= 1.79$ m/s for the 6 AK signal model. The standard deviation of the extra Gaussian white noise term $s = 1.66$ m/s. The mean measurement uncertainty was 0.674 m/s. Three signals (P = 17, 48.8, and $\sim 1100$ d) were initially classified as planetary although the there was a noticeable signature ($p(\omega) > 0.05$) of the 17 day signal in the FWHM control trace which resulted in a probable P classification. 
\begin{figure*}
\begin{center}
\includegraphics[width=120mm]{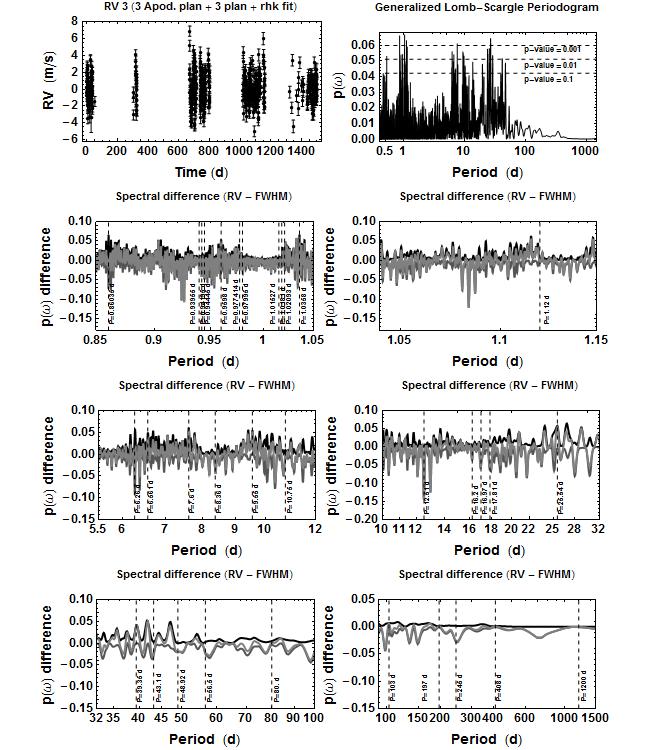}
\caption{The top left panel shows the RV 3 residuals for a model consisting of of 3 AK signals, 3 Keplerian signals plus the $\log(R'hk)$ regression. The other panels show the GLS periodogram (top right), and the differential periodogram for selected period ranges.\label{fig:RV3_3ApodRV3plan_difSpecGray}}
\end{center}
\end{figure*}
\begin{figure}
\begin{center}
\includegraphics[width=70mm]{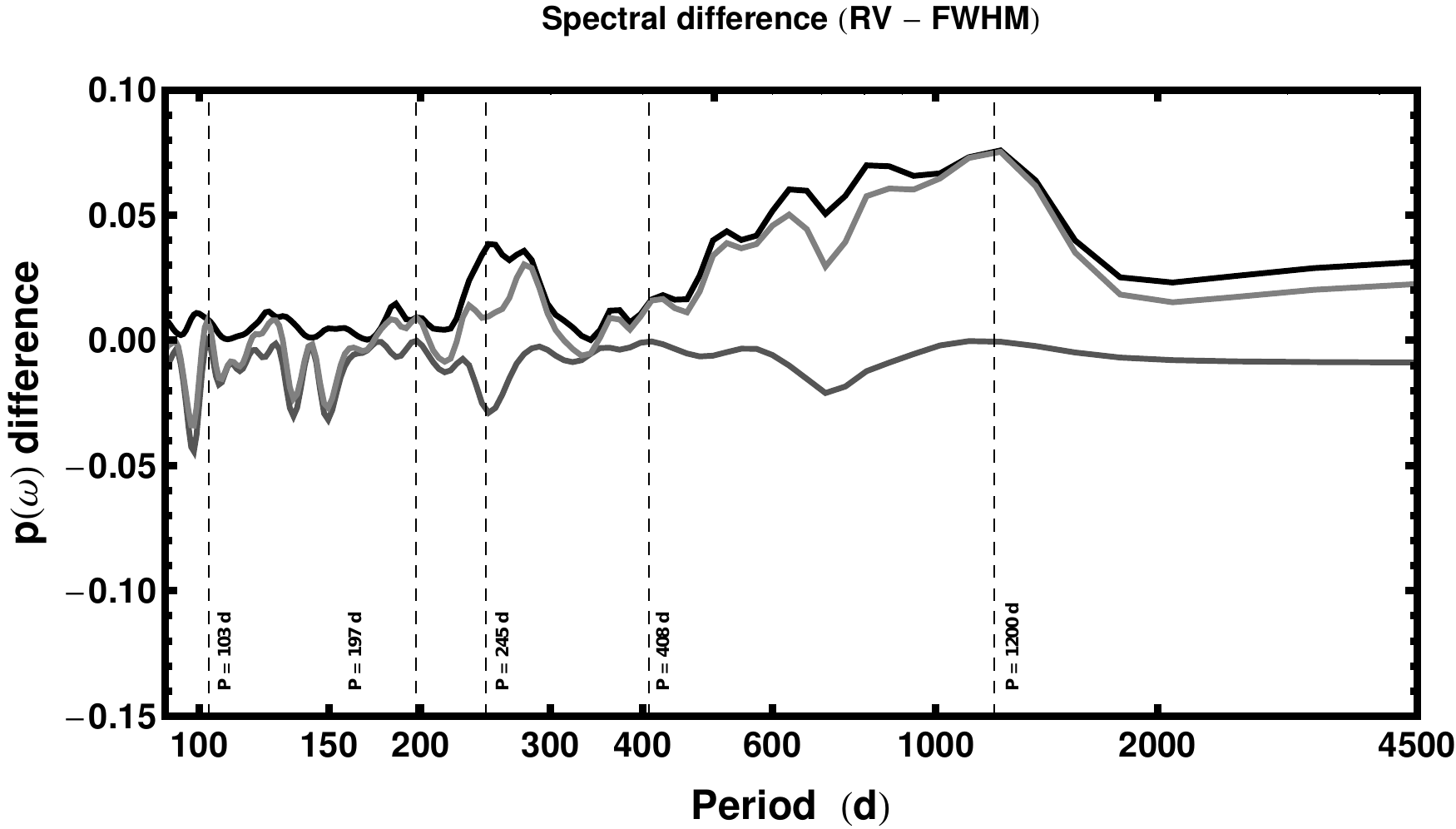}
\caption{A portion of the differential GLS periodogram illustrating the $\sim 1200$ d signal for the RV 3 data after the 12.5, 17, 48.8 and 333 d signals  plus the $\log(R'hk)$ regression were removed.\label{fig:1200d}}
\end{center}
\end{figure}
\begin{figure}
\begin{center}
\includegraphics[width=70mm]{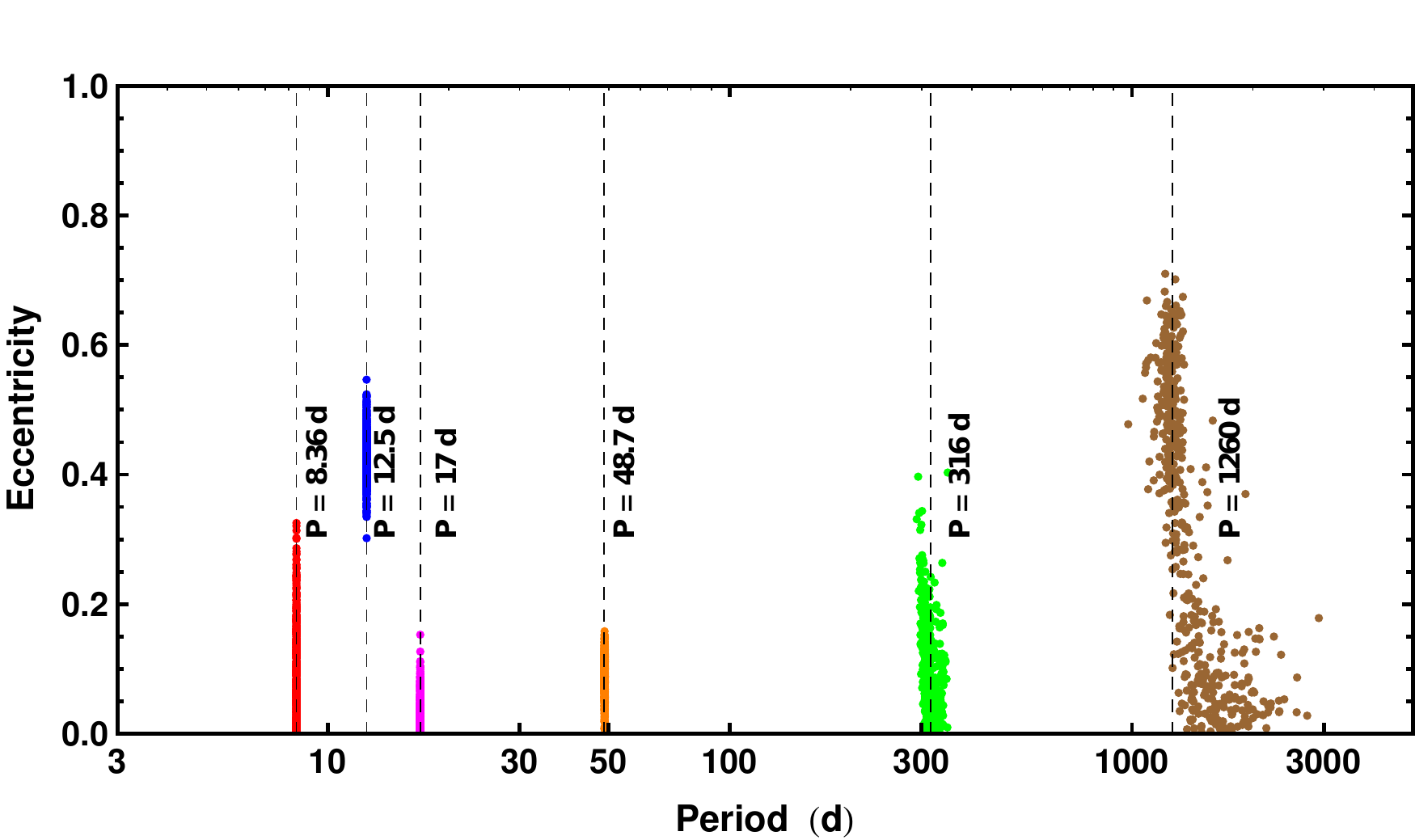}
\caption{Figure for RV 3 data shows the eccentricity versus period parameters for the 6 signals obtained using a model consisting of 3 AK signals, 3 Keplerian signals plus $\log(R'hk)$ regression.\label{fig:RV3_ApodKep6_grams}}
\end{center}
\end{figure} 

\begin{table}
\begin{center}
  \caption{Order of signals extracted from RV 3 together with approximate parameter values and designation of signal type as planetary (P) or stellar activity (SA) .}
  \label{tab:signalsRV3}
  \begin{tabular}{@{}llllll@{}}
  \hline
   Period   & K  & ecc & $\tau$ & $t_a$ & Signal\\
 (d) & (m/s) & & (d) & (d)& type\\
\hline
48.83 & 2.8 & 0.09 & 2579 & 578 & P \\
& & & & & \\
16.991 & 4.1 & 0.03 & 4432 & 164 & P? \\
& & & & & \\
12.501 & 4.7 & 0.43 & 571 & -609 & SA \\
& & & & & \\
333 & 14 & 0.29 & 500 & -1089 & SA \\
& & & & & \\
$\sim 1100$ & 2.3 & 0.5 & 1015 & 279 & P \\
& & & & & \\
8.365 & 2.8 & 0.1 & 872 & -1088 & SA \\
& & & & & \\
\hline
\end{tabular}
\end{center}
\end{table}
The $\sim 1100$ day signal only became prominent, as shown in Fig.~\ref{fig:1200d}, after signals at 12.5, 17, 48.8 and 333 d were first removed. 

The same set of periods was obtained using a model consisting of 3 AK signals plus 3 Keplerian signals and the $\log(R'hk)$ regression term. Fig.~\ref{fig:RV3_ApodKep6_grams} shows the eccentricity versus period parameters for the 6 signals. The one difference from the 6 AK signal fit is that now the signal near $P \sim 1100$ d has a low eccentricity tail extending to approximately double that period.

To test the significance of the first two planetary candidates, we computed Bayes factors comparing the following models: (a) zero Keplerian signals with only the $\log(R'hk)$ regression removed, (b) one 48 d Keplerian signal plus the $\log(R'hk)$ regression, (c) two (48 and 17 d) Keplerian signals plus $\log(R'hk)$ regression. The  Bayes factors relative to model (c) are 

$$2.7 \times 10^{-101}:1.2 \times 10^{-54}:1.0$$

The three planet model was compared to the 2 planet model by comparing the two models: (a) two apodized Keplerians ($P = 12.5, 333$ d) plus 2 Keplerians (48, 17 d) plus the $\log(R'hk)$ regression to (b) two apodized Keplerians ($P = 12.5, 333$ d) plus 3 Keplerians (48, 17, $\sim 1100$ d) plus the $\log(R'hk)$ regression. The Bayes factor relative to model (b) is $1.3 \times 10^{-13}$:1.0.

Final parameter estimates for the planetary candidates were derived from this 3 AK signals plus 3 Keplerian signals plus $\log(R'hk)$ regression model. The results are given in Table~\ref{tab:parerrorsRV3}. The last row gives the date of the periastron passage the occurs just prior to our reference date jdb $= 55855.6693$ d which is the unweighted mean observation time for RV 3.
\begin{table}
\begin{center}
  \caption{ RV 3 parameter estimates and 68\% credible boundaries for 3 planetary candidates discussed in the text.  The quoted eccentricity values are for the mode while for the other parameters we quote the median. The value immediately below in parenthesis is the MAP estimate. The errors on semi-major axis, $a$, and $ M \ sin i$ do not include the uncertainty in the mass of the host star.}
  \label{tab:parerrorsRV3}
  \begin{tabular}{@{}lllllll@{}}
  \hline
   Parameter  & planet 1 & planet 2 & planet 3  \\
\hline
$P$  (d) & $16.992_{-.003}^{+.003}$ & $48.82_{-.03}^{+.03}$ & $1306_{-133}^{+191}$  \\
& (16.992)& (48.83) & (1067)\\
& & & \\
$K$ (m/s) & $3.98_{-0.11}^{+0.14}$ & $5.0_{-0.2}^{+0.2}$ & $2.8_{-0.5}^{+0.4}$  \\
& (4.0) & (5.1) & (2.3) \\
& & & \\
$e$ & $0.00_{-0.0}^{+0.04}$ & $0.09_{-0.03}^{+0.03}$ & $0.00_{-0.00}^{+0.61}$   \\
& (0.00) & (0.10) & (0.57)  \\
& & &\\
$\omega$  (rad) & $5.4_{-1.3}^{+1.3}$ & $5.5_{-0.3}^{+0.3}$  & $3.3_{-0.5}^{+0.5}$ \\
& (5.3) & (5.5) & (3.1)  \\
& & &  \\
$a$  (au) & $0.11907_{-.00001}^{+.00001}$ & $0.24066_{-.00009}^{+.00009}$ & $2.15_{-.13}^{+.21}$\\
& (0.11907) & (0.24068)  & (1.88) \\
& & &\\
$M \sin i$  & $13.5_{-0.4}^{+0.5}$ & $24.2_{-0.7}^{+0.7}$  & $36_{-10}^{+9}$\\
($M_E$) & (13.6) & (24.5) & (25) \\
& & &\\
Periastron & $55845_{-3}^{+3}$ & $55822_{-2}^{+2}$  & $55145_{-201}^{+146}$ \\
\ passage &  (55847) & (55821) & (55364) \\
\hline
\end{tabular}
\end{center}
\end{table}

Table~\ref{tab:SIMsignalsRV3} shows the true $P, M, K, e$ parameters of the 7 planetary signals that were employed in the RV 3 simulation. 
\begin{table}
 \centering
  \caption{The true $P, M, K, e$ parameters of the 7 planetary signals that were employed in the RV 3 simulation.}
  \label{tab:SIMsignalsRV3}
  \begin{tabular}{@{}llll@{}}
  \hline
   Period (d)  & M & K (m/s) & ecc \\
\hline
1.1188 & 1.3205 & 0.96 & 0.0000 \\
& & & \\
17.0110 & 12.4176 & 3.68 & 0.14872\\
& & & \\
26.3000 & 1.500 & 0.38 & 0.0770 \\
& & & \\
48.7521 & 24.8920 & 5.14 & 0.0637 \\
& & & \\
201.5000 & 3.2000 & 0.42 & 0.2000 \\
& & & \\
595.9800 & 21.1860 & 1.91 & 0.1317 \\
& & & \\
2315.4400 & 67.2590 & 3.87 & 0.1555 \\
\hline
\end{tabular}
\end{table}

Clearly two planetary signals with $K > 1$ were recovered in our analysis and the $P/2$ harmonic of the $P = 2315$ d signal. There is evidence for a signal at $P = \sim 590$ d in the rhk corrected data (Fig.~\ref{fig:RV3rhkCorDetail}) before the removal of any periodic signals. It does not survive the removal of the larger $\sim 1200$ d signal, possibly because it is essential a harmonic of the $\sim 1200$ d signal. Fig.~\ref{fig:RV3_3ApodRV3plan_difSpecGray} shows the residuals, their GLS and differential GLS periodogram, for a model consisting of 3 AK signals plus 3 Keplerian signals plus $\log(R'hk)$ regression. The highest residual is at $P = 1.036$d and has a p-value $\sim 0.001$. Attempts to extract further unique signals were unsuccessful. Weak features are in evidence close to $P = 1.1188$ and $26.3$ d but nothing close to $201$ d.

\subsection{Fourth data set, RV 4}
\label{sec:RV4}

\begin{figure*}
\begin{center}
\includegraphics[width=130mm]{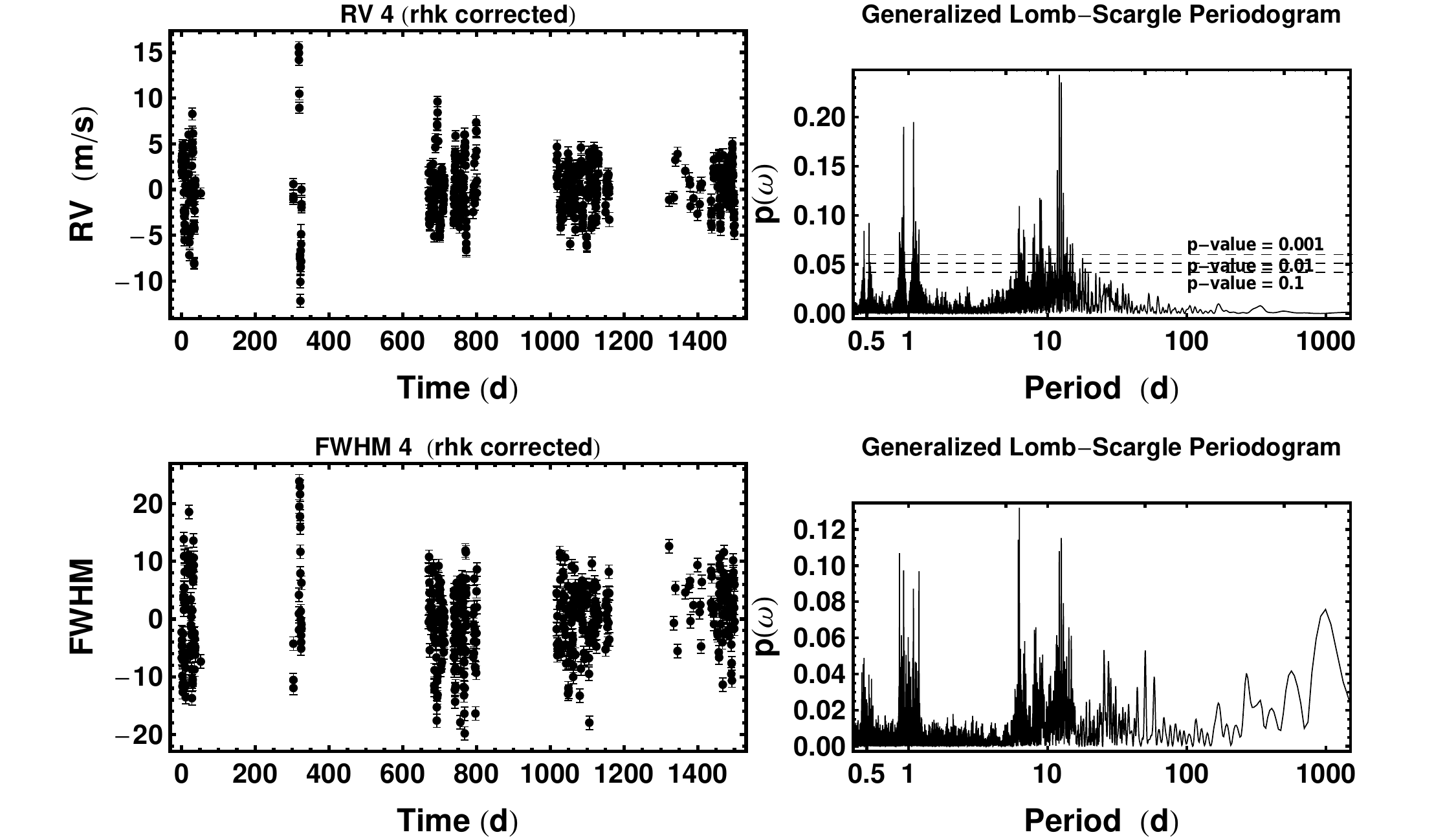}
\caption{The RV 4 data and FWHM (control) after removing the best linear regression fits to $\log(R'hk)$ together with their GLS periodograms on the right.\label{fig:RV4rhkCor}}
\end{center}
\end{figure*}

\begin{figure*}
\begin{center}
\includegraphics[width=130mm]{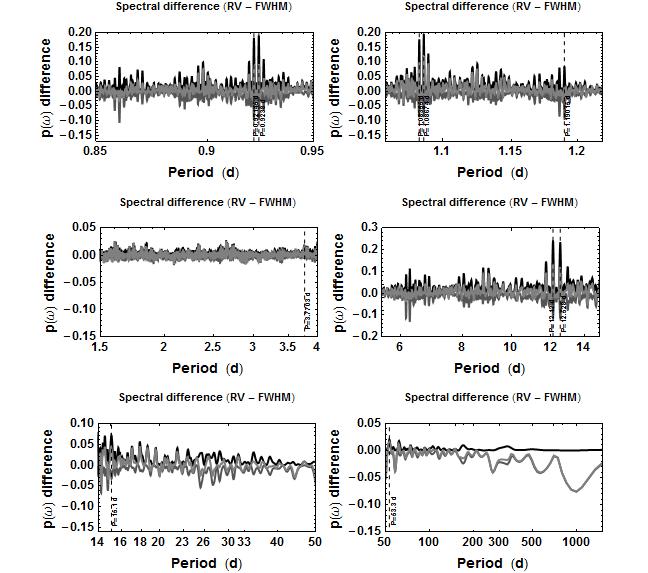}
\caption{The differential GLS periodogram for the rhk corrected RV 4 data for selected period ranges.\label{fig:RV4rhkCorDetail}}
\end{center}
\end{figure*}

Comparison of the raw RV and FWHM and $\log(R'hk)$ diagnostics indicated that there was a clear correlation between the three. The raw RV 4 data had a standard deviation of 8.30 m/s. After removing the best linear regression fit with $\log(R'hk)$ as the independent variable, the standard deviation was reduced to 3.32 m/s. The top two rows of  Fig.~\ref{fig:RV4rhkCor} show the rhk corrected RV data and FWHM control together with their GLS periodogram on the right. Fig.~\ref{fig:RV4rhkCorDetail} show the differential periodogram for the rhk corrected RV data for selected period ranges.  
\begin{table}
  \caption{Order of signals extracted from RV 4 together with approximate parameter values and designation of signal type as planetary (P) or stellar activity (SA) .}
  \label{tab:signalsRV4}
  \begin{tabular}{@{}llllll@{}}
  \hline
   Period   & K  & ecc & $\tau$ & $t_a$ & Signal\\
 (d) & (m/s) & & (d) & (d)& type\\
\hline
12.55 & 10.0 & 0.24 & 195 & -545 & SA \\
& & & & & \\
11.33 & 2.22 & 0.058 & 198 & 20 & SA \\
& & & & & \\
11.74 & 1.51 & 0.16 & 2646 & -547 & P? \\
& & & & & \\
9.25 & 2.4 & 0.054 & 588 & -609 & SA \\
& & & & & \\
6.26 & 1.59  & 0.056 & 894 & -794 & SA \\
& & & & & \\
27.34 & 2.9 & 0.59 & 175 & 36 & SA \\
& & & & & \\
0.9435 & 0.67 & 0.21 & 1505 & 576 & P? \\
& & & & & \\
36.32 & 2.2 & 0.73 & 721 & 904 & SA \\
& & & & & \\
\hline
\end{tabular}
\end{table}

\begin{figure*}
\begin{center}
\includegraphics[width=150mm]{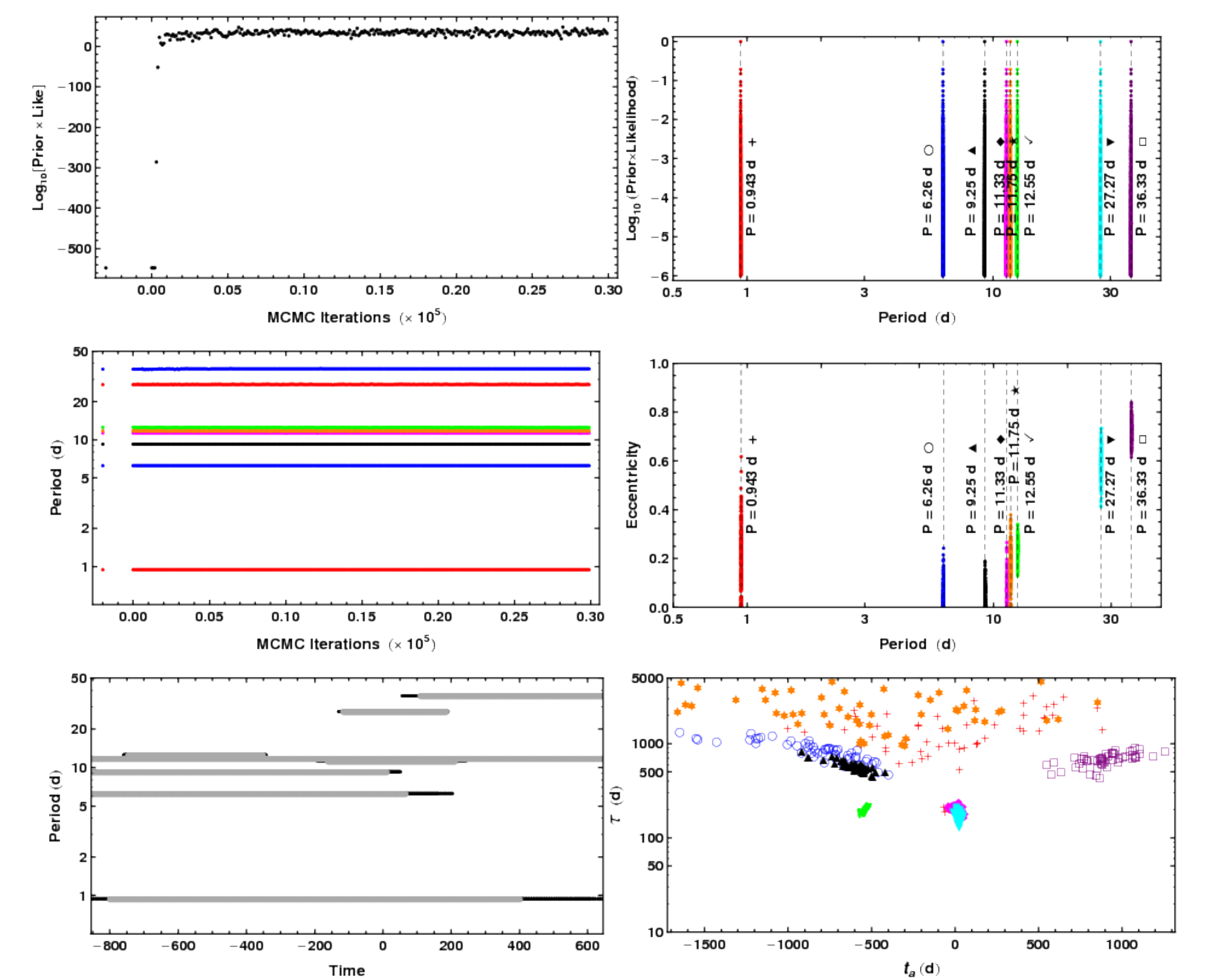}
\caption{The upper left panel is a plot of the Log$_{10}$[Prior $\times$ Likelihood] versus iteration for the 8 signal AK periodogram of the RV 4 data. The upper right shows Log$_{10}$[Prior $\times$ Likelihood] versus period showing the 8 periods detected. The middle left shows the values of the 8 unknown period parameters versus iteration number. The middle right shows the eccentricity parameters versus period parameters. The lower left shows the apodization window for each signal (gray trace for MAP values of $\tau$ and $t_a$, black for a representative set of samples which is mainly hidden below the gray). The lower right is a plot of the apodization time constant, $\tau$, versus apodization window center time, $t_a$.\label{fig:RV4_ApodKep8_grams}}
\end{center}
\end{figure*}

\begin{figure*}
\begin{center}
\includegraphics[width=130mm]{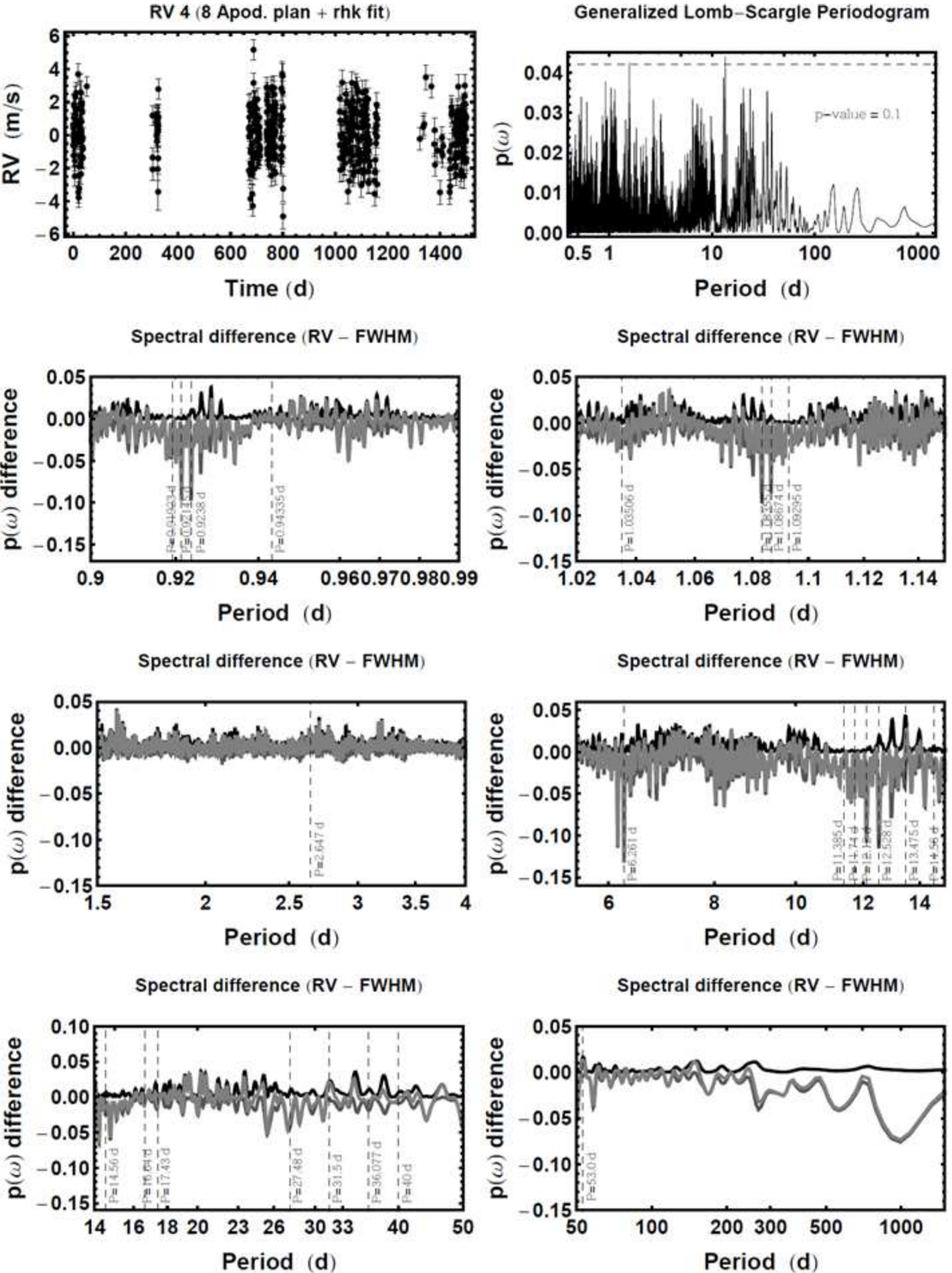}
\caption{The RV 4 residuals, their GLS periodogram and differential periodogram for the 8 AK signal model.\label{fig:RV4_8Apod_difSpecGray}}
\end{center}
\end{figure*}

Fig.~\ref{fig:RV4_ApodKep8_grams} shows FMCMC results for a eight AK signal model. The lower left panel shows the span of the apodization window within the overall data window for each signal (gray trace for MAP values of $\tau$ and $t_a$, black for a representative set of samples which is mainly hidden below the gray). The MAP apodization window for the $P = 11.75$ signal clearly  spans the duration of the data and on this basis alone it would be classified as a planetary candidate. However, the FWHM control trace (see Fig.~\ref{fig:RV4_8Apod_difSpecGray}) has a $p(\omega) = -0.06$ at the 11.75 d period, which is the main reason it was classified as (P?). Also, the proximity of the SA signals at 11.33 suggested that the 11.75 d signal might be another SA component. 

Based on the apodization window, the $P = 0.943$ d signal was also initially classified as a possibly planetary (P?). Bayesian model comparison (see below) indicates that the $P = 0.943$ d signal is not significant.

The weighted RMS residual $= 1.52$ m/s for the 8 AK signal model. The standard deviation of the extra Gaussian white noise term $s = 1.36$ m/s.  The mean measurement uncertainty was 0.674  m/s.

Table~\ref{tab:signalsRV4} lists (in order of extraction) the signals, their nominal parameter values and designation of signal type as planetary (P), (P?) or stellar activity (SA) based on the MCMC parameter estimates. The parameter values were taken from the 8 apodized Kepler signal fit.  Two signals (P = 11.75, and 0.943 d) which as mentioned above were initially classified as possibly planetary candidates.

To test the significance of the possible $P = 0.943$ d planetary candidate,  we computed the Bayes factor comparing the following models: (a) 6 apodized Keplerian signals plus the $\log(R'hk)$ regression, and (b)  6 apodized Keplerian signals plus one Keplerian (0.943 d) plus the $\log(R'hk)$ regression. The Bayes factor of (a) relative to model (b) are $64:1.0$ which clearly indicates that the 0.943 d signal is not significant. That left one possible planetary candidate at $P = 11.75$ d. When the results of the challenge were announced it turned out that RV 4 data set contained no planetary signal.

\newpage

\subsection{Fifth data set, RV 5}
\label{sec:RV5}

\begin{figure*}
\begin{center}
\includegraphics[width=150mm]{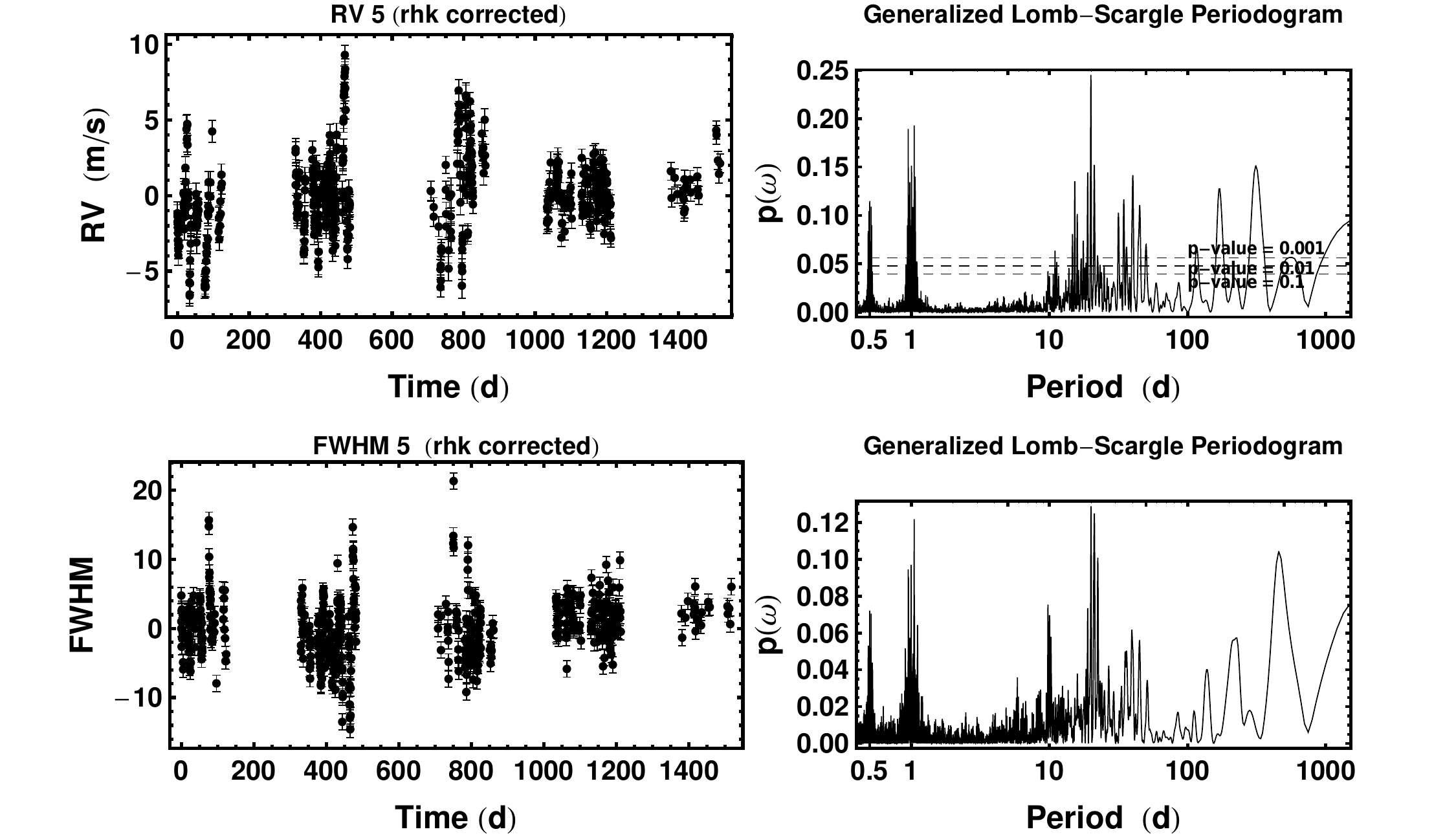}
\caption{The RV 5 data and FWHM (control) after removing the best linear regression fits to $\log(R'hk)$ together with their GLS periodograms on the right.\label{fig:RV5rhkCor}}
\end{center}
\end{figure*}

\begin{figure*}
\begin{center}
\includegraphics[width=150mm]{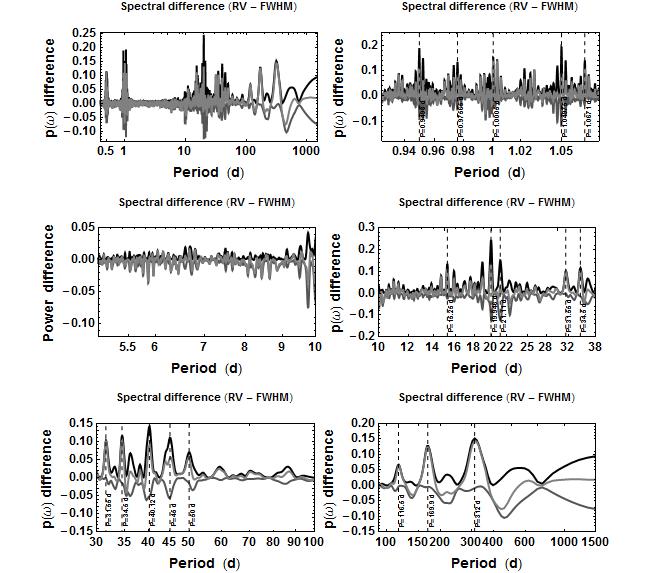}
\caption{The differential GLS periodogram for the rhk corrected RV 5 data for selected period ranges.\label{fig:RV5rhkCorDetail}}
\end{center}
\end{figure*}

Comparison of the raw RV and FWHM and $\log(R'hk)$ diagnostics indicated that there was a strong correlation between the three. The raw RV 5 data had a standard deviation of 8.92 m/s. After removing the best linear regression fit with $\log(R'hk)$ as the independent variable, the standard deviation was reduced to 2.59 m/s. The top two rows of  Fig.~\ref{fig:RV5rhkCor} show the rhk corrected RV data and FWHM control together with their GLS periodogram on the right. Fig.~\ref{fig:RV5rhkCorDetail} show the differential periodogram for the RV data (rhk corrected) for selected period ranges.  

\begin{figure*}
\begin{center}
\includegraphics[width=150mm]{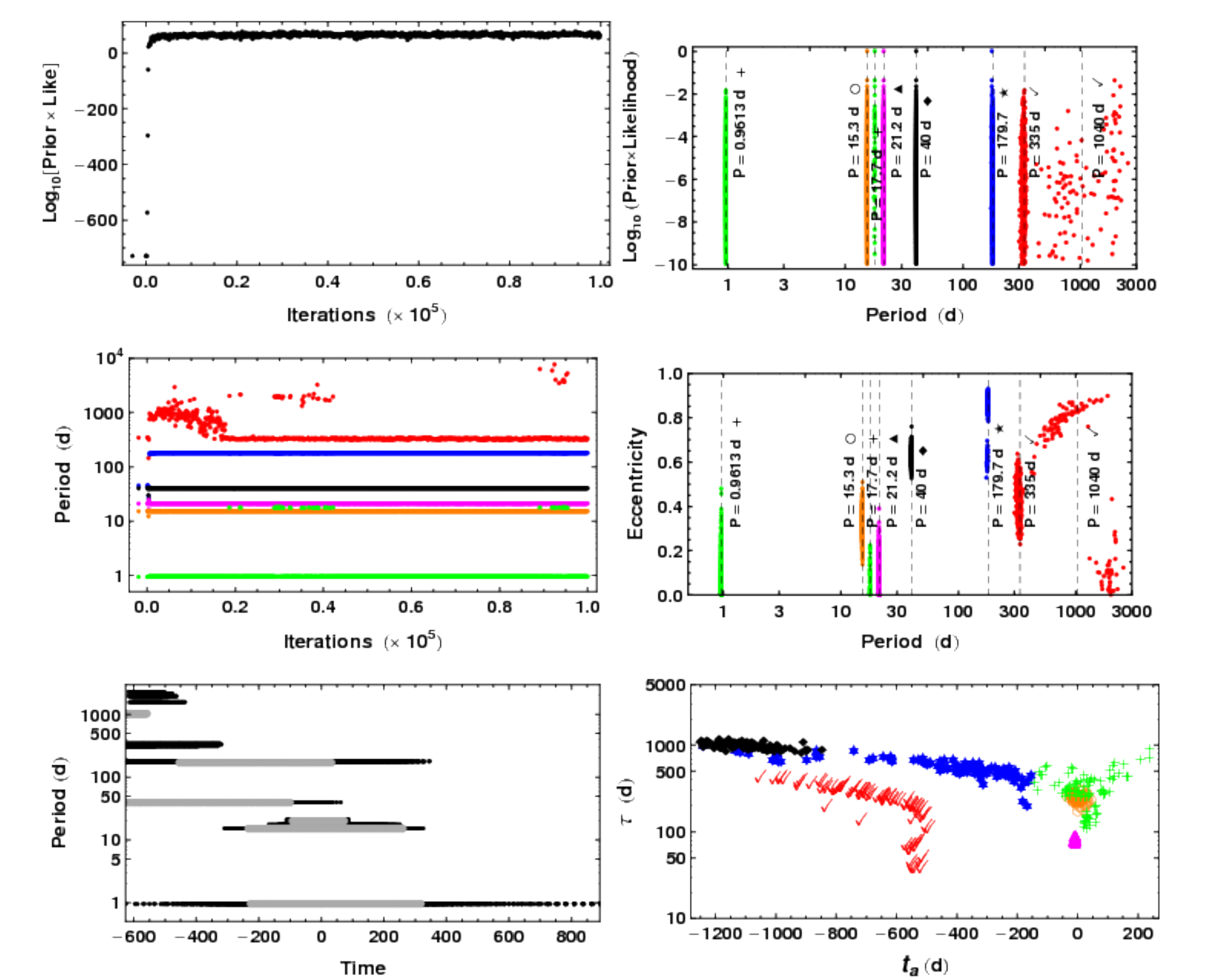}
\caption{The upper left panel is a plot of the Log$_{10}$[Prior $\times$ Likelihood] versus iteration for the 6 signal AK periodogram of the RV 5 data. The upper right shows Log$_{10}$[Prior $\times$ Likelihood] versus period indicating multiple 6 periods solutions. The middle left shows the values of the unknown period parameters versus iteration number. The middle right shows the eccentricity parameters versus period parameters. The lower left shows the apodization window for each signal (gray trace for MAP values of $\tau$ and $t_a$, black for a representative set of samples which is mainly hidden below the gray). The lower right is a plot of the apodization time constant, $\tau$, versus apodization window center time, $t_a$.\label{fig:RV5_ApodKep6_grams}}
\end{center}
\end{figure*}

Fig.~\ref{fig:RV5_ApodKep6_grams} shows FMCMC results for a six AK signal model. The upper left panel is a plot of the Log$_{10}$[Prior $\times$ Likelihood] versus iteration for the 6 signal AK periodogram of the RV 5 data. The upper right shows Log$_{10}$[Prior $\times$ Likelihood] versus period indicating multiple 6 periods solutions. The middle left panel which plots period versus MCMC iteration indicates that the $0.9613, 15.3, 21.2, 40,179.7, 335$ d solution is dominant. The middle right shows the eccentricity parameters versus period parameters. The lower left shows  the span of the apodization window within the overall data window for each signal (gray trace for MAP values of $\tau$ and $t_a$, black for a representative set of samples). The lower right is a plot of the apodization time constant, $\tau$, versus apodization window center time, $t_a$.

Table~\ref{tab:signalsRV5} lists (in order of extraction) the signals, their nominal parameter values and designation of signal type as planetary (P) or stellar activity (SA) based on the MCMC parameter estimates. The weighted RMS residual $= 1.18$ m/s for the 6 AK signal model. The standard deviation of the extra Gaussian white noise term $s = 0.98$ m/s. Only one signal (P = 0.9610 d) was initially classified as a possible planet although an examination of the apodization window span (lower left panel in Fig.~\ref{fig:RV5_ApodKep6_grams}) indicates this classification is very doubtful. Our Bayes factor analysis concluded that 0.9610 d signal is not significant. 
\begin{table}
 \centering
  \caption{Order of signals extracted from RV 5 together with approximate parameter values and designation of signal type as planetary (P) or stellar activity (SA) .}
  \label{tab:signalsRV5}
  \begin{tabular}{@{}llllll@{}}
  \hline
   Period   & K  & ecc & $\tau$ & $t_a$ & Signal\\
 (d) & (m/s) & & (d) & (d)& type\\
\hline
21.14 & 12.6 & 0.07 & 82 & -10 & SA \\
& & & & & \\
40.01 & 3.5 & 0.62 & 1024 & -1096 & SA \\
& & & & & \\
328 & 3.6 & 0.42 & 284 & -702 & SA \\
& & & & & \\
179 & 3.4 & 0.86 & 563 & -321 & SA \\
& & & & & \\
15.279 & 2.7  & 0.33 & 264 & 6 & SA \\
& & & & & \\
0.9610 & 0.73 & 0.09 & 406 & 30 & P? \\
& & & & & \\
\hline
\end{tabular}
\end{table}

To test the significance of a possible $P = 0.9610$ d planetary candidate,  we computed the Bayes factor comparing the following models: (a) 5 AK signals plus the $\log(R'hk)$ regression, and (b) 5 AK signals plus one Keplerian (0.961 d) plus the $\log(R'hk)$ regression. The periods of the 5 apodized planets are the first 5 listed in Table~\ref{tab:signalsRV5}. The Bayes factor of (a) relative to model (b) is $7.9 \times 10^5:1.0$ which clearly indicates that the 0.961 d signal is not significant. On the basis of this analysis, we cannot make a good case for any planetary signal in this data set. 

What can we learn looking back at our analysis from knowledge of the true planetary signals used in the simulation? Table~\ref{tab:SIMsignalsRV5} shows the true $P, M, K, e$ parameters of the 6 planetary signals that were employed in the RV 5 simulation. All of the injected signals have $K$ values significantly $< 1$. The 0.961 d signal is consistent with a sidereal alias of 26.4237 d which is close to the value of the true $26.2$ d period. If our signal at 179 d is a consequence of the true 173 d signal then it is apparent that for a true amplitude of 0.59 m/s the apodizing model is failing to discern this as a planetary signal. The weighted RMS residual $= 1.18$ m/s of the 6 AK signal model while the mean measurement uncertainty is 0.674 m/s. 
\begin{table}
 \centering
  \caption{The true $P, M, K, e$ parameters of the 6 planetary signals that were employed in the RV 5 simulation.}
  \label{tab:SIMsignalsRV5}
  \begin{tabular}{@{}llll@{}}
  \hline
   Period (d)  & M (Me) & K (m/s) & ecc \\
\hline
14.6632 & 2.1000 & 0.65 &  0.1680 \\
& & & \\
26.2000 & 1.7000 & 0.44 &0.2500\\
& & & \\
34.6548 &3.0380 & 0.69 &  0.0294 \\
& & & \\
173.1636 & 4.4290 & 0.59 & 0.0515 \\
& & & \\
283.1000 & 3.5000 & 0.41 & 0.3000 \\
& & & \\
616.3200 & 6.3360 & 0.55 & 0.0288 \\
\hline
\end{tabular}
\end{table}

\section{Discussion}
\label{sec:Discussion}

We now turn to consider how successful the AK method has been in the 5 challenge simulations that were analyzed in this way. Out of a total of 10 planetary signals with $K >1$ m/s, the method correctly identified 6 as class (P), 2 as (P?) and claimed another as (P) which turned out to be the first harmonic of a true planet with period longer than the data set. It also correctly identified the single planet in the initial test data set. It missed a true planet with a $P = 596$ d and $K = 1.91$  m/s. The AK method did not successfully detect any of the 7 planets with $K < 1$ m/s. Although the mean measurement error for all 5 of the data sets was 0.674 m/s, the fit residuals ranged from 1.2 to 1.7 m/s. 

On the other hand the AK method did not lead to any false detections. Two signals were initially listed as (P?) in data set 4 but one was strongly ruled out by the Bayes factor. For the other, the differential GLS periodogram played an important role in restricting the classification to (P?). Another signal was initially listed as (P?) in data set 5 but again was strongly ruled out by the Bayes factor. 

For the challenge data sets analyzed to date, the standard deviation of the data is clearly dominated by SA. Table~\ref{tab:summary} summarizes the performance of the AK approach regarding its ability to penetrate the fog of SA. The last column gives the ratio of the initial data standard deviation to the standard deviation of the final residuals which has an average value of $5.9$, however, the residuals are on average a factor of 2.3 times the mean measurement uncertainty. The average reduction that is due to the linear regression term alone is a factor of 2.5. 
\begin{table*}
 \centering
  \caption{Summary statistics.}
  \label{tab:summary}
  \begin{tabular}{@{}ccccccc@{}}
  \hline
Data set  &  \# Signals   & Initial  & Regression & Final Residual  & Mean meas. & $\sigma_I/\sigma_R$  \\
 & extracted  & $\sigma_I$  (m/s)  & residual (m/s)  & $\sigma_R$ (m/s) & error (m/s) &  \\
\hline
Test & 5 & 8.55 & 2.7 & 1.35 & 0.5 & 6.3 \\
& & & & & & \\
RV 1 & 6 & 5.6  & 3.0 & 1.44 & 0.67 & 3.8 \\
& & & & & & \\
RV 2 & 8 & 8.58 & 4.0 & 1.42 & 0.67 & 6.0 \\
& & & & & & \\
RV 3  & 6 & 10.85 & 5.5 & 1.79 & 0.67 & 6.1 \\
& & & & & & \\
RV 4 & 8 & 8.3 & 3.3 & 1.52 & 0.67 & 5.5 \\
& & & & & & \\
RV 5 & 6 & 8.92 & 2.6 & 1.18 & 0.67 & 7.6 \\
\hline
\end{tabular}
\end{table*}

\subsection{Accuracy of $K$ and $P$ parameter estimates}
\label{sec:bias}

\begin{figure}
\begin{center}
\includegraphics[width=75mm]{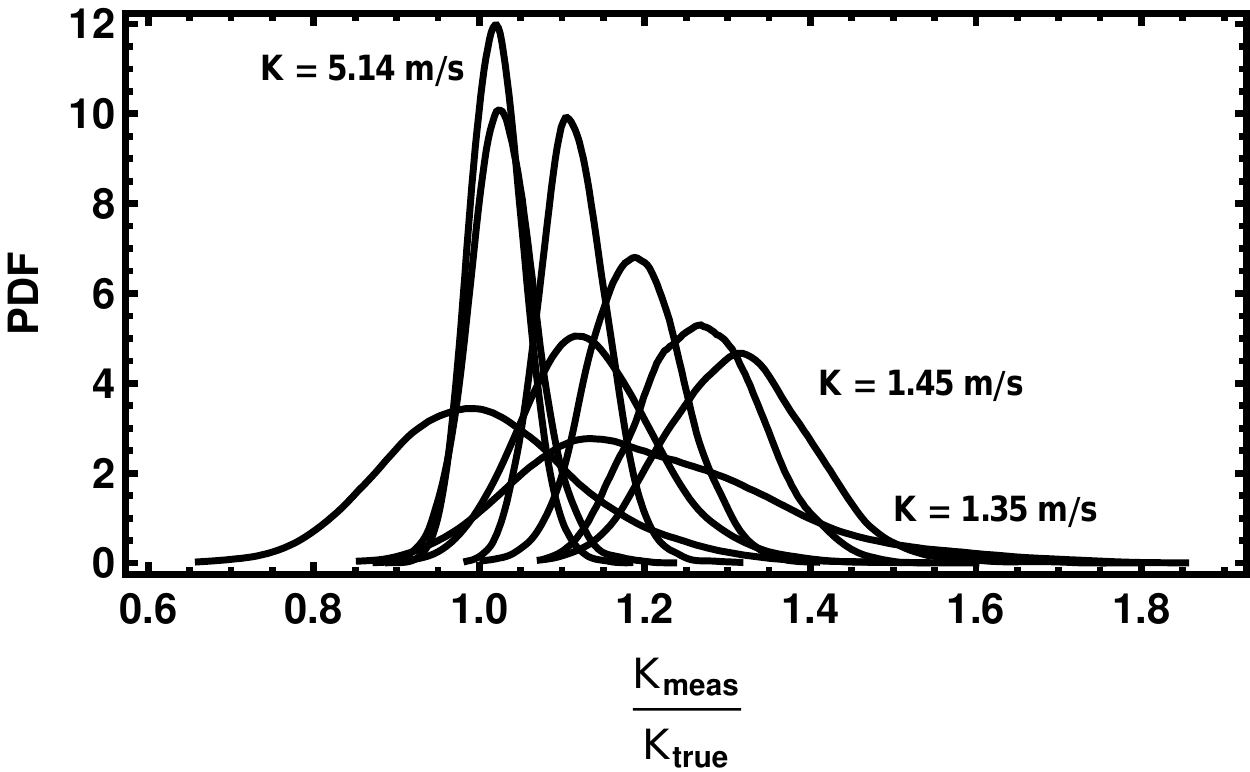}
\caption{Individual $K$ parameter marginal distributions for the 9 planets detected. For each data set, the results are based on model fits consisting of $m$ apodized Keplerians plus a $\log(R'hk)$ regression, where $m$ is the total number of signals fit. The $K$ parameter values have been divided by their true values.\label{fig:individMargKapodmeasKtrue}}
\end{center}
\end{figure}

As was mentioned in Section~\ref{sec:results}, the AK method was employed to distinguish between SA and planetary candidates. Fig.~\ref{fig:individMargKapodmeasKtrue} shows the individual $K$ parameter marginal distributions for the 8 planets detected plus one test data set planet based on AK model fits plus a $\log(R'hk)$ regression term. The $K$ parameter values have been divided by their true values. It is clear that the AK estimates have a bias to larger values.
 
\begin{figure}
\begin{center}
\includegraphics[width=75mm]{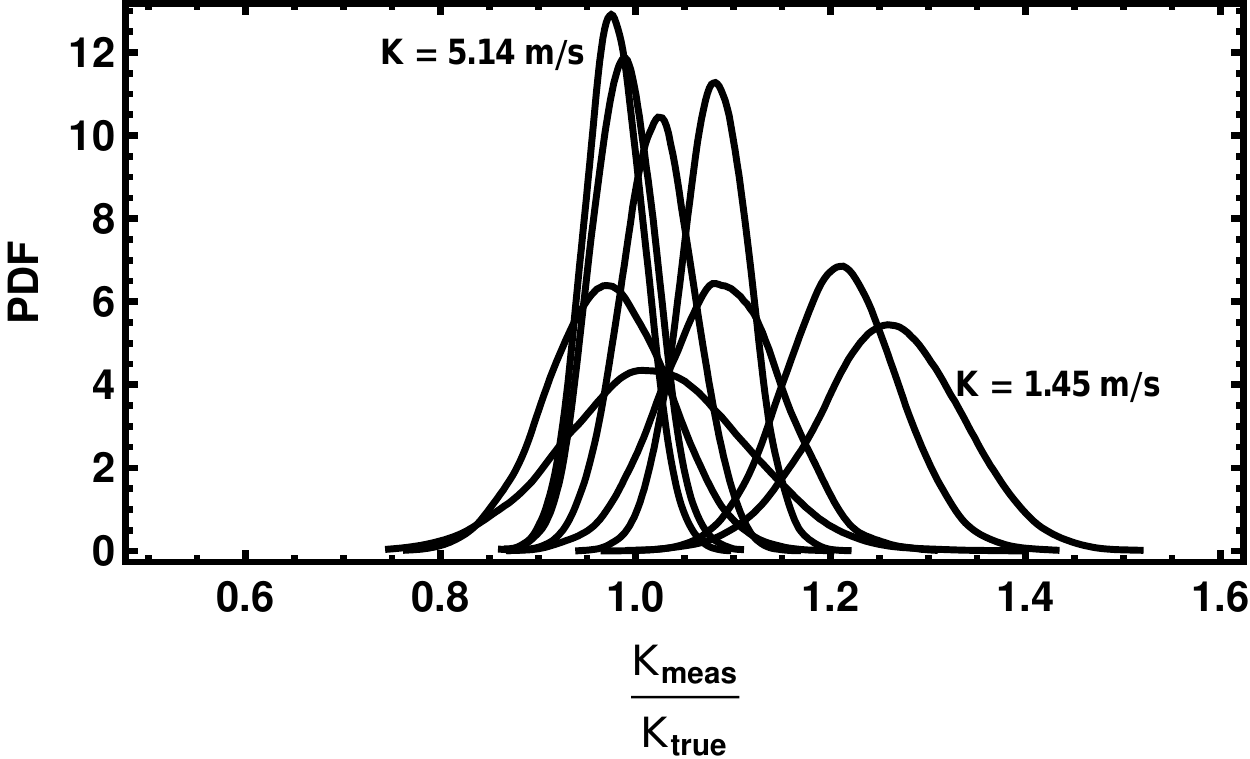}
\caption{Individual $K$ parameter marginal distributions for the 9 planets detected. For each data set, the results are based on model fits consisting of $k$ apodized Keplerians plus $(m-k)$ Keplerians plus a $\log(R'hk)$ regression, where $m$ is the total number of signals. The $K$ parameter values have been divided by their true values.\label{fig:individMargKmeasKtrue}}
\end{center}
\end{figure}

Fig.~\ref{fig:individMargKmeasKtrue} shows the individual $K$ parameter marginal distributions for the 8 planets detected plus the one test data set planet, based on model fits consisting of $k$ apodized Keplerians plus $m-k$ Keplerians plus a $\log(R'hk)$ regression. $m$ is the total number of signals and $(m-k)$ is the number classified as a planetary candidate in the analysis. In this case there is no longer evidence for a bias.
\begin{figure}
\begin{center}
\includegraphics[width=86mm]{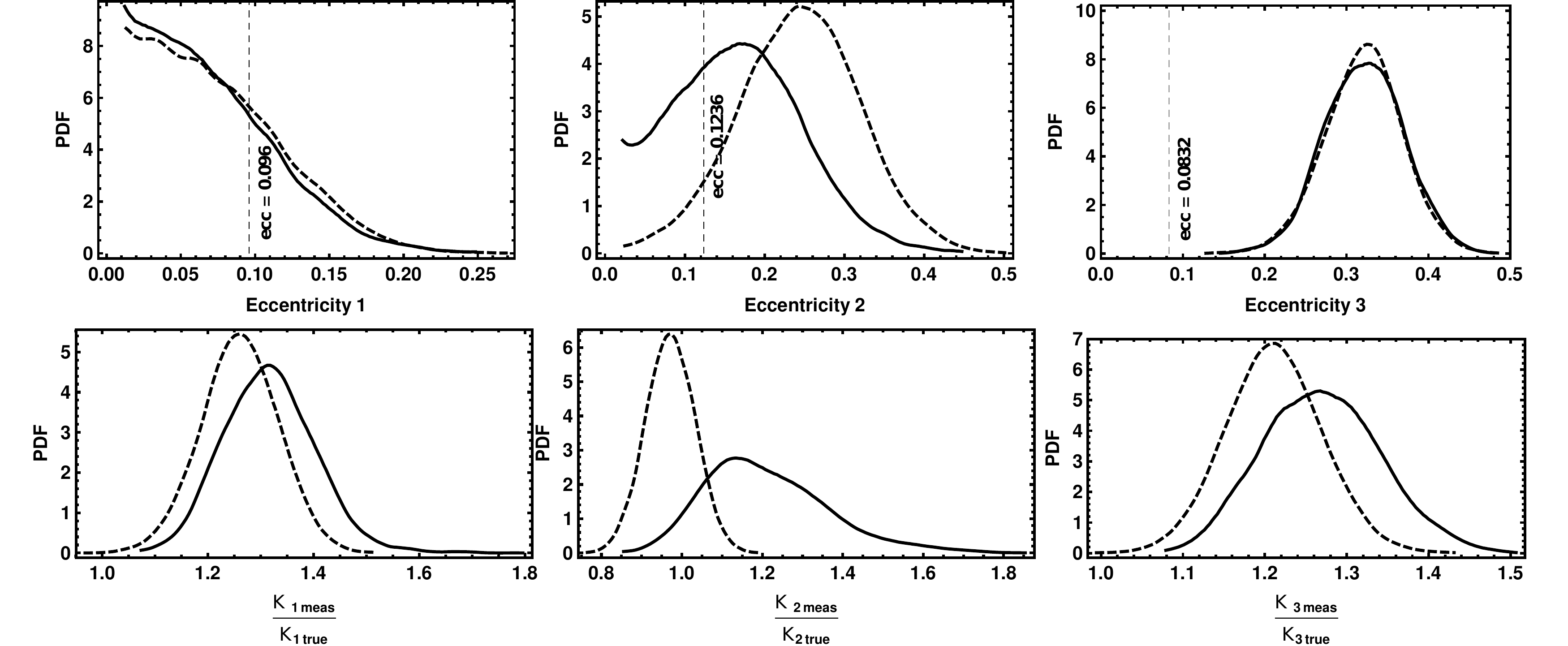}
\caption{The top row of panels of  show a comparison of the marginal eccentricity distributions for the three planets detected in the RV 1 data set. The solid curves are the distributions obtained with the AK model and the dashed are for the pure Kepler model. The true eccentricities are indicated by the dashed lines. The lower row of panels show the corresponding comparison for $K$ values.\label{fig:eccAKvseccK}}
\end{center}
\end{figure}

There is a known bias towards higher eccentricity values for Keplerian models of RV data, e.g., \cite{Shen2008}, \cite{Zakamska2011}.  Is it really true that the AK models are biased towards higher K values, or do they have a bias towards higher e, which would also inflate the K values? The top row of panels in Fig.~\ref{fig:eccAKvseccK} show a comparison of the marginal eccentricity distributions for the three planets detected in the RV 1 data set. The solid curves are the distributions obtained with the AK model and the dashed are for the pure Kepler model. If anything, the eccentricity bias of the AK model fits is towards lower values when compared to the pure Kepler fits. The lower row of panels show the corresponding comparison for $K$ values for the RV 1 planets.

\begin{figure}
\begin{center}
\includegraphics[width=75mm]{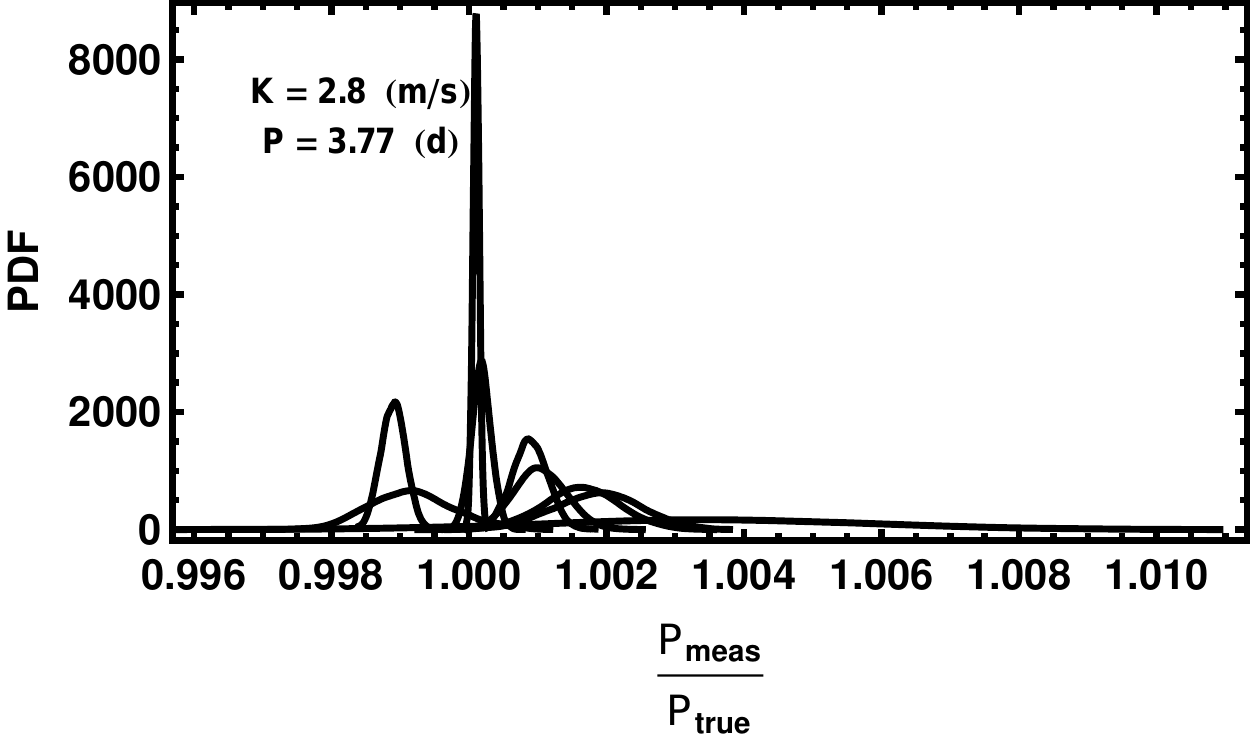}
\caption{Individual $P$ parameter marginal distributions for the 9 planets detected. For each data set, the results are based on model fits consisting of $m$ apodized Keplerians plus a $\log(R'hk)$ regression, where $m$ is the total number of signals fit. The $P$ parameter values have been divided by their true values.\label{fig:individMargPapodmeasPtrue}}
\end{center}
\end{figure}

\begin{figure}
\begin{center}
\includegraphics[width=75mm]{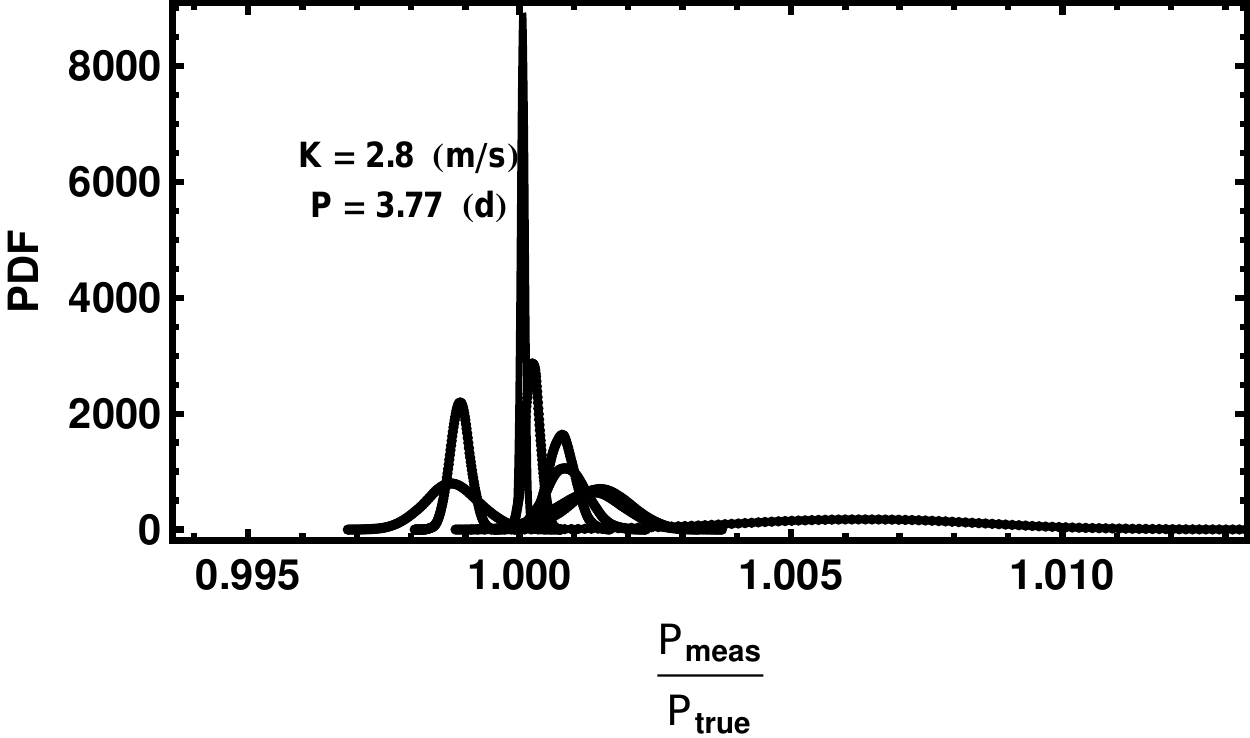}
\caption{Individual $P$ parameter marginal distributions for the 9 planets detected. For each data set, the results are based on model fits consisting of $k$ apodized Keplerians plus $(m-k)$ Keplerians plus a $\log(R'hk)$ regression, where $m$ is the total number of signals. The $P$ parameter values have been divided by their true values.\label{fig:individMargPmeasPtrue}}
\end{center}
\end{figure}

Figures~\ref{fig:individMargPapodmeasPtrue} and \ref{fig:individMargPmeasPtrue} show similar plots for the period parameter. In Figure~\ref{fig:individMargPmeasPtrue} only two of the 9 marginals overlap the true period within the $68\%$ credible region. This could be a consequence of the remaining undetected planets with smaller $K$ values and/or inadequate modeling of the SA signals, leading to correlated residuals.

\subsection{Inclusion of correlated residuals using a moving average term}
\label{sec:MA}

The RV Challenge winning team of Mikko Tuomi and Anglada Escude included a first order moving average (MA) in their Keplerian analysis along with a correlation with all SA observables. It therefore seemed logical to explore whether augmenting our apodized Keplerian models with a MA term would lead to a noticeable improvement in results.
To investigate we did several runs on the RV 3 data set that included a MA term \citep{Tuomi2013} up to third order.  Our general $p^{\rm th}$ order MA model RV is given by $z(t_i)$ in Equation~\ref{eq:lag}.
\begin{eqnarray}
z(t_i) & = & y(t_i) +\sum_{\eta=1}^p \Big[\gamma_{\eta} \times Exp \Big(-\frac{(t_i-t_{i-\eta})}{\lambda_{\eta}}\Big) \nonumber\\*
& \times & (v(t_{i-\eta}) - y(t_{i-\eta}))\Big],
\label{eq:lag}
\end{eqnarray}
where $y (t_i)$ is given by Equation~(\ref{eq:orbit1}), $v(t_i)$ is the measured radial velocity, $p$ is the order of the moving average, and $\gamma_{\eta}$ and $\lambda_{\eta}$ are the unknown MA parameters for $\eta = 1$ to $p$.
\begin{figure}
\begin{center}
\includegraphics[width=70mm]{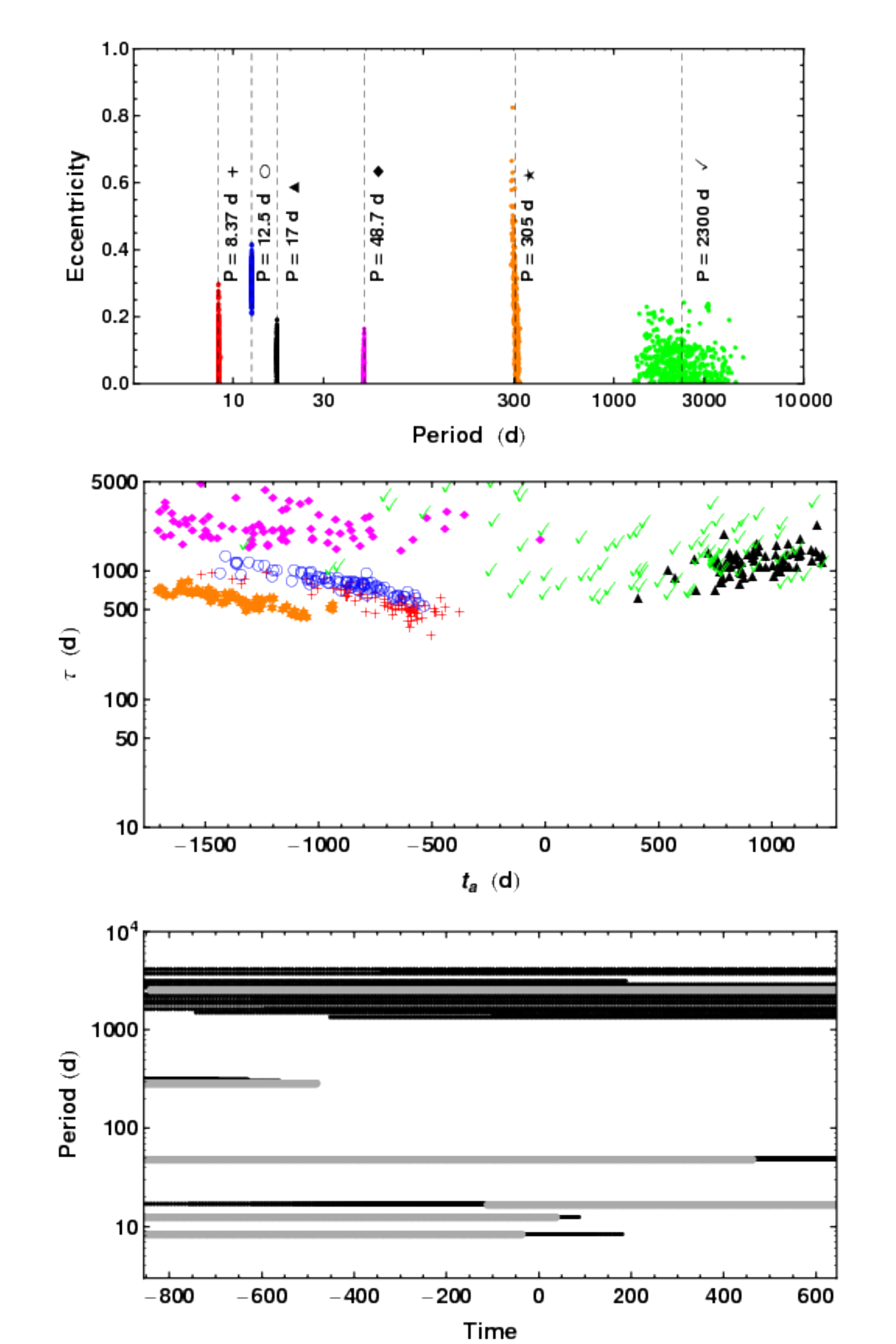}
\caption{The upper panel shows the eccentricity versus period parameters for an RV 3 six signal AK model plus the $\log(R'hk)$ regression and includes a second order moving average term. The middle panel shows a plot of the apodization time constant, $\tau$, versus apodization window center time, $t_a$, where the symbols refer to different signal periods specified in the panel above. The lower panel shows the span of the apodization window within the overall data window for each signal (gray trace for MAP values of $\tau$ and $t_a$, black for a representative set of samples which is partially hidden below the gray). \label{fig:RV3_ApodKep6MA2lag_grams}}
\end{center}
\end{figure}
\begin{figure}
\begin{center}
\includegraphics[width=70mm]{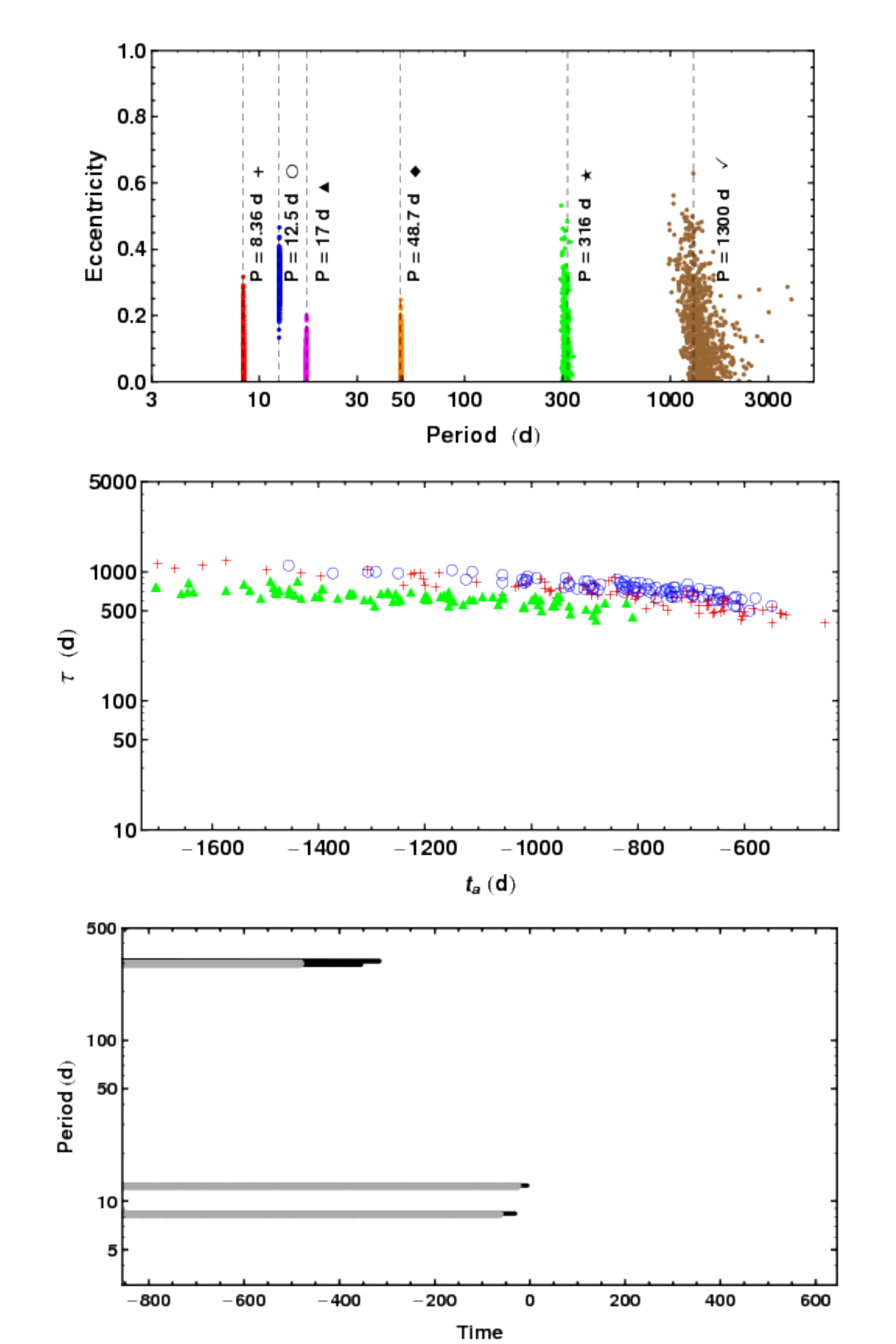}
\caption{The upper panel shows the eccentricity versus period parameters for an RV 3 six signal model consisting of 3 AK signals, 3 Keplerian signals plus the $\log(R'hk)$ regression and includes a second order moving average term. The middle panel shows a plot of the apodization time constant, $\tau$, versus apodization window center time, $t_a$, where the symbols refer to different signal periods specified in the panel above. The lower panel shows the apodization window for each signal (gray trace for MAP values of $\tau$ and $t_a$, black for a representative set of samples which is partially hidden below the gray). \label{fig:RV3_3AK3KepMA2lag_grams}}
\end{center}
\end{figure}

Of the three MA models, the second order MA was judged to yield the best results. The upper panel of Figure~\ref{fig:RV3_ApodKep6MA2lag_grams} shows the eccentricity versus period parameters for an RV 3 six signal AK model plus $\log(R'hk)$ regression which includes a second order moving average term. It exhibits a broad low eccentricity peak centered on a $P = 2300$ d. Recall the true period is 2315 d. The other periods our similar to our earlier results with no MA term.
The lower panel shows the span of the apodization window within the overall data window for each signal (gray trace for MAP values of $\tau$ and $t_a$, black for a representative set of samples which is partially hidden below the gray). Although, the apodization window for the $P = 2300$ d signal spans the data the coverage for the $P = 17$ and 48 d are judged to be not as good as that obtained for the no MA case. The middle is a plot of the apodization time constant, $\tau$, versus apodization window center time, $t_a$. 

The upper panel of Figure~\ref{fig:RV3_3AK3KepMA2lag_grams} shows the eccentricity versus period parameters for an RV 3 six signal model consisting of 3 AK signals, 3 Keplerian signals plus $\log(R'hk)$ regression and includes a second order moving average term. The results are very similar to the no MA case. 
The lower panel shows the span of the apodization window within the overall data window for each signal (gray trace for MAP values of $\tau$ and $t_a$, black for a representative set of samples which is partially hidden below the gray). The middle panel is a plot of the apodization time constant, $\tau$, versus apodization window center time, $t_a$. The values of the MA parameters and the extra noise parameter are: $\gamma_1 = 0.59 \pm 0.05, \lambda_1 = 3.0_{-1.4}^{+0.9}$ d, $\gamma_2 = 0.35 \pm 0.05, \lambda_2 = 68_{-63}^{39}$ d, and $s = 1.39 \pm 0.05$ m/s.

\begin{figure}
\begin{center}
\includegraphics[width=80mm]{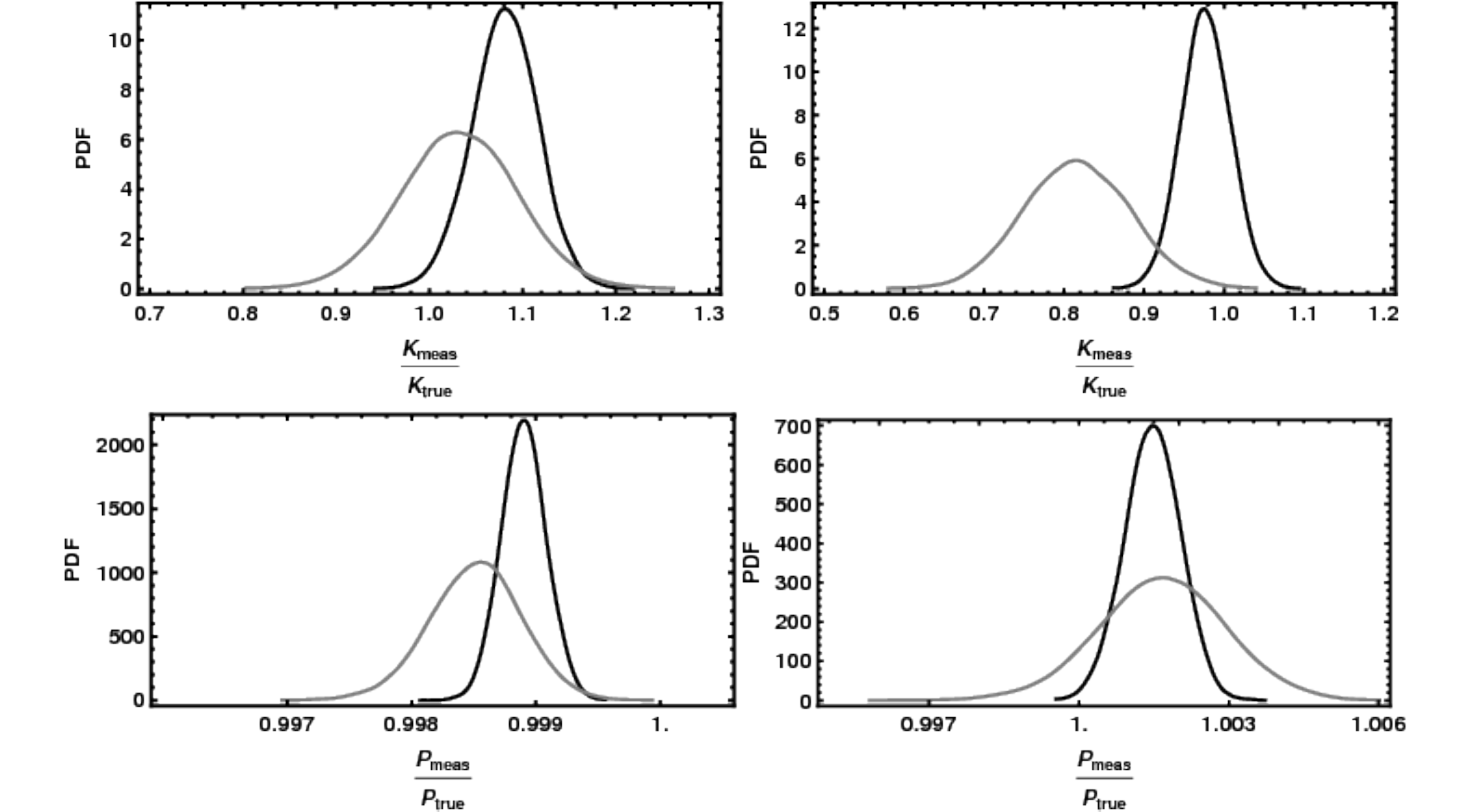}
\caption{The upper panels shows $K$ parameter marginal distributions for the 17 d and 48 d planets detected in RV3. The black trace is for the no MA case and the gray corresponds to the second order MA case. The lower panels show the same for the $P$ parameter. For each data set, the results are based on model fits consisting of $k$ apodized Keplerians plus $(m-k)$ Keplerians plus a $\log(R'hk)$ regression, where $m$ is the total number of signals. The $K$ parameter values have been divided by their true values. \label{fig:RV3_3AK3KepMA2lag_grams_compare}}
\end{center}
\end{figure}
Finally, we show a comparison of $K$ and $P$ parameter marginal distributions for the no MA case (black) and the second order MA case (gray). The upper panels shows $K$ parameter marginal distributions for the 17 d and 48 d planets detected in RV 3. The lower panels show the same for the $P$ parameter. For each data set, the results are based on model fits consisting of $k$ apodized Keplerians plus $(m-k)$ Keplerians plus a $\log(R'hk)$ regression, where $m$ is the total number of signals and $(m-k)$ is the number classified as a planetary candidate. The $K$ parameter values have been divided by their true values. As expected the marginal distributions are significantly broader for the MA case.

The second order MA analysis was repeated using a wider prior search range for $\tau$ extending from 10 d to eight times the data duration. This resulted in a change in period of one of the SA signals from 8.3 to 29 d (with a $\tau  = 10$ d) and a change in the parameters of the 17 d signal such that apodization window no longer spanned the data duration. Based on this very limited comparison, it is not clear that including a MA term has improved our ability to distinguish SA and planetary signals in this one particularly challenging data set. For the RV 3 dataset, the MA analysis suggests there is a significant correlation in the residuals of the six AK signal fit.

\section{Conclusions}
\label{sec:Conclusions}

The results reported in this paper indicate that an apodized Keplerian (AK) model provides a useful way to distinguish planetary signals from stellar activity (SA) induced signals in precision RV data. The general model for $m$ apodized Keplerian signals includes a linear regression term between RV and the stellar activity diagnostic $\log(R'hk)$, as well as an extra Gaussian noise term with unknown standard deviation. In the current implementation, the AK method achieved a reduction in SA noise by a factor of approximately 6. The analysis also employed a differential version of the Generalized Lomb-Scargle periodogram that uses a control diagnostic to provide an additional means of distinguishing SA signals and to help guide the choice of new periods. The AK model estimates for $K$ were found to have a bias to larger values. For this reason final parameter estimates for the planetary candidates were derived from fits that include AK signals to model the SA components and simple Keplerians to model the planetary candidates. The significance of planetary candidates was tested with Bayes factor model comparison. A full summary of the method is given in Section~\ref{sec:Summary}.

Preliminary results are also reported for AK models augmented by a moving average component that allows for correlations in the residuals. Based on a very limited comparison, it is not clear that including a MA term has improved our ability to distinguish SA and planetary signals.

The final residuals are still a factor of 2.3 times higher on average than the measurement uncertainties. A simple symmetrical Gaussian apodized Keplerian is only expected to provide an approximate model for some of the SA signals. Further tests employing other models are in progress.

\end{document}